\documentclass[twocolumn,tighten]{aastex61}
\usepackage{amsmath,amssymb,amstext}
\usepackage[all]{hypcap} 
\usepackage{amssymb,amsmath,amsthm}
\usepackage{longtable}
\usepackage{amssymb}
\usepackage{amsmath, xspace}
\usepackage{graphics,graphicx} 
\usepackage{rotating}
\usepackage{color}
\usepackage{multirow}

\renewcommand{\sc}{\footnotesize} 

\newcommand{\obsx}{\ensuremath{\mathcal{X}}}
\newcommand{\css}{\ensuremath{C_\mathrm{\obsx,SS}}}
\newcommand{\Mpiv}{\ensuremath{M_{\mathrm{piv}}}}
\newcommand{\zpiv}{\ensuremath{z_{\mathrm{piv}}}}

\newcommand{\Tx}{\ensuremath{{T_{\mathrm{X}}}}}
\newcommand{\Txe}{\ensuremath{{T_{\mathrm{X,cex}}}}}
\newcommand{\Txi}{\ensuremath{{T_{\mathrm{X,cin}}}}}

\newcommand{\Zi}{\ensuremath{Z_{\mathrm{X,cin}}}}
\newcommand{\Ze}{\ensuremath{Z_{\mathrm{X,cex}}}}
\newcommand{\Lx}{\ensuremath{L_{\mathrm{X}}}}
\newcommand{\Lxe}{\ensuremath{L_{\mathrm{X,cex}}}}
\newcommand{\Lxi}{\ensuremath{L_{\mathrm{X,cin}}}}
\newcommand{\Lxeb}{\ensuremath{L_{\mathrm{X,cex,bol}}}}
\newcommand{\Lxib}{\ensuremath{L_{\mathrm{X,cin,bol}}}}
\newcommand{\Yx}{\ensuremath{Y_{\mathrm{X}}}}
\newcommand{\Yxe}{\ensuremath{Y_{\mathrm{X,cex}}}}
\newcommand{\Yxi}{\ensuremath{Y_{\mathrm{X,cin}}}}
\newcommand{\Micm}{\ensuremath{M_{\mathrm{ICM}}}}
\newcommand{\Msun}{\ensuremath{M_{\odot}}}
\newcommand{\Mfiveoo}{\ensuremath{M_{500}}}
\newcommand{\Rfiveoo}{\ensuremath{R_{500}}}
\newcommand{\Xm}{\ensuremath{\mathcal{X}-M_{500}-z}}
\newcommand{\zm}{\ensuremath{\zeta-M_{500}-z}}
\newcommand{\tm}{\ensuremath{T_{\mathrm{X}}-M_{500}-z}}
\newcommand{\tme}{\ensuremath{T_{\mathrm{X,cex}}-M_{500}-z}}
\newcommand{\tmi}{\ensuremath{T_{\mathrm{X,cin}}-M_{500}-z}}
\newcommand{\mm}{\ensuremath{M_\mathrm{ICM}-M_{500}-z}}
\newcommand{\lm}{\ensuremath{L_\mathrm{X}-M_{500}-z}}
\newcommand{\lme}{\ensuremath{L_\mathrm{X,cex}-M_{500}-z}}
\newcommand{\lmi}{\ensuremath{L_{\mathrm{X,cin}}-M_{500}-z}}
\newcommand{\lmeb}{\ensuremath{L_{\mathrm{X,cex,bol}}-M_{500}-z}}
\newcommand{\lmib}{\ensuremath{L_{\mathrm{X,cin,bol}}-M_{500}-z}}
\newcommand{\ym}{\ensuremath{Y_{\mathrm{X}}-M_{500}-z}}
\newcommand{\yme}{\ensuremath{Y_{\mathrm{X,cex}}-M_{500}-z}}
\newcommand{\ymi}{\ensuremath{Y_{\mathrm{X,cin}}-M_{500}-z}}
\newcommand{\xmm}{{\it XMM-Newton}}
\newcommand{\chandra}{{\it Chandra}}

\def\apec {{\textit{apec\xspace}\  }}

\shorttitle{X-ray Scaling Relations}
\shortauthors{Bulbul et al.}

\definecolor{inon}{rgb}{1.00,0.27,0.00}

\def\ANL{Argonne National Laboratory, High-Energy Physics Division, 9700 S. Cass Avenue, Argonne, IL, USA 60439}
\def\KICPChicago{Kavli Institute for Cosmological Physics,
University of Chicago,
5640 South Ellis Avenue, Chicago, IL 60637}

\def\CfA{Harvard-Smithsonian Center for Astrophysics,
60 Garden Street, Cambridge, MA 02138}

\def\AIfA{Argelander-Institut f{\"u}r Astronomie, Auf dem H{\"u}gel 71, D-53121 Bonn, Germany}
\def\MIT{Kavli Institute for Astrophysics and Space
Research, Massachusetts Institute of Technology, 77 Massachusetts Avenue,
Cambridge, MA 02139}

\def\FNAL{Fermi National Accelerator Laboratory, Batavia, IL 60510-0500}
\def\AAUChicago{Department of Astronomy and Astrophysics,
University of Chicago,
5640 South Ellis Avenue, Chicago, IL 60637}
\def\LMU{Faculty of Physics, Ludwig-Maximilians-Universit\"{a}t, Scheinerstr.\ 1, 81679 Munich, Germany}
\def\ECUniverse{Excellence Cluster Universe, Boltzmannstr.\ 2, 85748 Garching, Germany}
\def\MPE{Max Planck Institute for Extraterrestrial Physics, Giessenbachstr. 1, 85748 Garching, Germany}
\def\UMKC{Department of Physics and Astronomy, University of Missouri, 5110 Rockhill Road, Kansas City, MO 64110}

\def\UCDavis{Physics Department, University of California, Davis, CA 95616}

\def\Berkeley{Department of Physics,
University of California, Berkeley, CA 94720}

\def\LSST{LSST, 950 North Cherry Avenue, Tucson, AZ 85719}
\def\Michigan{Department of Physics, University of Michigan, 450 Church Street, Ann  
Arbor, MI, 48109}

\def\ASIAA{Academia Sinica Institute of Astronomy and Astrophysics, 11F of AS/NTU Astronomy-Mathematics Building, No.1, Sec. 4, Roosevelt Rd, Taipei 10617, Taiwan}
\def\UdeM{D\'{e}partement de physique, Universit\'{e} de Montr\'{e}al, C.P. 6128 Succ. Centre-ville, Montr\'{e}al H3C 3J7, Canada}

\def\melbourne{School of Physics, University of Melbourne, Parkville, VIC 3010, Australia)}
\def\inaf{INAF-Osservatorio Astronomico di Trieste, via G. B. Tiepolo 11, I-34143 Trieste, Italy}
\begin{document}

\title{X-ray Properties of SPT Selected Galaxy Clusters at $0.2<\lowercase{z} <1.5$ Observed with \xmm}

\author{Esra~Bulbul}
\affiliation{\CfA}
\affiliation{\MIT}

\author{I-Non~Chiu}
\affiliation{\ASIAA}

\author{Joseph~J.~Mohr}
\affiliation{\LMU}
\affiliation{\MPE}
\affiliation{\ECUniverse}

\author{Michael~McDonald}
\affiliation{\MIT}

\author{Bradford~Benson}
\affiliation{\FNAL}
\affiliation{\AAUChicago}
\affiliation{\KICPChicago}

\author{Mark~W.~Bautz}
\affiliation{\MIT}

\author{Matthew~Bayliss}
\affiliation{\MIT}

\author{Lindsey~Bleem}
\affiliation{\ANL}
\affiliation{\KICPChicago}

\author{Mark~Brodwin}
\affiliation{\UMKC}

\author{Sebastian~Bocquet}
\affiliation{\LMU}

\author{Raffaella~Capasso}
\affiliation{\LMU}
\affiliation{\ECUniverse}

\author{J\"{o}rg~P.~Dietrich}
\affiliation{\LMU}
\affiliation{\ECUniverse}

\author{Bill~Forman}
\affiliation{\CfA}

\author{Julie~Hlavacek-Larrondo}
\affiliation{\UdeM}

\author{W.~L.~Holzapfel}
\affiliation{\Berkeley}

\author{Gourav~Khullar}
\affiliation{\KICPChicago}
\affiliation{\AAUChicago}

\author{Matthias Klein}
\affiliation{\LMU}

\author{Ralph~Kraft}
\affiliation{\CfA}

\author{Eric~D.~Miller}
\affiliation{\MIT}

\author{Christian~Reichardt}
\affiliation{\melbourne}

\author{Alex~Saro} 
\affiliation{\inaf}

\author{Keren~Sharon}
\affiliation{\Michigan}

\author{Brian~Stalder}
\affiliation{\LSST}

\author{Tim~Schrabback}
\affiliation{\AIfA}

\author{Adam~Stanford}
\affiliation{\UCDavis}

\correspondingauthor{Esra~Bulbul}
\email{ebulbul@cfa.harvard.edu}

\begin{abstract}
We present measurements of the X-ray observables of the intra-cluster medium (ICM), including luminosity $L_X$, ICM mass $M_{ICM}$, emission-weighted mean temperature $T_X$, and integrated pressure $Y_X$, that are derived from XMM-Newton X-ray observations of a Sunyaev-Zel'dovich Effect (SZE) selected sample of 59 galaxy clusters from the South Pole Telescope SPT-SZ survey that span the redshift range of $0.20 < z < 1.5$. We constrain the best-fit power law scaling relations between X-ray observables, redshift, and halo mass. The halo masses are estimated based on previously published SZE observable to mass scaling relations, calibrated using information that includes the halo mass function. Employing SZE-based masses in this sample enables us to constrain these scaling relations for massive galaxy clusters ($M_{500}\geq 3 \times10^{14}$ $M_\odot$) to the highest redshifts where these clusters exist without concern for X-ray selection biases. We find that the mass trends are steeper than self-similarity in all cases, and with $\geq 2.5{\sigma}$ significance in the case of $L_X$ and $M_{ICM}$. The redshift trends are consistent with the self-similar expectation, but the uncertainties remain large. Core-included scaling relations tend to have steeper mass trends for $L_X$. There is no convincing evidence for a redshift-dependent mass trend in any observable. The constraints on the amplitudes of the fitted scaling relations are currently limited by the systematic uncertainties on the SZE-based halo masses, however the redshift and mass trends are limited by the X-ray sample size and the measurement uncertainties of the X-ray observables.
\end{abstract}

%
\section{Introduction}
\label{sec:intro}
The evolution of the mass function of clusters of galaxies is dependent on cosmology, making clusters unique probes of fundamental cosmological parameters--- not only the normalization of the power spectrum $\sigma_{8}$ and mean matter density $\Omega_\mathrm{M}$, but also the equation of state parameter of the dark energy \citep{wang98,haiman01}. The ability to select clusters out to high redshifts and to measure their masses is particularly important for constraints on the dark energy equation of state and the growth rate of cosmic structure. 

The fully ionized intracluster medium (ICM) is heated to keV temperatures through gravitational acceleration and shocks as the cluster forms and grows. At these temperatures it emits X-rays through a combination of thermal bremsstrahlung and atomic line emission.  Serendipitous X-ray surveys with XMM enabled the detection of $z>1$ galaxy clusters \citep{fassbender11}, but the solid angle surveyed and the required optical and infrared imaging follow-up remain as challenges to this approach.  The all sky X-ray survey with ROSAT \citep[RASS;][]{voges99} has been used to define large samples of mostly low-redshift clusters \citep{boehringer04,piffaretti11}, and only now in combination with deep, large solid angle multi-wavelength optical surveys is beginning to deliver cluster samples extending to $z\approx1$ \citep{klein18}.

The ICM also distorts the cosmic microwave background (CMB) through inverse Compton scattering, known as the Sunyaev-Zel'dovich Effect \citep[SZE;][]{sunyaev72}. Large solid angle surveys employing the SZE have been carried out with the South Pole Telescope \citep[SPT;][]{carlstrom11}, {\it Planck} \citep{planck11-13}, and the Atacama Cosmology Telescope \citep[ACT;][]{fowler07}. The SZE-selected galaxy cluster sample from SPT is an approximately mass-selected sample ($\Mfiveoo\geq3\times10^{14}\Msun$) of over 500 clusters that extends to the highest redshifts at which these clusters exist \citep{bleem15}, and approximately 20~percent of the sample lies at $z>0.8$. To date, the highest redshift cluster identified in the 2500~deg$^2$ SPT-SZ survey has a redshift of $z =1.7\pm0.05$ (Strazzullo et al. in prep; Mantz et al. in prep).  

X-ray observations of SZE-selected clusters provide low-scatter mass proxies which can be used to aid in the calibration of the SZE-based cluster masses and in the cosmological analysis of the SZE cluster samples.  Pioneering observational studies have found low-scatter scaling relations that tie X-ray observables to cluster mass for X-ray selected low-redshift clusters \citep{mohr99,finoguenov01,reiprich02,arnaud07, pratt07, vikhlinin09b,mantz10b, maughan12}.  Through X-ray follow-up observations of these large samples of SZE-selected clusters, it has now become possible to extend these studies to high redshift. Moreover, by studying X-ray scaling relations in samples of SZE-selected clusters, it is possible to reduce the impact of selection-related biases that would have to be carefully corrected in studies of X-ray selected samples \citep{mantz10b}.

In this work, we leverage the previous cosmological analyses of the SPT-SZ sample to characterize the X-ray observable--mass scaling relations by utilizing \xmm\ follow-up observations of 59 SPT-selected clusters in the redshift range $0.20<z<1.5$. Here we focus on X-ray observables, which have direct implications for the structure evolution of the Universe. The halo masses we use in this analysis are derived from the observed SZE signal-to-noise ratio $\xi$ and redshift $z$ using the SZE mass--observable relation as calibrated within a self-consistent cosmological analysis that accounts for selection biases and systematic uncertainties on the masses \citep{bocquet15,deHaan16}. Employing SZE masses allows us to extend studies of scaling relations to higher redshifts, enabling more robust studies of the redshift trends in these scaling relations.
This cosmological analysis uses external mass information for a subset of clusters (i.e., weak lensing calibrated $Y_X$ measurements for 82 systems, as described in Section~\ref{sec:M500}), however inherently the cluster masses are based on the assumption that the cluster mass is well-described by the assumed functional form of the SZE-mass scaling relation and a general cluster mass function that can be well-fit to a $\Lambda$CDM cosmology.  In this context, our results are comparable to other works that have performed similar analyses that jointly constrain observable--mass scaling relations in the context of a cosmological model (e.g., \citet{mantz16}), but in our our case using a different observable (i.e., SZE vs X-ray) and cluster sample (i.e., SPT-SZ vs RASS).

In this work, we are interested in comparing the measured X-ray observable--mass scaling relations to other results in the literature, including: the self-similar expectation, results based on direct-mass measurements, and results that include cosmological information.  Agreement between results would indicate that cluster scaling relations are well-understood across a broad range of observables and assumptions, while differences could be indicative of tensions in the underlying assumptions or differences in the underlying cluster samples. 

Robust observations of cluster scaling relations and their comparison to scaling relations from structure formation simulations then allow the baryonic physics and subgrid physics in the simulations to be tested and constrained.  These constraints are crucial to accurately predicting the matter power spectrum \citep[e.g.,][]{springel18} and halo mass function \citep[e.g.,][]{bocquet16} needed to support forefront observational cosmological studies employing weak lensing, galaxy clustering and cluster counts.

The cluster sample and details of the \xmm\ data reduction are given in Section~\ref{sec:analysis}.  An explanation of the SZE-based halo masses and the measurements of the X-ray observables appears in Section~\ref{sec:xrayobs}.  In Section \ref{sec:sclrelformfit} we present our fitting procedure, and in Section~\ref{sec:sclrel} we present the X-ray scaling relations derived from this sample. Finally, we discuss our conclusions in Section \ref{sec:concl}.

All errors quoted throughout the paper correspond to $68\%$ (or $\Delta C$-stat=1) single-parameter confidence intervals unless otherwise stated.
Throughout the paper, we adopt a standard, flat $\Lambda$CDM cosmology with the latest cosmological results from \cite{deHaan16}---$H_{0} =67.74$ km s$^{-1}$ Mpc$^{-1}$, $\Omega_{M}$= 0.304, and $\sigma_{8}=$0.82.
In this work we refer to the cluster halo mass, \Mfiveoo, as the total mass within a sphere of radius \Rfiveoo. The overdensity radius \Rfiveoo\ is defined as the radius within which the mean mass density of the cluster is 500 times the critical density of the Universe at that redshift. 

\section{Sample Selection and Data Reduction}
\label{sec:analysis}
\subsection {Sample Selection}
SPT has detected 516 galaxy clusters via the SZE in the 2500 degree$^2$ SPT-SZ Survey at $0<z<1.8$ with masses $\Mfiveoo\geq 3\times10^{14}\Msun$ \citep{bleem15}. The redshifts of many of these clusters have also been reported in \citet{ruel14,bayliss16}. \xmm\ X-ray observations of 40 of these SZE-selected clusters have been performed through several programs (PIs: A. Andersson, B. Benson, J. Mohr, R. Suhada, E. Bulbul). An additional 33 clusters have been observed through various other non-SPT small programs. Five clusters have been excluded from this analysis, because one scattered below the detection threshold when better data were available (SPT-CL~J2343$-$5521), and four observations are dominated by background flares (SPT-CL~J0411$-$4819, SPT-CL~J0013$-$4906, SPT-CL~J0257$-$5732, SPT-CL~J2136$-$6307).  
\begin{figure}
\centering
\includegraphics[width=0.45\textwidth]{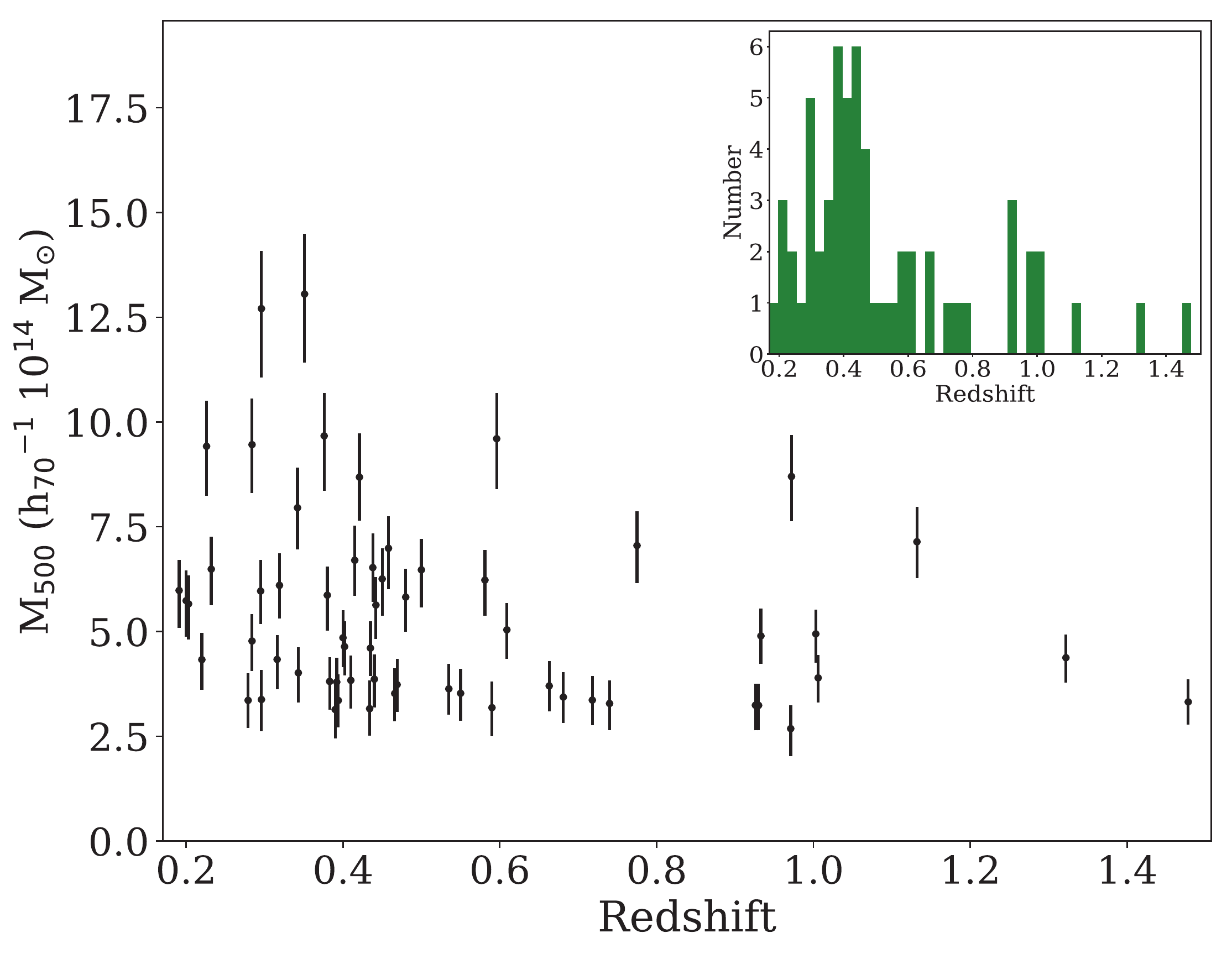}
\caption{The distribution in SZE halo mass and redshift of the SPT-selected galaxy clusters observed with \xmm\ is shown with each cluster appearing as a point with error bar. The inset shows the cluster redshift histogram.}
\label{fig:zhist}
\vspace{2mm}
\end{figure}

We exclude clusters at $z<0.20$ from the scaling relation analysis, because their SZE mass estimates obtained via the $\zeta$--\Mfiveoo\ relation (Section~\ref{sec:M500}) are impacted by the filtering adopted to remove signal from the primary CMB \citep[see e.g.,][]{benson13}. 
From this sample of 68 clusters, 59 are at redshift z$>$0.2 and have a total of 1000 or more filtered source counts in MOS observations and are therefore included in our final sample.  The details of the \xmm\ observations of these clusters are given in Table~\ref{table:obs}.

The final sample is shown in Figure~\ref{fig:zhist} in redshift-mass space with an inset redshift histogram.  This cluster sample is not a complete SZE signal-to-noise selected cluster sample.  It has a median mass and redshift of $\Mfiveoo=4.77\times10^{14}\Msun$ and $z_{\mathrm{med}}=0.45$, and five of the clusters lie at $z>1$.  Nevertheless, the sample we study here has similar median mass, median redshift, and fraction of $z>1$ clusters as the  SPT-SZ cosmology sample \cite{deHaan16} , which has a median mass
 $\Mfiveoo=4.57\times10^{14}\Msun$ (with roughly $6\%$ of the clusters at $z>1$),
although it has a lower median redshift, 0.45 vs 0.55.

\subsection {XMM-Newton Data Reduction}
\label{sec:reduction}

Our \xmm\ data reduction is described in detail in \citet{Bulbul12a}; here we summarize the main steps. 
\xmm\ EPIC-MOS data analysis is carried out with Science Analysis System (SAS) version 16.0.0 and the latest available calibration files from Feb 2017. The Extended Source Analysis (ESAS) tools are used to reduce the data and extract the final data products \citep{snowden08}. The event files are filtered from the periods with elevated backgrounds through light-curve filtering. The good time interval files are produced and used to create cleaned event lists. The net exposure time after filtering the event lists for good time intervals is given in Table \ref{table:obs}. There are three main detectors on board \xmm: MOS1, MOS2, and PN. The back illuminated PN observations can be more sensitive to proton flares compared to MOS observations\setcounter{footnote}{0}\footnote{https://heasarc.gsfc.nasa.gov/docs/xmm/uhb/epicextbkgd.html}. As a result, the majority of the PN observation of some clusters in the sample is lost due to background filtering. Additionally, \citet{schellenberger15} reports up to a $54\%$ bias in temperature measurements between {\it Chandra} and  PN temperature measurements in the soft 0.7--2~keV band where the bulk of detected photon flux from the high redshift clusters appears. To avoid creating potential biases in the X-ray observables, we only use MOS observations in this analysis.  We examine the individual chips which may be affected by an anomalous background level and exclude them from further analysis \citep{kuntz08}.

The images are created in the 0.5--2~keV band from the filtered event files and used to detect point sources within the MOS field-of-view (FOV). The images are examined carefully for point sources missed by the CIAO algorithm {\it wavdetect}. An exposure map is created for each MOS detector and each pointing to account for chip gaps and mirror vignetting. 
The quiescent particle background (QBP) image is created from the filter-wheel closed data as described in \citet{snowden08}. The images and exposure maps of MOS1 and MOS2 detectors are combined prior to the background subtraction. The CIAO tool {\it wavdetect} convolved with the \xmm's point-spread-function (PSF) is used on the background-subtracted and exposure corrected images to detect point sources within the MOS FOV.  All these point sources are  excluded from further analysis. 

We extract spectra using the ESAS tool {\it mos-spectra} within a radius of \Rfiveoo\ for each cluster (see Section \ref{sec:xrayobs} for the details of the \Rfiveoo\ calculation). Redistribution matrix files (RMFs) and ancillary response files (ARFs) are created with {\it rmfgen} and {\it arfgen}, respectively. QPB is subtracted from the total spectra prior to the fitting. The spectral fitting of the source is done in the spectral fitting package XSPEC 12.9.0 \citep{arnaud96} with ATOMDB version 3.0.8 \citep{Smith2001,Foster2012}. The adopted solar abundances are from \citet{lodders09}. The Galactic Column density is allowed to vary within $15\%$ of the measured \citet{kalberla05} LAB value in our fits, following the approach described in \citet{mcdonald16b}. We use {\it C-stat} as a goodness of the fit estimator in XSPEC. 

Spectra are extracted from two apertures of $<\Rfiveoo$ and 0.15~\Rfiveoo--\Rfiveoo\ (again, see Section~\ref{sec:xrayobs} for discussion of \Rfiveoo). The fits are performed in the 0.3--10~keV energy interval. The higher energy band 7--10~keV is used to constrain soft-proton contamination accurately. Soft-proton flares are largely removed by the light curve filtering. However, after the filtering some residuals may remain in the data. These are modeled by including an extra power-law model component to the total model and the MOS diagonal response matrices provided in the SAS distribution  \citep{snowden08}. The cluster emission is fit with an absorbed single temperature {\it apec} model with free metallicity and temperature. Constraining metallicity is challenging for low-count observations of some of our high-z clusters. 
In these cases, we fixed the metallicity at 0.3$Z_{\odot}$, the typical value at both low and high redshifts \citep{tozzi03}.

\startlongtable
\begin{deluxetable}{lcccc}\tabletypesize{\tiny}
\tablecaption{SPT clusters observed with \xmm. \label{table:obs}}
\tablehead{
\colhead{Name} & \colhead{z} &  \colhead{Obs. ID}  & \colhead{Exposure [ks]} & \colhead{Counts}\\
\colhead{} & \colhead{} &  \colhead{}  & \colhead{MOS1/MOS2} & \colhead{MOS1/MOS2}
}
\startdata
SPT-CLJ0114-4123 & 0.38 & 724770901 & 12.46 / 12.64 & 2350 / 2314 \\
				&  & 404910201 & 16.31 / 16.78  \\
  SPT-CLJ0205-5829 & 1.32 & 675010101 & 55.86 / 57.14 &1378 / 1299 \\
  SPT-CLJ0217-5245 & 0.34 & 652951401 & 8.91 / 14.52 & 1280 / 2052 \\
  SPT-CLJ0225-4155 & 0.22 & 692933401 & 12.40 / 11.92 & 7650 / 7257\\
  SPT-CLJ0230-6028 & 0.68 & 675010401 & 19.37 / 22.72 &  875 / 921\\
  SPT-CLJ0231-5403 & 0.59 & 204530101 & 17.38 / 22.17  &719 / 843 \\
  SPT-CLJ0232-4421 & 0.28 & 423403010 & 11.59 / 12.09 &  7917 / 8183 \\
  SPT-CLJ0233-5819 & 0.66 & 675010601 & 49.14 / 50.01 &  2396 / 2370\\
  SPT-CLJ0234-5831 & 0.42 & 674491001 & 12.53 / 13.47 &2182 / 2333\\
  SPT-CLJ0240-5946 & 0.40 & 674490101 & 14.38 / 14.03 &  1852 / 1733\\
  SPT-CLJ0243-4833 & 0.50 & 672090501 & 9.80 / 9.74 &2078 / 2002 \\
   &  & 723780801 & 12.40 / 11.31 &\\
  SPT-CLJ0254-5857 & 0.44 & 656200301 & 11.62 / 13.17& 3145 / 3501 \\
  				 &  & 674380300 & 11.62 / 13.17  \\
  SPT-CLJ0254-6051  & 0.44  & 692900201 & 16.01 / 15.65  & 1516 / 1399\\
  SPT-CLJ0257-5732 & 0.43 & 674491101 & 27.31 / 27.95 & 886 / 859 \\
  SPT-CLJ0304-4401 & 0.46 & 700182201 & 16.57 / 16.75 &3439 / 3429 \\
  SPT-CLJ0317-5935 & 0.47 & 674490501 & 7.83 / 10.82 &  1615 / 1572 \\
  				 &  & 724770401 & 14.73 / 14.80 \\
  SPT-CLJ0330-5228 & 0.44 & 400130101 & 69.12 / 67.81 & 24535 / 24023\\
  SPT-CLJ0343-5518 & 0.55 & 724770801 & 17.91 / 18.03 &988 / 918 \\
  SPT-CLJ0344-5452 & 1.00 & 675010701 & 48.74 / 48.94 & 769 / 735\\
  SPT-CLJ0354-5904 & 0.41 & 724770501 & 14.19 / 16.87 &1669 / 1931 \\
  SPT-CLJ0403-5719 & 0.46 & 674491201 & 18.40 / 19.94	& 1900 / 1977\\
  SPT-CLJ0406-5455 & 0.74 & 675010501 & 53.25 / 55.69 &  1611 / 1646  \\
  SPT-CLJ0417-4748 & 0.58 & 700182401 & 21.86 / 23.39 & 2590 / 2736\\
  SPT-CLJ0438-5419 & 0.42 & 656201601 & 17.87 / 17.87 & 4904 / 4857 \\
  SPT-CLJ0510-4519 & 0.20 & 692933001 & 12.80 / 13.05  &7838 / 7901\\
  SPT-CLJ0516-5430 & 0.29 & 205330301 & 10.41 / 10.67 &14848 / 14812\\
				 &  & 692934301 & 26.94 / 26.97 \\
  SPT-CLJ0522-4818 & 0.29 & 303820101 & 11.57 / 15.00 & 2631 / 3314 \\
    SPT-CLJ0549-6205 & 0.37 & 656201301 & 13.17 / 13.00  &7835 / 7688\\
  SPT-CLJ0559-5249 & 0.61 & 604010301 & 16.64 / 17.39 & 1800 / 1756\\
  SPT-CLJ0611-5938 & 0.39 & 658201101 & 12.91 / 13.18 &1629 / 1591 \\
  SPT-CLJ0615-5746 & 0.97 & 658200101 & 12.59 / 13.31 & 1587 / 1613\\
  SPT-CLJ0637-4829 & 0.20 & 692933101 & - / 11.81 & 6643 / 5859\\
  SPT-CLJ0638-5358 & 0.23 & 650860101 & 24.77 / 31.65  & 21985 / 27911 \\
  SPT-CLJ0658-5556 & 0.29 & 112980201 & 21.50 / 21.66 & 26213 / 26326\\
  SPT-CLJ2011-5725 & 0.28 & 744390401 & 17.07 / 17.53 & 3981 / 4065 \\
  SPT-CLJ2017-6258 & 0.53 & 674491501 & 25.43 / 25.42 & 1428 / 1322 \\
  SPT-CLJ2022-6323 & 0.38 & 674490601 & 14.33 / 14.21  & 2129 / 2015\\
  SPT-CLJ2023-5535 & 0.23& 069293370 & 2.93 / 4.31 & 1942 / 2627\\
  SPT-CLJ2030-5638 & 0.39 & 724770201 & 20.82 / 21.05 & 1393 / 1447\\
  SPT-CLJ2031-4037 & 0.34 & 690170701 & 10.25 / 10.20 & 3553 / 3447\\
  SPT-CLJ2032-5627 & 0.28 & 674490401 & 24.67 / 25.32 &10032 / 10221  \\
  SPT-CLJ2040-5725 & 0.93 & 675010201 & 75.08 / 76.75 & 1916 / 1919\\
  SPT-CLJ2040-4451 & 1.48 & 723290101 & 75.96 / 75.37 & 844 / 775 \\
  SPT-CLJ2056-5459 & 0.72 & 675010901 & 40.11 / 39.58 & 1060 / 990 \\
  SPT-CLJ2106-5844 & 1.13 & 744400101 & 39.10 / 45.70 & 3035 / 3501\\
  SPT-CLJ2109-4626 & 0.97 & 694380101 & 52.36 / 55.57  & 713 / 744 \\
  SPT-CLJ2124-6124 & 0.44 & 674490701 & 14.00 / 14.36 &  1365 / 1416\\
  SPT-CLJ2130-6458 & 0.31 & 069290010 &  6.3 / 8.5  & 1108 / 1450\\
  SPT-CLJ2131-4019 & 0.45 & 724770601 & 12.50 / 12.73 & 2598 / 2600 \\
  SPT-CLJ2136-6307 & 0.93 & 675010301 & 56.68 / 59.69  & 2465 / 2499\\
  SPT-CLJ2138-6008 & 0.32 & 674490201 & 12.80 / 14.12  & 2918 / 3253\\
  SPT-CLJ2145-5644 & 0.48 & 674491301 & 10.12 / 10.65 & 1442 / 1443\\
  SPT-CLJ2146-4633 & 0.93 & 744401301 & 70.40 / 74.13 & 2370 / 2434\\
  &  & 744400501 & 93.33 / 96.20  \\
  SPT-CLJ2200-6245  & 0.39 & 724771001 & 9.55 / 10.53  &  623 / 631 \\
  SPT-CLJ2248-4431 & 0.35 & 504630101 & 25.23 / 26.25 &  24646 / 25604\\
  SPT-CLJ2332-5358 & 0.40 & 604010101 & 6.82 / 6.82 & 1434 / 1443  \\
  SPT-CLJ2337-5942 & 0.77 & 604010201 & 18.36 / 19.32 & 1893 / 1952\\
  SPT-CLJ2341-5119 & 1.00 & 744400401 & 72.63 / - & 2788 / 3145\\
  SPT-CLJ2344-4243 & 0.59 & 722700101 & 108.58 / 110.77 &  32697 / 33314 \\
   & & 722700201 & 87.18 / 87.01\\
   & & 693661801 & 12.96 / 13.44  \\

\enddata
\vspace{0.5cm}
\end{deluxetable}

We also consider the X-ray foreground emission, including Galactic halo, local hot bubble, cosmic X-ray background due to unresolved extragalactic sources, and solar wind charge exchange. The ROSAT All-Sky Survey background spectra\footnote{https://heasarc.gsfc.nasa.gov/cgi\-bin/Tools/xraybg/xraybg.pl} extracted beyond $R_{\mathrm{vir}}$ (discussed in Section~\ref{sec:xrayobs}) are used to model the soft X-ray background as described in \cite{Bulbul12a}. The soft X-ray emission from the local hot bubble is modeled with a cool unabsorbed {\it apec} component with kT$\approx$0.1~keV, abundance of $Z_{\odot}$ at $z=0$, while the Galactic halo is modeled with a warmer absorbed thermal component kT$\approx$0.25~keV, abundance of $Z_{\odot}$ at $z=0$. The temperatures of the {\it apec} models are restricted, but the normalizations are allowed to vary in our fits. We model the cosmic X-ray background due to unresolved point sources using an absorbed power-law component with a spectral index of 1.4 \citep{Hickox2005} and normalization of $\approx$9$\times10^{-7}$~photons~keV$^{-1}$~cm$^{-2}$~s$^{-1}$ at $\approx$1~keV \citep{kuntz08,moretti2003}. The bright instrumental fluorescent lines Al--K (1.49~keV) and Si--K (1.74~keV) are not included in the MOS QBP files. Therefore, we model these instrumental lines by adding Gaussian models to our spectral fits to determine the best-fit energies, and normalizations.

Because of scattering in the \xmm\ mirrors, some of the flux that originates from one area of the sky is detected in a different area of the detector. This is not a major concern if the gradient in plasma temperature from core to outskirts is smooth; however, it may be important for clusters with a strong cool core. Additionally, for high redshift clusters, 0.15~$\Rfiveoo$ (discussed in Section~\ref{sec:xrayobs}) is comparable to the PSF for \xmm, so this PSF effect is crucial and must be accounted for when making spectral fits. This radial cross-talk or contamination effect is treated as an additional model component in XSPEC. The cross-talk ARFs for the contribution of X-rays originating from a region on the sky to the another region on the detector are created using the SAS tool {\it arfgen} \citep{snowden08}. The cross-talk correction is applied to eliminate PSF effects for all clusters in our sample. 

\section{Cluster Masses and X-ray Observables}
\label{sec:xrayobs}

The relationship between cluster X-ray observables (including emission-weighted mean temperature \Tx, integrated pressure \Yx, ICM mass \Micm\ and luminosity \Lx) and halo mass and cluster redshift exhibit a low scatter outside of the cluster center, 
where non-gravitational effects such as  heating and cooling processes are less important \citep{fabian94c,mohr97,ohara06, kravtsov06b, nagai07}. We, therefore, measure all the X-ray observables both with and without the core region (except for the ICM mass \Micm\ where the core has no impact). Specifically, we extract observables within an aperture (0.15--1)\Rfiveoo\ (core-excised marked as $cex$) and (0--1)\Rfiveoo\ (core-included marked as $cin$). The cluster radius \Rfiveoo\ is determined using the SZE-based halo mass \Mfiveoo\ using  
\begin{equation}
\Rfiveoo = \left( \frac{3 \Mfiveoo}{4\pi \times 500 \rho_{\mathrm{crit}}(z)}\right)^{1/3},
\end{equation}
where the \Mfiveoo\ masses are described in the next section, and $\rho_{\mathrm{crit}}(z)$ is the critical density of the Universe at the cluster redshift.

\subsection{SZE-based mass $M_{500}$}
\label{sec:M500}

We derive the cluster mass \Mfiveoo\ based on the SZE signal-to-noise ratio $\xi$ and redshift $z$ as determined by SPT. The measured signal-to-noise $\xi$ is a biased observable subject to Gaussian noise that is extracted through a matched filter approach that employs a $\beta$ model with three degrees of freedom:  sky location ($\alpha,\delta$) and core-radius $\theta_\mathrm{C}$.  The mean value of the signal-to-noise $\left\langle \xi \right\rangle$ is related to the underlying unbiased signal-to-noise $\zeta$ as follows.
\begin{equation}
\label{eq:sz_sr3}
\left\langle \xi \right\rangle = \sqrt{\zeta^{2} + 3} \, ,
\end{equation}
for $\zeta>2$ \citep{deHaan16}.  The $\zeta$--mass scaling relation is parametrized as follows:
\begin{equation}
\label{eq:sz_sr1}
\zeta = A_{\mathrm{SZ}}\left( \frac{\Mfiveoo}{4.3\times10^{14}\Msun} \right)^{B_{\mathrm{SZ}}} 
\left( \frac{E(z)}{E(z_{\mathrm{piv}})} \right)^{C_{\mathrm{SZ}}} \, ,
\end{equation}
where the normalization is $A_{\mathrm{SZ}}$, the mass trend parameter is $B_{\mathrm{SZ}}$, the redshift trend parameter is $C_{\mathrm{SZ}}$, and there is log-normal intrinsic scatter in the observables at fixed mass of $\sigma_{\ln\zeta}$.

In this work we marginalize over the parameters of the $\zeta$--mass relation while fitting the parameters of the X-ray scaling relations that are investigated.  This ensures that the final uncertainties in the X-ray observable--mass--redshift scaling relations include the systematic uncertainties associated with the imperfectly known SZE-based halo masses.  In the interest of focusing on the X-ray scaling relations, we adopt priors on the parameters of the $\zeta$--mass relation that correspond to the fully marginalized posterior distributions reported in \cite{deHaan16} (Table~\ref{tab:priors}).  This approach does not capture any covariances among the $\zeta$--mass scaling relation parameters, but these are indeed small \citep[see][Figure~5]{deHaan16}.  The advantage is that the likelihood we must calculate in each iteration of the Markov chain involves our X-ray observables and the simple priors on the SZE $\zeta$--mass relation parameters.

\begin{table}
\centering
\caption{Gaussian priors $\mathcal{N}(\mu,\sigma^2)$ on the SZE observable-mass relation parameters appear first followed by uniform priors $\mathcal{U(\mathrm{min},\mathrm{max})}$ on the X-ray observable--mass relation parameters.}
\label{tab:priors}
\begin{tabular}{ll}
\hline\hline
Parameters   &  Priors \\[3pt]
\hline
\multicolumn{2}{l}{SZE \zm\ parameters}\\
$A_{\mathrm{SZ}}$ &$\mathcal{N}(4.842, 0.913^2)$\\
$B_{\mathrm{SZ}}$ &$\mathcal{N}(1.668, 0.083^2)$\\
$C_{\mathrm{SZ}}$ &$\mathcal{N}(0.550, 0.315^2)$\\
$\sigma_{\ln\zeta}$ &$\mathcal{N}(0.199, 0.069^2)$\\
\hline
\multicolumn{2}{l}{X-ray \Xm\ parameters}\\
$A_{\Tx}$ &$\mathcal{U}(0.1, 20)$~keV \\
$A_{\Micm}$ & $\mathcal{U}(10^{12}, 2\times10^{14})$~\Msun \\
$A_{\Yx}$ &$\mathcal{U}(5\times10^{12}, 2\times10^{15})$~keV\Msun \\
$A_{\Lx}$ &$\mathcal{U}(2\times10^{43}, 1.2\times10^{45})$~ergs s$^{-1}$ \\
$B_{\obsx}$  &$\mathcal{U}(0.1, 3.5)$\\
$C_{\obsx}$  &$\mathcal{U}(-4, 4)$\\
$\sigma_{\ln\obsx}$  &$\mathcal{U}(0.005, 1.5)$\\
$\gamma_{\obsx}$   &$\mathcal{U}(-4, 4)$\\
$\delta_{\obsx}$  &$\mathcal{U}(-4,4)$\\
\hline\hline
\end{tabular}
\end{table}

The priors we adopt on the SZE observable mass relation are shown in Table~\ref{tab:priors}, where $\mathcal{N}(\mu, \sigma^2)$ corresponds to a Gaussian with mean $\mu$ and dispersion $\sigma$.  These SZE $\zeta$--mass parameter constraints emerge from a joint cosmology and mass calibration analysis that uses as input:  (1) the SPT cluster distribution in $\xi$ and z (i.e. the number counts), (2) mass information from externally weak lensing calibrated \Yx\ measurements for 82 systems, and (3) external cosmological parameter constraints \citep[for more extensive discussion of SPT mass calibration see, e.g.,][]{bocquet15,chiu17}.  For the baseline priors listed above, the external cosmological priors include a prior on the  Hubble parameter \citep{riess11} and a prior on the baryon density parameter from Big Bang Nucleosynthesis \citep{cooke14}.

Although the mass calibration presented in \citet{deHaan16} includes information from Chandra X-ray observations of 82 clusters, we stress that the mass information is dominated by the cluster distribution in $\xi$ and redshift (i.e., the halo mass function information). That is, the $\zeta-$mass$-$redshift relation used to calculate SPT-SZ masses does not simply follow the employed $\Yx-$mass$-$redshift relation, because it is a subdominant component of the mass information.
Moreover, we adopt the resulting posteriors of the $\zeta-$mass relation as the priors in this work, effectively marginalizing over the systematic uncertainties of all ingredients used in calibrating the cluster mass.  Modeling these priors as independent Gaussian distributions is appropriate, given the lack of strong covariances in the joint parameter constraints presented in \citet[][see Figure 5]{deHaan16}.
It is important to note that the correlated intrinsic scatter between the mass proxies of SZE and X-ray does not impact the mass calibration with the current sample size \citep{deHaan16,dietrich17}, therefore, we can use the existing $\zeta-$mass$-$redshift relation with marginalized systematic uncertainties to investigate the X-ray observable-to-mass scaling relations.

To foreshadow an additional set of results that we present, we also adopt a separate set of priors derived from the second results column of Table 3 in \citet{deHaan16}, which include also an external cosmological prior coming from BAO distance measurements \citep{anderson14}.  This set of results is consistent with the baseline results, but has smaller uncertainties (because the cosmological uncertainties typically dominate the posterior distributions of the SZE $\zeta$--mass parameters) and has a shift of $\Delta C_\mathrm{SZ}=+0.3$ that translates into a corresponding shift in the redshift trend parameters in the X-ray scaling relations.

Because we adopt similar four-parameter scaling relations for both the SZE and X-ray observables, we denote the targeted X-ray scaling relation (e.g., equation~\ref{eq:FormA}) as $r_\obsx = (A_\obsx, B_\obsx, C_\obsx, \sigma_{\ln \obsx})$ and the one used for estimating \Mfiveoo\ as $r_{\zeta}=(A_{\mathrm{SZ}}, B_{\mathrm{SZ}}, C_{\mathrm{SZ}}, \sigma_{\ln\zeta})$.
The notation $r_\obsx$ can be similarly extended to the five-parameter scaling relations for the X-ray observables, for which we define the functional forms in Section~\ref{sec:form}.

We stress that the cluster masses in our work include corrections for selection biases (e.g. the Eddington bias, the Malmquist bias) and therefore they reflect the unbiased distribution of cluster mass \Mfiveoo\ given the observable $\xi$ and redshift $z$ measured for each SZE-selected cluster.

\startlongtable
\begin{deluxetable*}{lrccccccccccc}
\tablecaption{
Measurements of the X-ray observables and cluster masses. \label{table:observables}
}
\tablehead{
&\colhead{\Rfiveoo}&\colhead{\Lxi} & \colhead{\Lxib} & \colhead{\Txi } &  \colhead{\Zi}  &  \colhead{\Lxeb} & \colhead{\Lxe} & \colhead{\Txe} &  \colhead{\Ze} &  \colhead{\Micm} & \colhead{\Yxi} &  \colhead{\Mfiveoo} \\
\colhead{Cluster}  & \colhead{[kpc]} & \colhead{[$10^{44}$erg~s$^{-1}$]}&\colhead{[$10^{44}$erg~s$^{-1}$]} &  \colhead{[keV]} & \colhead{[$Z_{\odot}$]} & \colhead{[$10^{44}$erg~s$^{-1}$]} & \colhead{[$10^{44}$erg~s$^{-1}$]}& \colhead{[keV]} & \colhead{[$Z_{\odot}$]} & \colhead{[$10^{13}\Msun$]} & \colhead{[$10^{14}\Msun\mathrm{keV}$]} & \colhead{[$10^{14}\Msun$] }
}
\startdata
SPT-CLJ0114-4123 & 1241   &     3.33$\pm$0.28 & 11.60$\pm$1.58   &      5.62$^{+0.47}_{-0.53}$  &      0.3$^{*}$              & 8.53$\pm$0.34     &      2.56$\pm$0.36 &       5.01$^{+0.86}_{-0.62}$  &       0.3$^{*}$              &       8.02$^{+0.90}_{-0.89}$  &      4.50$\pm$0.64 &       5.86$^{+0.85}_{-0.69}$  \\
  SPT-CLJ0205-5829 & 759    &     4.64$\pm$0.94 & 17.70$\pm$2.64   &      6.29$^{+1.34}_{-1.13}$  &      0.31$^{+0.19}_{-0.17}$ & 13.70$\pm$3.05    &      3.73$\pm$1.18 &       6.07$^{+2.17}_{-0.65}$  &       0.30$^{+0.29}_{-0.12}$ &       5.27$^{+0.61}_{-0.61}$  &      3.31$\pm$0.75 &       4.37$^{+0.59}_{-0.55}$  \\
  SPT-CLJ0217-5245 & 1110   &     1.40$\pm$0.15 & 6.19$\pm$0.58    &      10.43$^{+4.66}_{-1.64}$ &      0.3$^{*}$              & 5.35$\pm$1.22    &      1.18$\pm$0.23 &       8.13$^{+2.95}_{-1.95}$  &       0.3$^{*}$              &       4.40$^{+0.41}_{-0.40}$  &      4.59$\pm$1.45 &       4.01$^{+0.71}_{-0.61}$  \\
  SPT-CLJ0225-4155 & 1144   &     3.57$\pm$0.25 & 12.60$\pm$1.07   &      6.0$^{+0.23}_{-0.33}$   &      0.21$_{-0.17}^{+0.27}$ & 9.04$\pm$0.55     &      2.47$\pm$0.11 &       6.54$^{+0.27}_{-0.41}$  &       0.19$_{-0.13}^{+0.27}$ &       6.76$^{+0.81}_{-0.80}$  &      4.06$\pm$0.52 &       4.33$^{+0.72}_{-0.64}$  \\
  SPT-CLJ0230-6028 & 909    &     3.39$\pm$0.63 & 11.10$\pm$0.94    &      4.81$^{+0.70}_{-0.77}$  &      0.3$^{*}$              & 7.36$\pm$0.43     &      2.24$\pm$0.48 &       4.86$^{+1.20}_{-1.12}$  &       0.3$^{*}$              &       4.83$^{+1.11}_{-1.04}$  &      2.32$\pm$0.62 &       3.43$^{+0.61}_{-0.58}$  \\
  SPT-CLJ0231-5403 & 921    &     1.50$\pm$0.27 & 5.26$\pm$0.90    &      5.34$^{+1.68}_{-1.09}$  &      0.50$^{+0.21}_{-0.24}$ & 4.43$\pm$1.31    &      1.21$\pm$0.22 &       5.84$^{+2.24}_{-1.51}$  &       0.49$^{+0.37}_{-0.27}$ &       3.02$^{+0.49}_{-0.48}$  &      1.61$\pm$0.49 &       3.18$^{+0.67}_{-0.62}$  \\
  SPT-CLJ0232-4421 & 1507   &     7.24$\pm$0.30 & 27.80$\pm$1.24   &      7.03$^{+0.21}_{-0.41}$  &      0.35$^{+0.06}_{-0.05}$ & 14.90$\pm$0.92     &      3.83$\pm$0.2 &       7.19$^{+0.46}_{-0.50}$  &       0.31$^{+0.05}_{-0.05}$ &       16.66$^{+0.80}_{-0.81}$ &      11.71$\pm$0.76 &       9.45$^{+1.16}_{-1.10}$  \\
  SPT-CLJ0233-5819 & 940    &     2.16$\pm$0.31 & 7.40$\pm$0.40    &      5.12$^{+0.50}_{-0.51}$  &      0.31$^{+0.13}_{-0.10}$ & 6.08$\pm$0.30     &      1.79$\pm$0.19 &       5.04$^{+0.63}_{-0.66}$  &       0.34$^{+0.08}_{-0.07}$ &       4.41$^{+0.65}_{-0.61}$  &      2.26$\pm$0.39 &       3.70$^{+0.61}_{-0.59}$  \\
  SPT-CLJ0234-5831 & 1273   &     6.14$\pm$0.46 & 20.00$\pm$1.59   &      4.67$^{+0.34}_{-0.25}$  &      0.35$^{+0.07}_{-0.05}$ & 9.75$\pm$1.26    &      2.86$\pm$0.32 &       5.21$^{+0.55}_{-0.43}$  &       0.50$^{+0.19}_{-0.17}$ &       6.72$^{+0.47}_{-0.47}$  &      3.13$\pm$0.29 &       6.70$^{+0.84}_{-0.82}$  \\
  SPT-CLJ0240-5946 & 1155   &     2.18$\pm$0.22 & 9.18$\pm$1.42   &      8.60$^{+1.17}_{-0.86}$  &      0.25$^{+0.10}_{-0.17}$ & 4.60$\pm$0.48     &      1.15$\pm$0.14 &       7.65$^{+1.89}_{-1.45}$  &       0.3$^{*}$              &       4.30$^{+0.48}_{-0.50}$  &      3.69$\pm$0.61 &       4.85$^{+0.70}_{-0.65}$  \\
  SPT-CLJ0243-4833 & 1220   &     5.71$\pm$0.55 & 21.50$\pm$1.49   &      6.26$^{+0.41}_{-0.71}$  &      0.48$^{+0.12}_{-0.13}$ & 13.70$\pm$0.71     &      3.47$\pm$0.46 &       6.85$^{+0.70}_{-1.02}$  &       0.47$^{+0.20}_{-0.17}$ &       9.46$^{+1.99}_{-2.04}$  &      5.92$\pm$1.37 &       6.47$^{+0.90}_{-0.73}$  \\
  SPT-CLJ0254-6051 & 1053   &     1.51$\pm$0.22 & 5.03$\pm$0.26    &      5.13$^{+1.12}_{-0.64}$  &      0.37$^{+0.19}_{-0.16}$ & 4.49$\pm$0.62     &      1.36$\pm$0.28 &       5.12$^{+1.18}_{-0.93}$  &       0.31$^{+0.21}_{-0.11}$ &       4.49$^{+0.33}_{-0.33}$  &      2.30$\pm$0.40 &       3.86$^{+0.67}_{-0.58}$  \\
  SPT-CLJ0254-5857 & 1250   &     5.41$\pm$0.21 & 21.70$\pm$0.93    &      7.62$^{+0.25}_{-0.25}$  &      0.30$^{+0.02}_{-0.04}$ & 18.90$\pm$0.54     &      4.73$\pm$0.27 &       7.60$^{+0.29}_{-0.37}$  &       0.31$^{+0.05}_{-0.06}$ &       10.95$^{+3.73}_{-3.28}$ &      8.34$\pm$2.68 &       6.52$^{+0.81}_{-0.81}$  \\
  SPT-CLJ0257-5732 & 981    &     0.35$\pm$0.10 & .97$\pm$0.03    &      3.48$^{+1.31}_{-0.96}$  &      0.3$^{*}$              & .86$\pm$0.04     &      0.32$\pm$0.08 &       3.31$^{+1.06}_{-0.91}$  &       0.3$^{*}$              &       2.25$^{+0.42}_{-0.40}$  &      0.78$\pm$0.29 &       3.15$^{+0.64}_{-0.69}$  \\
  SPT-CLJ0304-4401 & 1274   &     3.40$\pm$0.43 & 11.90$\pm$0.95    &      5.36$^{+0.50}_{-0.33}$  &      0.42$^{+0.11}_{-0.11}$ & 9.70$\pm$0.34     &      2.57$\pm$0.19 &       6.40$^{+0.74}_{-0.87}$  &       0.46$^{+0.16}_{-0.15}$ &       8.88$^{+0.86}_{-0.86}$  &      4.75$\pm$0.59 &       6.98$^{+0.97}_{-0.77}$  \\
  SPT-CLJ0317-5935 & 1022   &     2.34$\pm$0.29 & 7.50$\pm$0.37    &      4.61$^{+0.39}_{-0.59}$  &      0.29$^{+0.10}_{-0.12}$ & 5.76$\pm$0.26     &      1.98$\pm$0.2 &       3.72$^{+0.34}_{-0.48}$  &       0.28$^{+0.14}_{-0.15}$ &       5.19$^{+0.68}_{-0.67}$  &      2.39$\pm$0.40 &       3.73$^{+0.64}_{-0.61}$  \\
  SPT-CLJ0330-5228 & 1193   &     8.22$\pm$0.24 & 25.20$\pm$0.84    &      4.22$^{+0.15}_{-0.06}$  &      0.13$^{+0.03}_{-0.03}$ & 25.00$\pm$0.63     &      7.63$\pm$0.24 &       4.48$^{+0.10}_{-0.10}$  &       0.10$^{+0.02}_{-0.03}$ &       3.32$^{+0.39}_{-0.37}$  &      1.38$\pm$0.17 &       5.63$^{+0.81}_{-0.66}$  \\
  SPT-CLJ0343-5518 & 975    &     1.57$\pm$0.51 & 4.81$\pm$0.51    &      4.09$^{+0.90}_{-0.61}$  &      0.3$^{*}$              & 4.36$\pm$0.49     &      1.32$\pm$0.29 &       4.87$^{+0.91}_{-0.98}$  &       0.19$^{+0.25}_{-0.19}$ &       3.07$^{+0.49}_{-0.47}$  &      1.25$\pm$0.30 &       3.52$^{+0.65}_{-0.58}$  \\
  SPT-CLJ0344-5452 & 827    &     2.02$\pm$0.50 & 6.49$\pm$0.22    &      4.45$^{+0.96}_{-0.60}$  &      0.3$^{*}$              & 4.52$\pm$0.18     &      1.37$\pm$0.32 &       4.67$^{+0.99}_{-0.73}$  &       0.3$^{*}$              &       2.88$^{+0.47}_{-0.43}$  &      1.28$\pm$0.30 &       3.89$^{+0.58}_{-0.54}$  \\
  SPT-CLJ0354-5904 & 1063   &     1.60$\pm$0.17 & 5.28$\pm$0.26    &      4.72$^{+0.62}_{-0.49}$  &      0.54$^{+0.18}_{-0.16}$ & 4.66$\pm$0.83     &      1.37$\pm$0.25 &       5.20$^{+0.83}_{-0.77}$  &       0.40$^{+0.21}_{-0.21}$ &       4.02$^{+0.37}_{-0.35}$  &      1.89$\pm$0.28 &       3.83$^{+0.67}_{-0.59}$  \\
  SPT-CLJ0403-5719 & 1008   &     2.68$\pm$0.26 & 8.30$\pm$0.78    &      4.16$^{+0.20}_{-0.29}$  &      0.43$^{+0.10}_{-0.11}$ & 4.96$\pm$0.37     &      1.56$\pm$0.18 &       4.26$^{+0.39}_{-0.30}$  &       0.80$^{+0.27}_{-0.20}$ &       3.32$^{+0.39}_{-0.37}$  &      1.38$\pm$0.17 &       3.52$^{+0.66}_{-0.59}$  \\
  SPT-CLJ0406-5455 & 878    &     1.09$\pm$0.18 & 4.36$\pm$0.30    &      7.23$^{+2.14}_{-1.35}$  &      0.41$^{+0.26}_{-0.22}$ & 3.81$\pm$0.20     &      0.95$\pm$0.13 &       7.26$^{+2.89}_{-1.92}$  &       0.63$^{+0.42}_{-0.21}$ &       2.57$^{+0.32}_{-0.31}$  &      1.86$\pm$0.50 &       3.28$^{+0.63}_{-0.54}$  \\
  SPT-CLJ0417-4748 & 1164   &     6.28$\pm$0.36 & 23.60$\pm$1.28   &      6.17$^{+0.48}_{-0.34}$  &      0.45$^{+0.10}_{-0.08}$ & 14.10$\pm$2.55    &      3.6$\pm$0.44 &       6.78$^{+1.49}_{-0.84}$  &       0.51$^{+0.18}_{-0.13}$ &       6.26$^{+0.48}_{-0.48}$  &      3.85$\pm$0.39 &       6.22$^{+0.85}_{-0.71}$  \\
  SPT-CLJ0438-5419 & 1385   &     8.36$\pm$0.37 & 34.80$\pm$2.03   &      8.09$^{+0.48}_{-0.39}$  &      0.33$^{+0.04}_{-0.04}$ & 19.90$\pm$1.78    &      5.09$\pm$0.31 &       7.06$^{+0.61}_{-0.41}$  &       0.28$^{+0.10}_{-0.09}$ &       12.53$^{+0.52}_{-0.52}$ &      10.13$\pm$0.69 &       8.68$^{+1.03}_{-1.03}$  \\
  SPT-CLJ0510-4519 & 1323   &     2.98$\pm$0.14 & 10.40$\pm$0.68    &      5.93$^{+0.28}_{-0.21}$  &      0.24$^{+0.05}_{-0.05}$ & 6.29$\pm$0.43     &      1.81$\pm$0.08 &       5.87$^{+0.37}_{-0.36}$  &       0.36$^{+0.06}_{-0.09}$ &       7.06$^{+0.30}_{-0.30}$  &      4.18$\pm$0.24 &       5.73$^{+0.85}_{-0.72}$  \\
  SPT-CLJ0516-5430 & 1292   &     4.38$\pm$0.22 & 17.40$\pm$0.91    &      7.64$^{+0.23}_{-0.23}$  &      0.28$^{+0.03}_{-0.03}$ & 16.00$\pm$1.04    &      4.38$\pm$0.22 &       7.55$^{+0.25}_{-0.25}$  &       0.24$^{+0.02}_{-0.04}$ &       9.64$^{+2.51}_{-2.38}$  &      7.37$\pm$1.88 &       5.96$^{+0.78}_{-0.74}$  \\
  SPT-CLJ0522-4818 & 1062   &     1.54$\pm$0.18 & 5.67$\pm$0.60    &      6.26$^{+1.02}_{-0.63}$  &      0.41$^{+0.16}_{-0.13}$ & 3.70$\pm$0.49     &      0.96$\pm$0.12 &       6.90$^{+1.49}_{-1.01}$  &       0.48$^{+0.27}_{-0.23}$ &       2.73$^{+0.31}_{-0.30}$  &      1.71$\pm$0.29 &       3.37$^{+0.75}_{-0.71}$  \\
  SPT-CLJ0549-6205 & 1470   &     11.6$\pm$0.48 & 48.80$\pm$1.31   &      8.60$^{+0.42}_{-0.35}$  &      0.38$^{+0.06}_{-0.05}$ & 21.20$\pm$1.27    &      4.96$\pm$0.25 &       8.97$^{+1.27}_{-0.42}$  &       0.54$^{+0.17}_{-0.13}$ &       11.60$^{+0.46}_{-0.45}$ &      9.97$\pm$0.59 &       9.66$^{+1.31}_{-1.03}$  \\
  SPT-CLJ0559-5249 & 1072   &     3.53$\pm$0.46 & 13.60$\pm$1.99   &      6.64$^{+1.17}_{-1.17}$  &      0.28$^{+0.10}_{-0.15}$ & 10.90$\pm$1.34    &      2.87$\pm$0.44 &       6.59$^{+1.40}_{-0.96}$  &       0.25$^{+0.18}_{-0.19}$ &       7.09$^{+1.14}_{-1.17}$  &      4.71$\pm$1.13 &       5.03$^{+0.69}_{-0.63}$  \\
  SPT-CLJ0611-5938 & 992    &     1.27$\pm$0.20 & 4.12$\pm$0.77    &      4.62$^{+0.73}_{-0.78}$  &      0.33$^{+0.22}_{-0.19}$ & 3.00$\pm$0.21     &      1.11$\pm$0.21 &       4.30$^{+0.74}_{-0.76}$  &       0.45$^{+0.30}_{-0.25}$ &       3.21$^{+0.86}_{-0.87}$  &      1.48$\pm$0.40 &       3.13$^{+0.68}_{-0.67}$  \\
  SPT-CLJ0615-5746 & 1098   &     15.9$\pm$1.48 & 88.80$\pm$6.94   &      14.16$^{+2.04}_{-1.32}$ &      0.65$^{+0.22}_{-0.25}$ & 56.90$\pm$7.09    &      10.9$\pm$1.58 &       12.50$^{+1.60}_{-1.99}$ &       0.36$^{+0.26}_{-0.21}$ &       11.19$^{+1.06}_{-1.05}$ &      15.86$\pm$2.40 &       8.69$^{+1.07}_{-0.99}$  \\
  SPT-CLJ0637-4829 & 1258   &     1.01$\pm$0.14 & 3.80$\pm$0.75    &      6.53$^{+1.50}_{-1.38}$  &      0.21$_{-0.09}^{+0.35}$ & 2.75$\pm$0.53     &      0.87$\pm$0.11 &       5.01$^{+1.69}_{-0.95}$  &       0.24$_{-0.10}^{+0.40}$ &      6.29$\pm$0.13        &      4.11$\pm$0.94 &       5.66$^{+0.81}_{-0.68}$  \\
  SPT-CLJ0638-5358 & 1459   &     6.43$\pm$0.15 & 26.10$\pm$0.99    &      8.38$^{+0.24}_{-0.29}$  &      0.32$_{-0.28}^{+0.36}$ & 13.10$\pm$0.80     &      3.24$\pm$0.14 &       8.44$^{+0.86}_{-0.48}$  &       0.33$_{-0.24}^{+0.39}$ &      9.77$\pm$0.21        &      8.19$\pm$0.32 &       9.42$^{+1.18}_{-1.09}$  \\
  SPT-CLJ0658-5556 & 1664   &     13.3$\pm$0.46 & 62.40$\pm$3.48   &      12.40$^{+0.32}_{-0.54}$ &      0.28$^{+0.03}_{-0.02}$ & 45.00$\pm$4.22    &      9.43$\pm$0.51 &       13.44$^{+1.14}_{-0.32}$ &       0.29$^{+0.07}_{-0.07}$ &       20.08$^{+1.05}_{-1.04}$ &      24.90$\pm$1.56 &       12.70$^{+1.64}_{-1.38}$ \\
  SPT-CLJ2011-5725 & 1067   &     2.12$\pm$0.15 & 6.49$\pm$0.14    &      4.13$^{+0.15}_{-0.13}$  &      0.39$^{+0.08}_{-0.07}$ & 4.44$\pm$0.16     &      1.52$\pm$0.1 &       3.65$^{+0.21}_{-0.21}$  &       0.56$^{+0.14}_{-0.12}$ &       3.39$^{+0.37}_{-0.34}$  &      1.40$\pm$0.15 &       3.35$^{+0.65}_{-0.64}$  \\
  SPT-CLJ2017-6258 & 986    &     1.55$\pm$0.21 & 4.55$\pm$0.32    &      3.65$^{+0.61}_{-0.76}$  &      0.25$^{+0.09}_{-0.08}$ & 3.65$\pm$0.31     &      1.29$\pm$0.28 &       3.39$^{+0.81}_{-0.75}$  &       0.3$^{*}$              &       7.78$^{+0.09}_{-0.10}$  &      4.56$\pm$0.16 &       4.03$^{+0.68}_{-0.64}$  \\
  SPT-CLJ2022-6323 & 1073   &     0.94$\pm$0.14 & 3.90$\pm$0.84    &      8.45$^{+4.91}_{-3.29}$  &      0.3$^{*}$              & 3.02$\pm$0.78     &      0.83$\pm$0.15 &       5.73$^{+2.28}_{-1.03}$  &       0.3$^{*}$              &       3.45$^{+0.55}_{-0.54}$  &      2.91$\pm$1.48 &       3.80$^{+0.67}_{-0.57}$  \\
  SPT-CLJ2023-5535 & 1309   &     3.28$\pm$0.26 & 14.80$\pm$1.54   &      10.93$^{+2.00}_{-1.55}$ &      0.43$_{-0.25}^{+0.65}$ & 10.00$\pm$0.99     &      2.47$\pm$0.26 &       8.31$^{+1.69}_{-1.15}$  &       0.29$_{-0.06}^{+0.54}$ &       8.43$^{+0.71}_{-0.68}$  &      9.22$\pm$1.68 &       6.49$^{+0.81}_{-0.71}$  \\
  SPT-CLJ2030-5638 & 1018   &     1.06$\pm$0.17 & 2.99$\pm$0.21    &      3.46$^{+0.39}_{-0.33}$  &      0.3$^{*}$              & 2.42$\pm$0.14     &      0.81$\pm$0.13 &       3.88$^{+0.45}_{-0.45}$  &       0.3$^{*}$              &       2.62$^{+0.27}_{-0.29}$  &      0.91$\pm$0.13 &       3.35$^{+0.64}_{-0.62}$  \\
  SPT-CLJ2031-4037 & 1389   &     5.02$\pm$0.24 & 20.50$\pm$0.56    &      8.14$^{+1.22}_{-0.75}$  &      0.29$^{+0.10}_{-0.09}$ & 10.50$\pm$0.41     &      2.8$\pm$0.21 &       6.67$^{+0.89}_{-1.01}$  &       0.38$^{+0.18}_{-0.16}$ &       7.79$^{+0.41}_{-0.40}$  &      6.34$\pm$0.83 &       7.95$^{+0.99}_{-0.95}$  \\
  SPT-CLJ2032-5627 & 1204   &     3.29$\pm$0.12 & 10.80$\pm$0.57    &      4.99$^{+0.19}_{-0.18}$  &      0.24$^{+0.03}_{-0.05}$ & 9.66$\pm$0.42     &      2.92$\pm$0.11 &       5.07$^{+0.28}_{-0.34}$  &       0.23$^{+0.03}_{-0.03}$ &       5.93$^{+0.36}_{-0.37}$  &      2.95$\pm$0.21 &       4.77$^{+0.71}_{-0.63}$  \\
  SPT-CLJ2040-5725 & 803    &     3.68$\pm$0.59 & 10.80$\pm$0.88    &      3.71$^{+0.32}_{-0.26}$  &      0.23$^{+0.10}_{-0.05}$ & 7.18$\pm$0.29     &      2.24$\pm$0.39 &       4.61$^{+0.58}_{-0.52}$  &       0.3$^{*}$              &       3.68$^{+0.34}_{-0.34}$  &      1.36$\pm$0.10 &       3.23$^{+0.59}_{-0.51}$  \\
  SPT-CLJ2040-4451 & 649    &     1.92$\pm$0.57 & 5.83$\pm$1.43   &      3.75$^{+0.85}_{-0.65}$  &      0.3$^{*}$              & 8.39$\pm$1.79    &      2.92$\pm$1.59 &       4.78$^{+1.50}_{-1.51}$  &       0.53$^{+0.26}_{-0.27}$ &       3.34$^{+0.55}_{-0.54}$  &      1.17$\pm$0.31 &       3.31$^{+0.53}_{-0.54}$  \\
  SPT-CLJ2056-5459 & 889    &     1.91$\pm$0.27 & 6.01$\pm$0.25    &      4.22$^{+0.42}_{-0.45}$  &      0.52$^{+0.22}_{-0.17}$ & 5.07$\pm$0.09     &      1.62$\pm$0.16 &       4.19$^{+0.46}_{-0.65}$  &       0.63$^{+0.29}_{-0.26}$ &       3.64$^{+0.31}_{-0.30}$  &      1.53$\pm$0.20 &       3.36$^{+0.60}_{-0.57}$  \\
  SPT-CLJ2106-5844 & 963    &     12.2$\pm$0.85 & 55.80$\pm$5.16   &      9.43$^{+0.70}_{-1.67}$  &      0.3$^{*}$              & 47.50$\pm$3.48    &      10.5$\pm$0.97 &       9.19$^{+0.88}_{-1.06}$  &       0.3$^{*}$              &       11.73$^{+0.38}_{-0.39}$ &      11.05$\pm$1.43 &       7.14$^{+0.86}_{-0.83}$  \\
  SPT-CLJ2109-4626 & 737    &     1.81$\pm$0.46 & 5.24$\pm$0.66    &      3.52$^{+0.51}_{-0.36}$  &      0.51$^{+0.31}_{-0.13}$ & 3.50$\pm$0.13     &      1.22$\pm$0.18 &       3.46$^{+0.73}_{-0.53}$  &       0.85$^{+0.55}_{-0.56}$ &       2.55$^{+0.45}_{-0.44}$  &      0.90$\pm$0.19 &       2.68$^{+0.65}_{-0.56}$  \\
  SPT-CLJ2124-6124 & 1113   &     1.01$\pm$0.27 & 3.54$\pm$0.69    &      5.66$^{+1.56}_{-1.05}$  &      0.3$^{*}$              & 4.71$\pm$0.22     &      1.44$\pm$0.17 &       4.72$^{+0.93}_{-0.65}$  &       0.24$^{+0.25}_{-0.16}$ &       5.84$^{+0.84}_{-0.79}$  &      3.30$\pm$0.89 &       4.60$^{+0.66}_{-0.63}$  \\
  SPT-CLJ2130-6458 & 1151   &     1.88$\pm$0.27 & 5.84$\pm$0.64    &      4.26$^{+0.55}_{-0.42}$  &      0.26$^{+0.18}_{-0.15}$ & 3.83$\pm$0.12     &      1.2$\pm$0.13 &       4.44$^{+0.73}_{-0.52}$  &       0.30$^{+0.14}_{-0.25}$ &       4.48$^{+0.92}_{-0.85}$  &      1.90$\pm$0.43 &       4.33$^{+0.71}_{-0.58}$  \\
  SPT-CLJ2131-4019 & 1232   &     6.00$\pm$0.33 & 24.30$\pm$1.98   &      7.64$^{+0.50}_{-0.59}$  &      0.37$^{+0.09}_{-0.05}$ & 14.10$\pm$1.90    &      3.27$\pm$0.34 &       8.79$^{+1.79}_{-2.15}$  &       0.43$^{+0.18}_{-0.15}$ &       8.17$^{+0.67}_{-0.68}$  &      6.24$\pm$0.68 &       6.25$^{+0.88}_{-0.72}$  \\
   SPT-CLJ2136-6307 & 804    &     1.53$\pm$0.60 & 3.57$\pm$0.29    &      2.58$^{+1.03}_{-0.62}$  &      0.3$^{*}$              & 1.42$\pm$0.39     &      3.54$\pm$0.53 &       2.04$^{+0.79}_{-0.40}$  &       0.3$^{*}$              &       4.23$^{+1.41}_{-1.30}$  &      1.09$\pm$0.49 &       3.24$^{+0.60}_{-0.51}$  \\
  SPT-CLJ2138-6008 & 1283   &     2.83$\pm$0.14 & 10.50$\pm$0.52    &      6.59$^{+0.42}_{-0.29}$  &      0.18$^{+0.09}_{-0.08}$ & 6.54$\pm$0.16     &      1.9$\pm$0.13 &       5.43$^{+0.66}_{-0.46}$  &       0.29$^{+0.13}_{-0.15}$ &       6.25$^{+0.41}_{-0.39}$  &      4.11$\pm$0.34 &       6.10$^{+0.79}_{-0.76}$  \\
  SPT-CLJ2145-5644 & 1188   &     4.07$\pm$0.38 & 15.40$\pm$1.58   &      6.36$^{+0.57}_{-0.71}$  &      0.48$^{+0.17}_{-0.14}$ & 10.30$\pm$1.21    &      2.85$\pm$0.3 &       5.82$^{+0.71}_{-0.72}$  &       0.41$^{+0.22}_{-0.11}$ &       8.64$^{+1.54}_{-1.51}$  &      5.49$\pm$1.11 &       5.82$^{+0.82}_{-0.67}$  \\
 SPT-CLJ2146-4633 & 921    &     2.94$\pm$0.47 & 9.73$\pm$0.26    &      4.64$^{+0.42}_{-0.24}$  &      0.63$^{+0.07}_{-0.13}$ & 9.04$\pm$1.21    &      2.74$\pm$0.65 &       4.98$^{+0.18}_{-0.43}$  &       0.83$^{+0.09}_{-0.17}$ &       4.92$^{+1.75}_{-1.65}$  &      2.28$\pm$0.80 &       4.89$^{+0.66}_{-0.65}$  \\
  SPT-CLJ2200-6245 & 1067   &     1.08$\pm$0.33 & 2.63$\pm$0.30    &      2.26$^{+0.33}_{-0.38}$  &      0.32$^{+0.07}_{-0.14}$ & 1.99$\pm$0.09     &      0.84$\pm$0.24 &       2.10$^{+0.37}_{-0.31}$  &       0.30$^{+0.55}_{-0.19}$ &       3.38$^{+0.86}_{-0.85}$  &      0.76$\pm$0.22 &       3.79$^{+0.67}_{-0.57}$  \\
  SPT-CLJ2248-4431 & 1633   &     16.8$\pm$0.96 & 77.70$\pm$6.91   &      11.46$^{+0.28}_{-0.63}$ &      0.26$^{+0.02}_{-0.06}$ & 44.00$\pm$3.95    &      9.44$\pm$0.48 &       11.90$^{+0.97}_{-0.69}$ &       0.22$^{+0.07}_{-0.07}$ &       19.46$^{+0.90}_{-0.89}$ &      22.30$\pm$1.30 &       13.05$^{+1.64}_{-1.44}$ \\
  SPT-CLJ2332-5358 & 1137   &     2.24$\pm$0.19 & 8.56$\pm$0.99    &      7.63$^{+0.97}_{-0.97}$  &      0.3$^{*}$              & 5.75$\pm$1.19    &      1.58$\pm$0.25 &       6.17$^{+0.93}_{-0.84}$  &       0.3$^{*}$              &       4.01$^{+0.58}_{-0.57}$  &      3.06$\pm$0.58 &       4.63$^{+0.68}_{-0.60}$  \\
  SPT-CLJ2337-5942 & 1112   &     8.2$\pm$0.81 & 36.40$\pm$2.42   &      9.11$^{+0.60}_{-1.01}$  &      0.3$^{*}$              & 24.40$\pm$1.35    &      5.59$\pm$0.76 &       8.60$^{+1.35}_{-1.33}$  &       0.3$^{*}$              &       9.21$^{+1.05}_{-1.05}$  &      8.38$\pm$1.21 &       7.05$^{+0.89}_{-0.81}$  \\
  SPT-CLJ2341-5119 & 902    &     4.76$\pm$0.39 & 19.40$\pm$1.38   &      7.47$^{+0.71}_{-0.88}$  &      0.14$^{+0.09}_{-0.08}$ & 12.10$\pm$0.31     &      3.48$\pm$0.59 &       5.34$^{+1.51}_{-0.87}$  &       0.3$^{*}$              &       5.26$^{+0.37}_{-0.36}$  &      3.93$\pm$0.50 &       4.94$^{+0.68}_{-0.58}$  \\
  SPT-CLJ2344-4243 & 1330   &     26.8$\pm$0.47 & 145.00$\pm$3.29   &      14.89$^{+0.32}_{-0.17}$ &      1.05$^{+0.02}_{-0.02}$ & 45.30$\pm$2.24    &      9.09$\pm$0.29 &       12.23$^{+0.31}_{-0.65}$ &       0.45$^{+0.06}_{-0.04}$ &       14.83$^{+0.31}_{-0.30}$ &      22.08$\pm$0.58 &       9.60$^{+1.20}_{-1.09}$  \\
\enddata
\vspace{0.0cm}
\tablecomments{
\sc
X-ray observables of the sample measured in core-included ($\mathrm{cin}$, $r<\Rfiveoo$) and core-excised ($\mathrm{cex}$, $0.15\Rfiveoo<r<\Rfiveoo$) apertures.
Parameters marked with $^{*}$ are fixed to the indicated values. From left to right is the cluster name, \Rfiveoo, bolometric and soft-band luminosity, emission-weighted mean temperature and metallicity, given first for the core-included and then for the core-excised measurements. Measured ICM masses \Micm, X-ray derived integrated Compton-$y$ \Yxi, and halo mass \Mfiveoo\ determined from the SZE observations are then listed for each cluster.
}
\end{deluxetable*}

\subsection{X-ray Observables}
\label{sec:X-rayObservables}

We measure the temperature, metallicity, and luminosity by fitting the spectra extracted in the apertures of the core-included region ($cin$, $r<\Rfiveoo$) and core-excised region ($cex$, $0.15\Rfiveoo<r<\Rfiveoo$) with a single temperature thermal {\apec} model. The best-fit core-included temperatures (\Txi), metallicity (\Zi), and luminosities (\Lxi) and core-excised temperatures (\Txe), metallicity (\Ze), and luminosities (\Lxe) are given in Table~\ref{table:observables}. In some clusters, the statistics of the observations are too poor to allow a determination of the global metallicity. In these cases, the metallicity is fixed to $0.3Z_{\odot}$ \citep{tozzi03,mcdonald16b}.
The metallicity constraints, and their evolution with redshift in this sample is extensively discussed in \citet{mcdonald16b}.

\begin{figure*}
\centering
\includegraphics[width=0.49\textwidth]{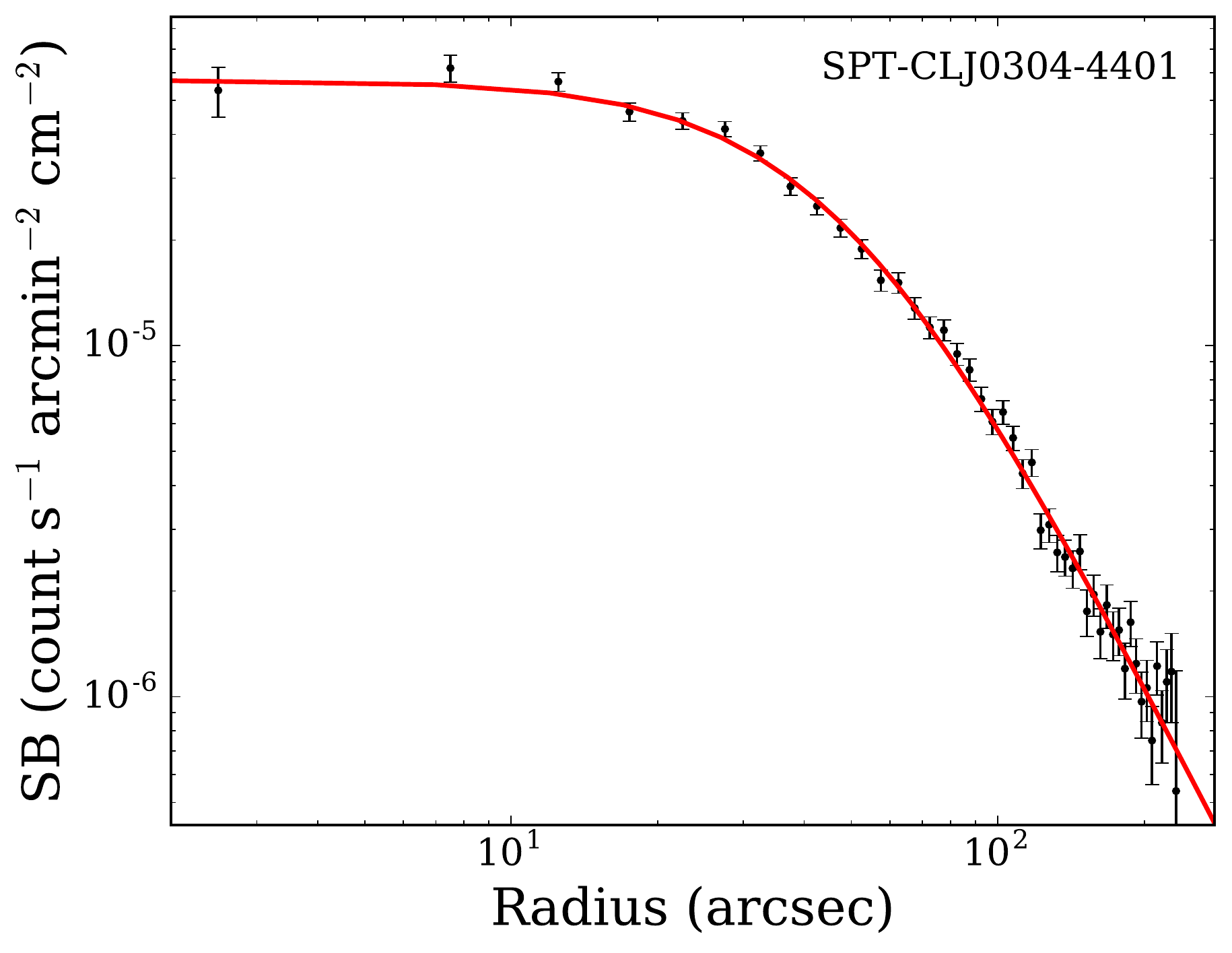}
\includegraphics[width=0.49\textwidth]{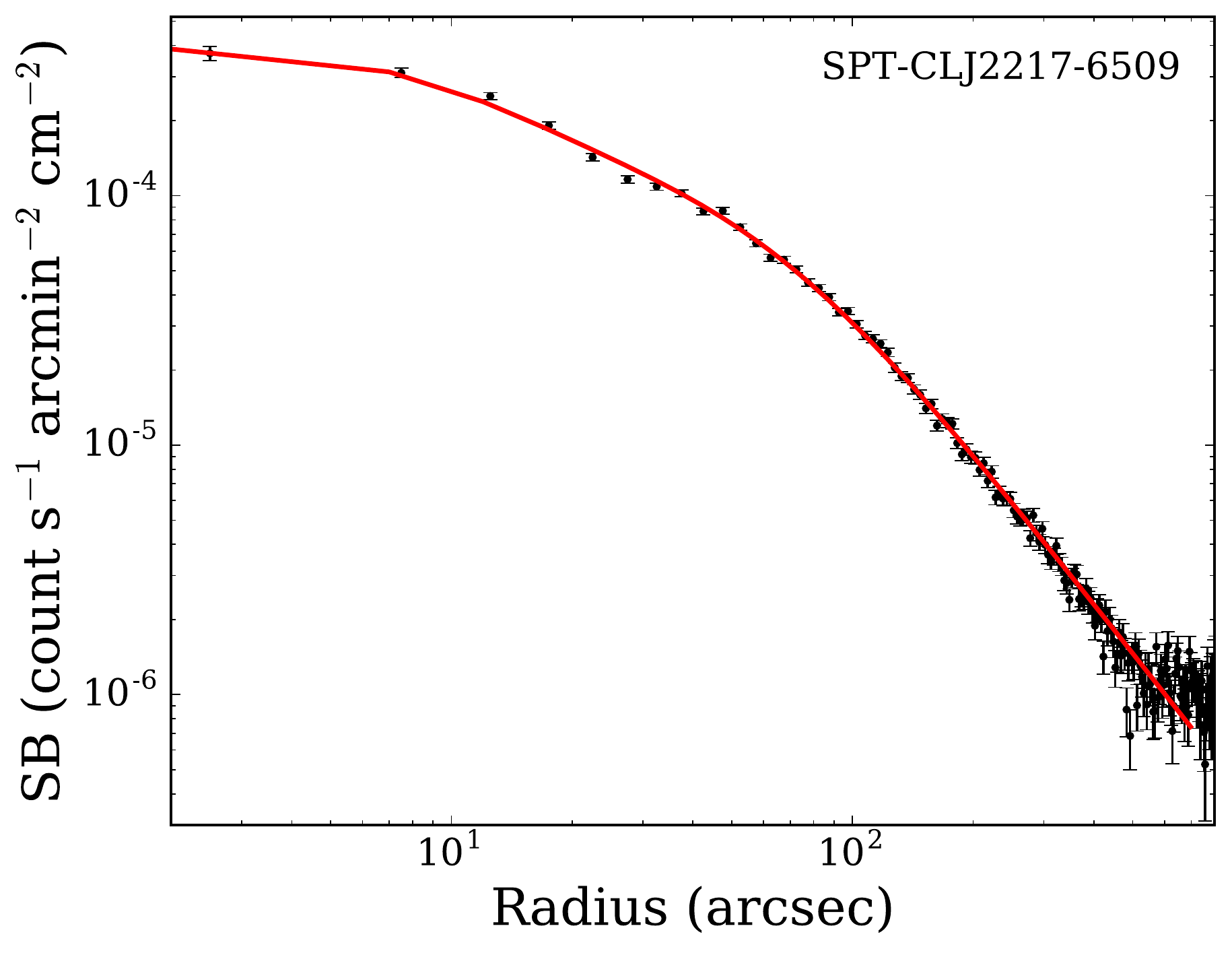}
\caption{\xmm\ MOS surface brightness profile of a non-cool core cluster SPT-CL~J0304$-$4401 (left) and a cool-core cluster  SPT-CL~J0403$-$5719 (right). The red line shows the best-fit  models convolved with the \xmm\ PSF. 
}
\label{fig:sbfit}
\vspace{2mm}
\end{figure*}

The X-ray surface brightness is extracted from background-subtracted, exposure-corrected images within 1.5\Rfiveoo\ in the fitting environment Sherpa in {\it CIAO} \citep{Freeman2001,doe07}.  We fit a 2-dimensional $\beta$ profile to determine the cluster centroids within software package {\it Sherpa}. This method also allows for precise measurements of X-ray centroids of the clusters in the sample.
The X-ray surface brightness $S_{\mathrm{X}}$ (in units of erg~s$^{-1}$~cm$^{-2}$~steradian$^{-1}$), produced by thermal Bremsstrahlung and line emission, is expressed as
\begin {equation}
S_\mathrm{X} = \frac{1}{4\pi(1+z)^{4}}\int n_\mathrm{e} n_\mathrm{H} \Lambda_\mathrm{eH} (T_{\mathrm{X}}, Z)\ dl,
\end{equation}

where $\Lambda_\mathrm{eH}(T_{\mathrm{X}}, Z)$ is the band averaged emissivity which is dependent on plasma temperature and metallicity, $dl$ is the integral along the line of sight, and $z$ is the cluster redshift.
The electron and Hydrogen number densities ($n_\mathrm{e}$ and $n_\mathrm{H}$) have only weak dependence on plasma temperature and assumed abundance when derived from surface brightness in the 0.5--2~keV band \citep{mohr99}.

We fit the surface brightness profiles using an analytic density model \citep[][Bu10 hereafter]{Bulbul2010}:
\begin{equation}
n_{\mathrm{e}}(r) =n_{\mathrm{e}0}\left(\frac{1}{\left(\beta-2\right)} \frac{(1+r/r_{\mathrm{s}})^{\beta-2}-1}{r/r_{\mathrm{s}}(1+r/r_{\mathrm{s}})^{\beta-2}} \right)^{n}
\end{equation}
where $n_{\mathrm{e}0}$ is the normalization of the electron density profile, $r_{\mathrm{s}}$ is the scale radius, $n$ is the slope of the density profile, and $\beta$ is the slope of the dark matter potential. We assume that the dark matter halos of the SPT selected sample follows the Navarro-Frenk-White (NFW) profile with a slope of $\beta=2$ \citep{navarro1997} and provides a good description of the electron density \citep[Bu10;][]{Bonamente2012}. Application of the  L'Hospital rule gives an electron density profile under the assumption of a NFW-like matter profile,
\begin{equation}
n_{\mathrm{e}}(r) = \tau_\mathrm{cool}(r) \ n_{\mathrm{e}0}\left(\frac{\ln (1+r/r_{\mathrm{s}})}{r/r_{\mathrm{s}}} \right)^{n}.
\end{equation}
The Bu10 density profile has been used for fitting both X-ray and SZE data \citep{Landry2013,Romero2017}. The core taper function $\tau_\mathrm{cool}(r)$ is used to fit the surface brightness profiles of cool-core clusters \citep{vikhlinin06}
\begin{equation}
\tau_\mathrm{cool}(r) = \frac{\alpha+(r/r_{\mathrm{cool}})^{\gamma}}{1+(r/r_{\mathrm{cool}})^{\gamma}}
\end{equation}
For non-cool core clusters the parameter $\alpha$ is set to 1. 

The Bu10 density model is projected along the line-of-sight and fit to the surface brightness profile obtained from background subtracted and exposure corrected X-ray images. The fitting is performed using a Markov chain Monte Carlo (MCMC) sampler within the {\it emcee} package in python \citep{foreman13}. The best-fit parameter values and their 1$\sigma$
uncertainties for non-cool core clusters (e.g., $n_{\mathrm{e}0}$, $n$, $r_{\mathrm{s}}$) and cool-core clusters (e.g., $n_{\mathrm{e}0}$, $n$, $\alpha$, $r_{\mathrm{s}}$, and $r_{cool}$) are determined using a maximum likelihood method. The surface brightness profile fit to the MOS observations of a non-cool core cluster SPT-CL~J0304$-$4401 and a cool-core cluster SPT-CL~J2217$-$6509 are shown in Figure~\ref{fig:sbfit}. 

To compute the ICM mass of a cluster within a given aperture of \Rfiveoo, we use the enclosed ICM mass obtained by integrating the best-fit 3D ICM density profile,
\begin{equation}
M_{\mathrm{ICM}} = 4\pi \mu_\mathrm{e} m_\mathrm{p}\int_{0}^{\Rfiveoo} n_\mathrm{e}(r)\ r^{2}\ dr,
\end{equation}
\noindent where $\mu_\mathrm{e}$ is the mean molecular weight of the electrons, and $m_\mathrm{p}$ is the proton mass. The ICM mass measurements within \Rfiveoo\ for each cluster in the sample are given in Table~\ref{table:observables}. We use  $\mu_\mathrm{e}= 1.17$  when determining the cluster ICM mass.
The integrated Compton-$y$ parameter is the product of the ICM mass and temperature
\begin{equation}
\Yx = \Micm\times \Tx \, ,
\end{equation}
where \Tx\ is the projected temperature measured within a 2D aperture either with or without the core and \Micm\ is integrated within a 3D sphere of radius \Rfiveoo.

As described already in Section~\ref{sec:M500}, there are remaining uncertainties in the SZE-based halo masses.  This means that the extraction radius \Rfiveoo\  used above is not a single value for each cluster, but a distribution of values.  To include these uncertainties, we marginalize over them when studying the X-ray scaling relations.  As described in Section~\ref{sec:priors}, this means that we evaluate the X-ray observable
at a range of radii \Rfiveoo\ consistent with the SZE observable $\xi$ and redshift $z$.  Specifically, 
we use the best-fit density profile to calculate the ICM mass in each fit iteration. 
For \Lx\ we extract the X-ray luminosity at a single radius---the baseline \Rfiveoo---in this work, because we find the change in \Lx\ due to the radial range in the surface brightness fit is negligible.
For \Tx\ we have in general too few photons to make spectral fits beyond the baseline \Rfiveoo, and so we adopt only a single radius for the temperature extraction.  This means that for \Yx\ we are properly including the variation of the \Micm\ component with \Rfiveoo\ but not the \Tx\ component.

\section{Scaling Relation Form and Fitting}
\label{sec:sclrelformfit}

Self-similar models, based on gravitational collapse in clusters, predict simple power-law relations between cluster properties \citep[][]{kaiser86} that have been observed \citep[using, e.g., ICM temperature, luminosity, ICM mass, X-ray isophotal size and total halo mass;][]{smith79,mushotzky97,mohr97, mohr99,arnaud99}. As previously noted, the observed scaling relations often depart from self-similiar behavior, and this has been interpreted as evidence of  feedback into the ICM from star formation and AGN as well as radiative cooling in the cluster cores.

In this section, we describe how we determine the best-fit parameters of the X-ray observable--halo mass--redshift scaling relations for the sample of 59 SPT selected galaxy clusters observed with \xmm\ at $0.2<z<1.5$.

\subsection{Three Forms of each Scaling Relation}
\label{sec:form}
We use three functional forms to characterize the X-ray observable--mass--redshift scaling relations. In all cases, there are pivot masses and redshifts that should be chosen to be near the median values of the sample to reduce artificial covariances between the amplitude parameter and the mass and redshift trend parameters.  For the X-ray observable \obsx\ to mass scaling relations, the pivot mass and pivot redshift are $\Mpiv=6.35\times10^{14}\Msun$ and $\zpiv=0.45$, respectively.  
 
The first form, similar to that used in \cite{vikhlinin09a} and many publications since, is defined as follows:
\begin{equation}
\label{eq:FormA}
\obsx = A_{\obsx}\left( \frac{\Mfiveoo}{\Mpiv} \right)^{B_{\obsx}} 
\left( \frac{E(z)}{E(\zpiv)} \right)^{C_{\obsx}},
\end{equation}
where the normalization and trend parameters in mass and redshift are $A_{\obsx}$, $B_{\obsx}$ and $C_{\obsx}$, respectively, for the observable $\obsx$.  Note that the redshift trend in this formulation is expressed as a function of the Hubble parameter $H(z)=H_0E(z)$, where $E^2(z)=\Omega_\mathrm{M}(1+z)^3+\Omega_\Lambda$ at late times in a flat $\Lambda$CDM Universe.  That is, in this parametrization, the redshift evolution of the X-ray observable--mass relation is attributed an explicit cosmological dependence.  In the case where the redshift evolution has a different cosmological dependence than adopted here (e.g., the evolution is non-self similar), then assuming this form will lead to biases in cosmological analyses.  We refer to equation~(\ref{eq:FormA}) as Form I hereafter. 

The second form includes the expected self-similar evolution of the observable with redshift, which depends on the cosmologically dependent evolution of the critical density, while modeling departures of the observable from self-similar evolution with a function $(1+z)^{\gamma_\obsx}$.  With this form we are adopting the view that the departures from self-similar evolution do not have a clearly understood cosmological dependence.  Therefore, we model the departures with the cosmologically agnostic form $(1+z)^{\gamma_\obsx}$ that has been adopted in many previous works \citep[e.g.][]{lin06}.  This form is defined as follows:
\begin{eqnarray}
\label{eq:FormB}
\obsx &= &A_{\obsx}\left( \frac{\Mfiveoo}{\Mpiv} \right)^{B_{\obsx}} 
\left(\frac{E(z)}{E(\zpiv)}\right)^{C_{\mathrm{\obsx,SS}}}
\left( \frac{1+z}{1+\zpiv} \right)^{\gamma_{\obsx}}
\end{eqnarray}
where the normalization and mass trend are similarly characterized by the parameters $A_{\obsx}$ and $B_{\obsx}$, respectively.  The redshift trend is modeled with \css\ fixed to the self-similar expectation along with the factor $(1+z)^{\gamma_{\obsx}}$ to describe the departure of the redshift trend from the self-similar expectation. For instance, $\css = \frac{2}{3}$ for the X-ray temperature--mass--redshift relation. In this way, the parameter $\gamma_{\obsx}$ directly quantifies the deviation from the self-similar redshift trend.  This form of the scaling relation is easily distinguishable, because it has a parameter $\gamma_\obsx$ rather than $C_\obsx$. We refer to equation~(\ref{eq:FormB}) as Form II hereafter. 

The third form we adopt is much like Form II above, but it includes a cross-term between cluster mass and redshift to characterize the possibility of having a redshift-dependent mass trend.  Specifically, the third functional form is
\begin{eqnarray}
\label{eq:FormC}
\obsx &= &A_{\obsx}\left( \frac{\Mfiveoo}{\Mpiv} \right)^{B'_{\obsx}} 
\left(\frac{E(z)}{E(\zpiv)}\right)^{C_{\mathrm{\obsx,SS}}}
\left( \frac{1+z}{1+\zpiv} \right)^{\gamma_{\obsx}}
\end{eqnarray}
where the mass trend $B'_{\obsx}=B_{\obsx} + \delta_{\obsx}\ln \left(\frac{1+z}{1+\zpiv} \right)$ has a characteristic value of $B_{\obsx}$ at the pivot redshift and an additional rate of variation $\delta_{\obsx}$ with redshift.  The  normalization parameter $A_{\obsx}$ and the redshift trend $\gamma_{\obsx}$ are defined as in Form II.  Specifically, the redshift trend is structured to capture the departures from the expected self-similar redshift evolution of the X-ray observable. 
It is worth mentioning that $\delta_{\obsx} = 0$ or statistically consistent with zero indicates there is no evidence for a redshift-dependent mass trend.
We refer to equation~(\ref{eq:FormC}) as Form III hereafter. 

For all three functional forms, we adopt log-normal intrinsic scatter in the observable at fixed mass, defined as
\begin{equation}
\label{eq:intrinsic_scatter}
\sigma_{\ln\obsx}    \equiv    \sigma_{ \ln\left( \obsx   | \Mfiveoo \right) } .
\end{equation}
In this way, each observable \obsx\ to mass scaling relation is parametrized by either $(A_{\obsx}, B_{\obsx}, C_{\obsx}, \sigma_{\ln\obsx})$, $(A_{\obsx}, B_{\obsx}, C_\mathrm{\obsx,SS}, \gamma_{\obsx}, \sigma_{\ln\obsx})$  or $(A_{\obsx}, B_{\obsx}, C_\mathrm{\obsx,SS}, \gamma_{\obsx}, \sigma_{\ln\obsx}, \\ \delta_{\obsx})$, and we denote these parameter sets by $r_{\obsx}$ hereafter for simplicity.  Note that the expected self-similar redshift evolution parameter $C_\mathrm{\obsx,SS}$ is fixed, and so the first two parameterizations have four free parameters, and the last parametrization has five.

\subsection{Fitting Procedure}
\label{sec:fitting}

We briefly introduce the likelihood and fitting framework below and refer the reader to previous publications for more details \citep{liu15a,chiu16c}. 
This likelihood is designed to obtain the parameters of the targeted X-ray observable--mass--redshift scaling relation (e.g., $r_{\obsx}$) for a given sample that is selected using another observable (e.g., the SPT signal-to-noise $\xi$), for which the  observable--mass--redshift relation is already known (e.g., equation~(\ref{eq:sz_sr1}) used in this work).
Specifically, the $i$-th term in the likelihood $\mathcal{L}_{i}$ contains the probability of obtaining
the X-ray observable $\obsx_{i}$ for the $i$-th cluster at redshift $z_{i}$ with SZE signal-to-noise $\xi_{i}$, given the scaling relations $r_\obsx$  and $r_\mathcal{\xi}$.
\begin{equation}
\label{eq:l15_likelihood1}
\begin{split}
\mathcal{L}_{i}(r_{\obsx}, r_{\zeta}) &= 
P(\obsx_{i} | \xi_{i}, z_{i}, r_{\obsx}, r_{\zeta} )    \\
              &= 
\frac{
\int\mathrm{d}\Mfiveoo\ P(\obsx_{i}, \xi_{i}| z_{i}, r_{\obsx}, r_{\zeta} )\ n(\Mfiveoo, z_{i})
}{
\int\mathrm{d}\Mfiveoo\ P( \xi_{i}| z_{i}, r_{\obsx}, r_{\zeta} ) \ n(\Mfiveoo, z_{i})} \, ,
\end{split}
\end{equation}
where $n(\Mfiveoo, z_{i})$ is the mass function whose inclusion allows the Eddington bias correction to be included when determining the mass corresponding to the SZE observable $\xi$ at redshift $z$.  The integrals are over the relevant range of the mass $M_{500}$ used in the mass function.  
The \citet{tinker08} mass function is used with fixed cosmological parameters in calculations of $n(\Mfiveoo, z_{i})$, although given the mass range of the SPT sample the use of a mass function determined from hydrodynamical simulations would make no difference \citep{bocquet16}.

We ignore correlated scatter between the X-ray observable $\obsx$ and SZE observable $\xi$ in our analysis. 
This will not impact our results, because in previous studies of the SPT-SZ sample no evidence of correlated scatter between the X-ray \Yx, X-ray based \Micm\ and the SZE signal-to-noise has emerged \citep{deHaan16,dietrich17}. In future analyses with much larger X-ray samples, we plan to explore again the evidence for correlated scatter in the X-ray and SZE properties of the clusters. As discussed in \citet{liu15a}, in this limit of no correlated X-ray and SZE scatter, there are no selection effects to be accounted for in the X-ray scaling relation.

Based on Bayes' Theorem, the best-fit scaling relation parameters $r_{\mathcal{X}}$ and $r_{\mathcal{X}}$ are obtained by maximizing the 
probability,
%
\begin{equation}
\label{eq:prob_baye}
P(r_{\obsx}, r_{\zeta}) \propto \mathcal{L}(r_{\obsx}, r_{\zeta}) \mathcal{P}(r_{\obsx}, r_{\zeta}) \, ,
\end{equation}
where $\mathcal{P}(r_{\obsx}, r_{\zeta})$ is the prior on $r_{\obsx}$ and $r_{\zeta}$ (see Table~\ref{tab:priors}), and the likelihood $\mathcal{L}(r_{\obsx}, r_{\zeta})$ is evaluated using equation (\ref{eq:l15_likelihood1}) as follows.

\begin{equation}
\label{eq:full_like}
\mathcal{L}(r_{\obsx}, r_{\zeta}) = \prod_{i=1}^{N_{\mathrm{cl}}}~\mathcal{L}_{i}(r_{\obsx}, r_{\zeta}) \, ,
\end{equation}
where $i$ runs over the $N_{\mathrm{cl}}$ clusters. 
We use the python package \texttt{emcee} to explore the parameter space.  The intrinsic scatter and measurement uncertainties of $\xi_{i}$ for each cluster are taken into account while evaluating equation~(\ref{eq:full_like}). 
We have verified that this likelihood recovers unbiased scaling relation parameters by testing it against large mocks ($>1300$ clusters). Moreover, it has been further optimized in the goal of obtaining the parameters of scaling relations in a high dimensional space \citep{chiu16c}.

We note that in each iteration of the chain we use the current value of \Rfiveoo\ for each cluster to recalculate the \Micm\ (see Section~\ref{sec:X-rayObservables}).  For the temperature \Tx\ and the luminosity \Lx\ we extract only once at the \Rfiveoo\ appropriate for the model \zm\ parameter values in our priors (see Table~\ref{tab:priors}), because the impact of adjusting \Rfiveoo\ at each iteration is small.

\subsection{Priors adopted during fitting}
\label{sec:priors}

As discussed in Section~\ref{sec:M500}, we marginalize over the parameters of the $\zeta$--\Mfiveoo-$z$ while fitting the X-ray observable \obsx--\Mfiveoo-$z$ relations (i.e., $r_{\zeta}$ and $r_{\mathcal{X}}$, respectively).  Specifically, we adopt informative priors on $r_{\zeta}$, which have been obtained in a joint cosmology and mass calibration analysis described in \cite[][see Table 3]{deHaan16}.  Our baseline priors on $r_{\zeta}$ are listed in Table~\ref{tab:priors} and correspond to the posterior distributions for each parameter reported in the first results column of \citet[][Table 3 ]{deHaan16}.  We explore a second set of priors on $r_{\zeta}$ corresponding to the posterior parameter distributions from the second results column of \citet[][Table 3]{deHaan16}, and we report those results in Table~\ref{tab:sclrel_col2}.

Our approach allows us to effectively marginalize over the remaining uncertainties in our \Mfiveoo\ estimates. In each iteration of the chain, each cluster has a different halo mass \Mfiveoo\ and associated radius \Rfiveoo. The X-ray observables \Micm\ and \Yx\ defined in Section~\ref{sec:X-rayObservables} are then extracted at this radius \Rfiveoo\ and used to determine the likelihood for this iteration. Final uncertainties on the X-ray observable scaling relation parameters $r_{\mathcal{X}}$ therefore include not only those due to measurement uncertainties but also due to the (largely systematic) uncertainties in the underlying halo masses.

In the fitting, we apply the uniform priors listed in Table~\ref{tab:priors} on $r_{\obsx}$ during the likelihood maximization.  With this approach we evaluate the scaling relations Form I, II, and III.  In Table~\ref{tab:priors} we present the parameter in the first column and the form of the prior in column two.  In this table $\mathcal{N}$ denotes a normal or Gaussian distribution, and $\mathcal{U}$ represents a uniform or flat distribution between the two values presented.

In a final step, we report the parameters of the Form III relation also while fixing the parameters of $r_{\zeta}$ to the best-fit values derived in \citealt[][]{deHaan16} (i.e., the central values listed in Table~\ref{tab:priors}).  Through the comparison to the results when marginalizing over the posterior distributions with those when fixing the $r_{\zeta}$ parameters we can gauge the impact of the remaining systematic uncertainties on the SZE-based halo masses.

We note that the \cite{deHaan16} priors we adopt when estimating cluster halo masses are derived using the cluster mass function information (distribution of clusters in signal-to-noise $\xi$ and $z$) together with a sample of 82 \Yx\ measurements that have been calibrated to mass first through hydrostatic masses \citep{vikhlinin09a} and later through weak lensing \citep{hoekstra15}.  We note that the mass information from the \Yx\ measurements is subdominant in comparison to that from the mass function information \citep[see prior and posterior distributions on $r_\mathcal{\xi}$ parameters in Figure~5 of ][]{deHaan16}.  

Moreover, the follow-up studies using weak lensing masses of 32 SPT-SZ clusters \citep{dietrich17} and using dynamical masses from 110 SPT-SZ clusters \citep{capasso17} have provided independent mass calibration of the \zm\ relation and cross-checks of cluster masses, and they are all in excellent agreement with the cluster masses in \cite{deHaan16} as we adopt for our study.  
Ongoing work with DES weak lensing will further improve our knowledge of the \zm\ relation, allowing even more accurate cluster halo mass estimates in the future \citep[e.g.,][]{stern18}.

\begin{figure*}
\centering
\includegraphics[width=0.95\textwidth]{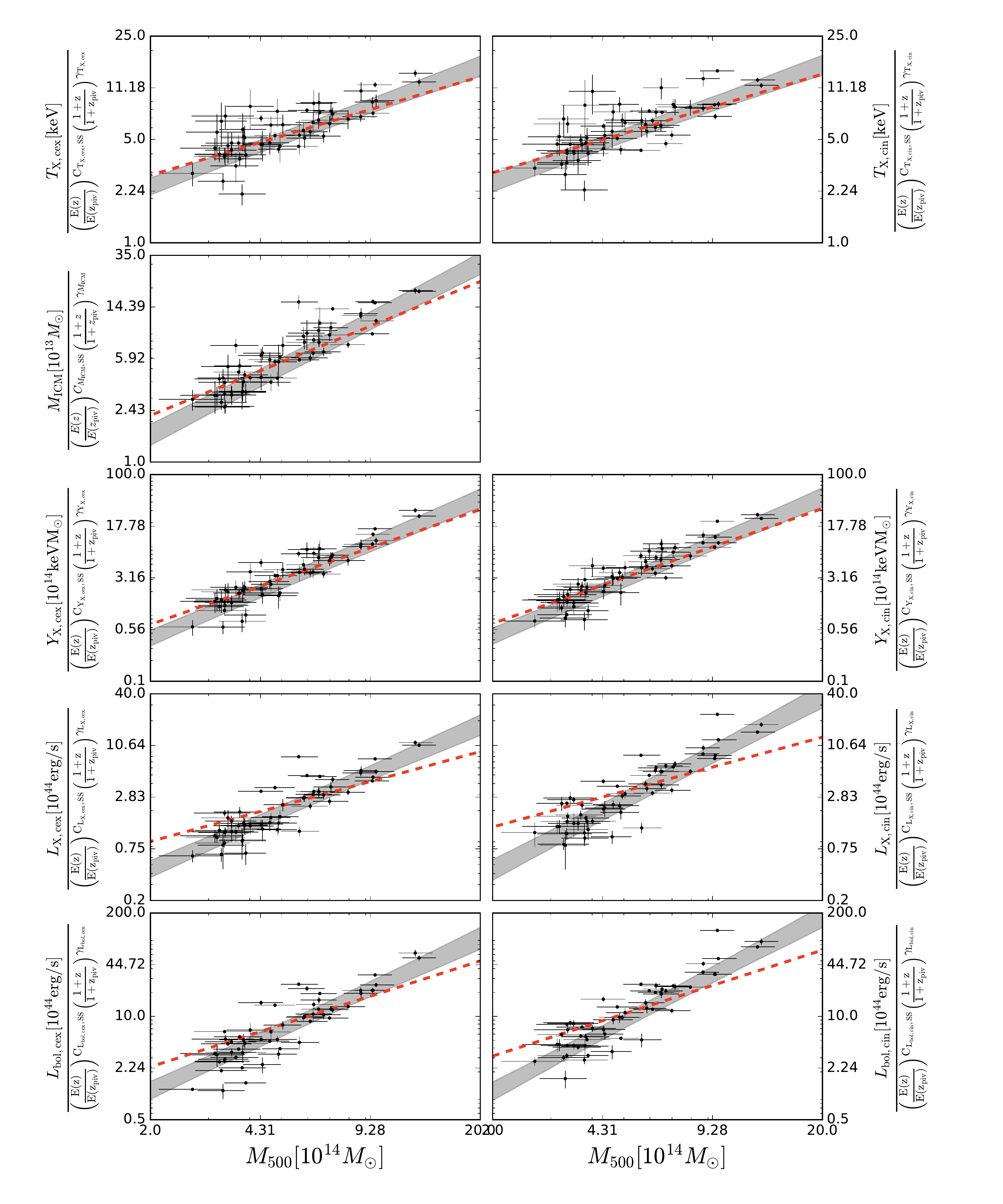}
\vskip-0.2in
\caption{The power law trends of X-ray observables in mass corrected to the pivot redshift $\zpiv=0.45$ using the best-fit redshift trend from the Form II scaling relation (equation~\ref{eq:FormB}) for each observable.  From top to bottom are \Tx, \Micm, \Yx\ and \Lx\, with the core-excised observables (left) and included (right).   The best-fit power law parameters and $1\sigma$ confidence intervals given in Table \ref{tab:sclrel} are shown in the shaded region.
For each row, the red dashed lines represent the best-fit normalizations at the pivotal mass with the mass scaling predicted by the self-similar trend in mass.
}
\label{fig:mtrends}
\vspace{2mm}
\end{figure*}
\begin{figure*}
\centering
\vspace{-2mm}
\includegraphics[width=0.95\textwidth]{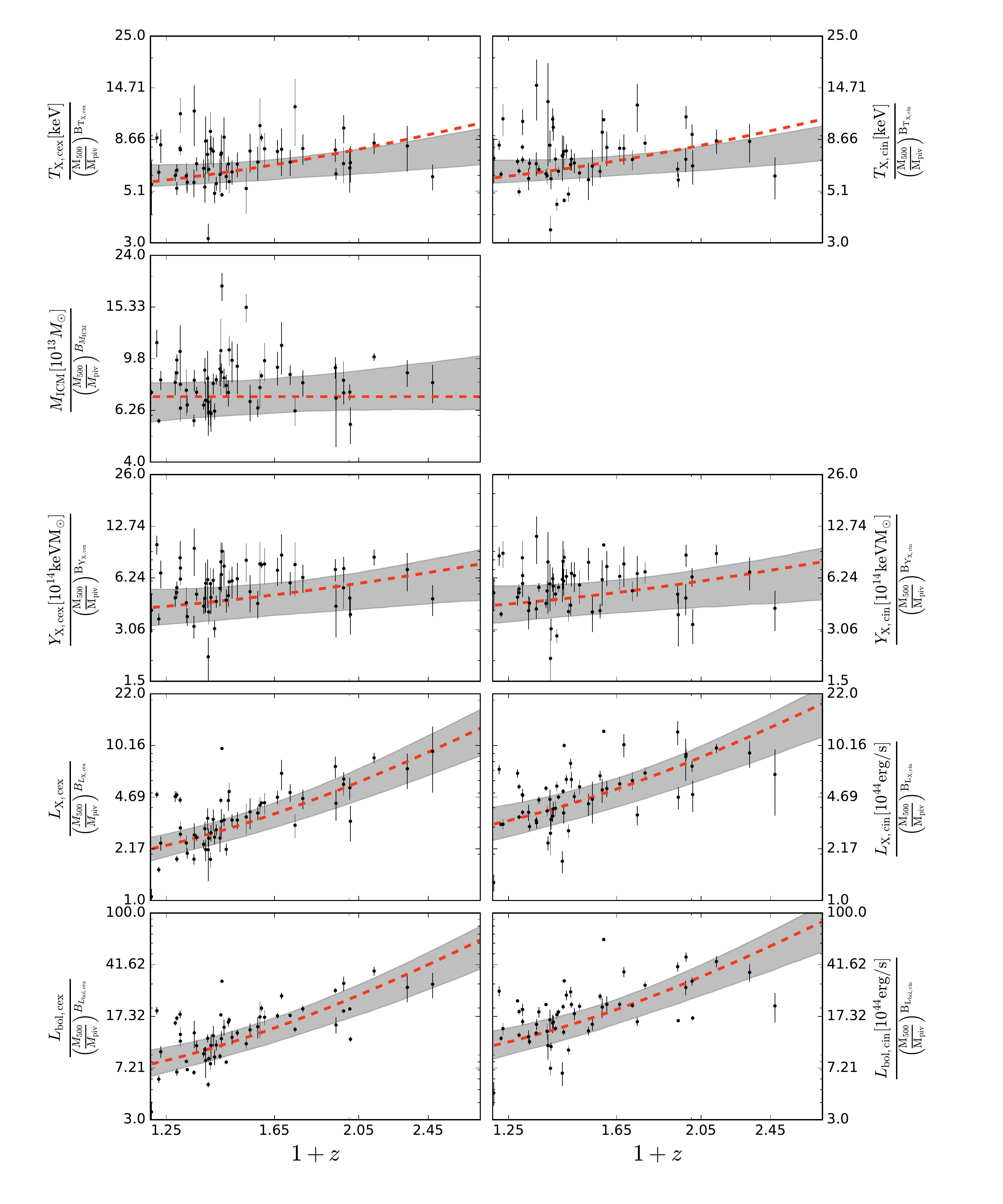}
\vskip-0.2in
\caption{The power law trends of X-ray observables in redshift corrected to the pivot mass $\Mpiv=6.35\times10^{14}\Msun$ using the best-fit mass trend from the Form II scaling relation (equation~\ref{eq:FormB}) for each observable.  From top to bottom are \Tx, \Micm, \Yx\ and \Lx\, with the core-excised observables (left) and included (right).   The best-fit power law parameters and $1\sigma$ confidence intervals given in Table \ref{tab:sclrel} are shown in the shaded region.
For each row, the red dashed lines represent the best-fit normalizations at the pivotal mass with the redshift scaling predicted by the self-similar evolution.
}
\label{fig:ztrends}
\vspace{2mm}
\end{figure*}
\begin{table*}\scriptsize
\caption{\scriptsize
        The best-fit parameters for the various X-ray observable--halo mass--redshift scaling relations.
        The first column contains the scaling relation identifier.
        Thereafter, the next six columns show best-fit parameters and associated fully marginalized $1\sigma$ uncertainties of the scaling relation
        normalization $A_\obsx$, mass trend $B_\obsx$, $E(z)$ redshift trend $C_\obsx$, log-normal intrinsic scatter $\sigma_{\ln\obsx}$, departure 
        from self-similarity in redshift trend $(1+z)^{\gamma_\obsx}$ and redshift dependence $\delta_\obsx$ of the mass trend.  
       }
        \label{tab:sclrel}
	\centering
        \begin{tabular}{lcccccc}
        \hline
        	Scaling Relation  & $A_{\obsx}$ & $B_{\obsx}$ & $C_{\obsx}$ & $\sigma_{\ln\obsx}$ 
        	& $\gamma_{\obsx}$ & $\delta_{\obsx}$
        	 \\[3pt]
		\hline
		\multicolumn{1}{l}{\tmi\ Relation} & & $B_\mathrm{\obsx,SS}={2\over3}$ & $C_\mathrm{\obsx,SS}={2\over3}$ \\
		I: $\obsx(M,z)\propto \Mfiveoo^{B_{\obsx}}E(z)^{C_\obsx}$
		& $6.36^{+0.70}_{-0.64}$ & $0.80\pm 0.10$ & $0.33\pm 0.27$ & $0.18\pm 0.04$ & -- & --
		\\[3pt]
		II: $\obsx(M,z)\propto \Mfiveoo^{B_{\obsx}}E(z)^{2\over3}(1+z)^{\gamma_\obsx}$
		& $6.41^{+0.64}_{-0.66}$ & $0.79^{+0.08}_{-0.12}$ & $2\over3$ & $0.18^{+0.05}_{-0.04}$ & $-0.36^{+0.27}_{-0.26}$ & --
		\\[3pt] 
		III: as II with $B'_\obsx=B_{\obsx} + \delta_{\obsx}\ln\left(\frac{1+z}{1+\zpiv} \right)$ 
		& $6.48^{+0.58}_{-0.69}$ & $0.79^{+0.09}_{-0.10}$ & $2\over3$ & $0.18^{+0.04}_{-0.04}$ & $-0.22^{+0.29}_{-0.35}$ & $0.81^{+0.56}_{-0.46}$
		\\[3pt]
		III with fixed SZE params 
		& $6.41\pm 0.22$ & $0.78^{+0.08}_{-0.09}$ & $2\over3$ & $0.16^{+0.04}_{-0.03}$ & $-0.20^{+0.23}_{-0.25}$ & $0.77^{+0.57}_{-0.47}$
		\\[3pt]
        		\hline
        		\multicolumn{1}{l}{\tme\ Relation}  & & $B_\mathrm{\obsx,SS}={2\over3}$ & $C_\mathrm{\obsx,SS}={2\over3}$\\
		I: $\obsx(M,z)\propto \Mfiveoo^{B_{\obsx}}E(z)^{C_\obsx}$ 
		& $6.17^{+0.71}_{-0.63}$ & $0.83^{+0.09}_{-0.10}$ & $0.28^{+0.28}_{-0.23}$ & $0.13^{+0.05}_{-0.05}$ & -- & --
		\\[3pt]
		II: $\obsx(M,z)\propto \Mfiveoo^{B_{\obsx}}E(z)^{2\over3}(1+z)^{\gamma_\obsx}$ 
		& $6.09^{+0.76}_{-0.51}$ & $0.80^{+0.11}_{-0.08}$ & $2\over3$ & $0.13^{+0.04}_{-0.05}$ & $-0.33^{+0.23}_{-0.28}$ & --
		\\[3pt]
		III: as II with $B'_\obsx=B_{\obsx} + \delta_{\obsx}\ln\left(\frac{1+z}{1+\zpiv} \right)$ 
		& $6.31^{+0.57}_{-0.69}$ & $0.81^{+0.09}_{-0.08}$ & $2\over3$ & $0.13^{+0.05}_{-0.04}$ & $-0.30^{+0.27}_{-0.28}$ & $0.35^{+0.53}_{-0.41}$ 
		\\[3pt]
		III with fixed SZE params
		& $6.17^{+0.20}_{-0.17}$ & $0.79^{+0.10}_{-0.06}$ & $2\over3$ & $0.12^{+0.04}_{-0.03}$ & $-0.29^{+0.19}_{-0.25}$ & $0.38^{+0.41}_{-0.43}$
		\\[3pt]
        \hline
        \multicolumn{1}{l}{\mm\ Relation}  & & $B_\mathrm{\obsx,SS}=1$ & $C_\mathrm{\obsx,SS}=0$\\
		I: $\obsx(M,z)\propto \Mfiveoo^{B_{\obsx}}E(z)^{C_\obsx}$ 
		& $6.80^{+1.09}_{-0.87}$ & $1.260^{+0.10}_{-0.11}$ & $0.17^{+0.28}_{-0.29}$ & $0.12^{+0.04}_{-0.08}$ & -- & --
		\\[3pt]
		II: $\obsx(M,z)\propto \Mfiveoo^{B_{\obsx}}E(z)^{0}(1+z)^{\gamma_\obsx}$ 
		& $7.37^{+0.76}_{-1.35}$ & $1.26^{+0.12}_{-0.09}$ & $0$ & $0.10^{+0.04}_{-0.07}$ & $0.18^{+0.30}_{-0.31}$ & --
		\\[3pt]
		III: as II with $B'_\obsx=B_{\obsx} + \delta_{\obsx}\ln\left(\frac{1+z}{1+\zpiv} \right)$ 
		& $7.09^{+0.91}_{-1.11}$ & $1.26^{+0.11}_{-0.09}$ & $0$ & $0.10^{+0.05}_{-0.07}$ & $0.16^{+0.33}_{-0.31}$ & $0.16^{+0.47}_{-0.44}$
		\\[3pt]
		III with fixed SZE params 
		& $7.02^{+0.21}_{-0.27}$ & $1.26^{+0.09}_{-0.07}$ & $0$ & $0.07\pm 0.05$ & $0.20^{+0.20}_{-0.22}$ & $0.26^{+0.42}_{-0.51}$
		\\[3pt]
        \hline
        \multicolumn{1}{l}{\ymi\ Relation}   & & $B_\mathrm{\obsx,SS}={5\over3}$ & $C_\mathrm{\obsx,SS}={2\over3}$ \\
		I: $\obsx(M,z)\propto \Mfiveoo^{B_{\obsx}}E(z)^{C_\obsx}$ 
		& $4.70\pm 1.1$ & $2.00^{+0.19}_{-0.14}$ & $0.44^{+0.46}_{-0.54}$ & $0.15^{+0.05}_{-0.12}$ & -- & --
		\\[3pt]
		II: $\obsx(M,z)\propto \Mfiveoo^{B_{\obsx}}E(z)^{2\over3}(1+z)^{\gamma_\obsx}$ 
		& $4.60\pm 1.1$ & $1.99^{+0.17}_{-0.15}$ & $2\over3$ & $0.16^{+0.05}_{-0.12}$ & $-0.21^{+0.50}_{-0.45}$ & --
		\\[3pt]
		III: as II with $B'_\obsx=B_{\obsx} + \delta_{\obsx}\ln\left(\frac{1+z}{1+\zpiv} \right)$ 
		& $4.52^{+1.23}_{-0.91}$ & $2.00^{+0.16}_{-0.17}$ & $2\over3$ & $0.16^{+0.07}_{-0.10}$ & $-0.28^{+0.56}_{-0.40}$ & $0.77^{+0.74}_{-0.53}$
		\\[3pt]
		III with fixed SZE params 
		& $4.57^{+0.25}_{-0.21}$ & $1.98^{+0.16}_{-0.10}$ & $2\over3$ & $0.07^{+0.09}_{-0.05}$ & $-0.09^{+0.34}_{-0.32}$ & $1.01^{+0.61}_{-0.71}$
		\\[3pt]
        \hline
        \multicolumn{1}{l}{\yme\ Relation}  & & $B_\mathrm{\obsx,SS}={5\over3}$ & $C_\mathrm{\obsx,SS}={2\over3}$ \\
		I: $\obsx(M,z)\propto \Mfiveoo^{B_{\obsx}}E(z)^{C_\obsx}$ 
		& $4.31\pm 0.96$ & $2.01^{+0.20}_{-0.13}$ & $0.44\pm 0.49$ & $0.16^{+0.04}_{-0.11}$ & -- & --
		\\[3pt]
		II: $\obsx(M,z)\propto \Mfiveoo^{B_{\obsx}}E(z)^{2\over3}(1+z)^{\gamma_\obsx}$ 
		& $4.50^{+1.0}_{-1.1}$ & $2.02^{+0.16}_{-0.17}$ & $2\over3$ & $0.11^{+0.07}_{-0.08}$ & $-0.17^{+0.47}_{-0.50}$ & --
		\\[3pt]
		III: as II with $B'_\obsx=B_{\obsx} + \delta_{\obsx}\ln\left(\frac{1+z}{1+\zpiv} \right)$ 
		& $4.54^{+1.09}_{-0.98}$ & $2.01^{+0.18}_{-0.14}$ & $2\over3$ & $0.13^{+0.07}_{-0.08}$ & $-0.20^{+0.52}_{-0.47}$ & $0.55^{+0.78}_{-0.56}$
		\\[3pt]
		III with fixed SZE params 
		& $4.40^{+0.23}_{-0.22}$ & $2.04^{+0.10}_{-0.15}$ & $2\over3$ & $0.04^{+0.08}_{-0.03}$ & $-0.14^{+0.33}_{-0.32}$ & $0.67^{+0.66}_{-0.74}$
		\\[3pt]
        \hline
        \multicolumn{1}{l}{\lmi\ Relation}  & & $B_\mathrm{\obsx,SS}={1}$ & $C_\mathrm{\obsx,SS}={2}$\\
		I: $\obsx(z)\propto E(z)^{C_\obsx}$ 
		& $4.20^{+0.91}_{-0.92}$ & $1.93^{+0.16}_{-0.20}$ & $1.72^{+0.53}_{-0.46}$ & $0.25^{+0.10}_{-0.10}$ & -- & --
		\\[3pt]
		II: $\obsx(z)\propto E(z)^{2}(1+z)^{\gamma_\obsx}$ 
		& $4.12^{+0.91}_{-0.94}$ & $1.89^{+0.23}_{-0.13}$ & $2$ & $0.27^{+0.08}_{-0.12}$ & $-0.20^{+0.51}_{-0.49}$ & --
		\\[3pt] 
		III: as II with $B'_\obsx=B_{\obsx} + \delta_{\obsx}\ln\left(\frac{1+z}{1+\zpiv} \right)$ 
		& $4.39^{+0.82}_{-0.99}$ & $1.93^{+0.19}_{-0.18}$ & $2$ & $0.28^{+0.07}_{-0.11}$ & $-0.13^{+0.63}_{-0.46}$ & $0.71^{+0.89}_{-0.72}$
		\\[3pt]
		III with fixed SZE params 
		& $3.96^{+0.22}_{-0.24}$ & $1.95^{+0.14}_{-0.18}$ & $2$ & $0.24^{+0.08}_{-0.06}$ & $-0.02^{+0.32}_{-0.48}$ & $0.84^{+0.81}_{-0.80}$
		\\[3pt]
        \hline
        \multicolumn{1}{l}{\lme\ Relation}   & & $B_\mathrm{\obsx,SS}={1}$ & $C_\mathrm{\obsx,SS}={2}$\\
		I: $\obsx(M,z)\propto \Mfiveoo^{B_{\obsx}}E(z)^{C_\obsx}$ 
		& $2.84^{+0.60}_{-0.53}$ & $1.60^{+0.17}_{-0.13}$ & $1.86^{+0.47}_{-0.43}$ & $0.27^{+0.07}_{-0.10}$ & -- & -- 
		\\[3pt]
		II: $\obsx(M,z)\propto \Mfiveoo^{B_{\obsx}}E(z)^{2}(1+z)^{\gamma_\obsx}$ 
		& $2.84^{+0.53}_{-0.50}$ & $1.60^{+0.16}_{-0.15}$ & $2$ & $0.27^{+0.07}_{-0.11}$ & $-0.10^{+0.47}_{-0.42}$ & --
		\\[3pt]
		III: as II with $B'_\obsx=B_{\obsx} + \delta_{\obsx}\ln\left(\frac{1+z}{1+\zpiv} \right)$ 
		& $2.89^{+0.55}_{-0.51}$ & $1.56^{+0.18}_{-0.16}$ & $2$ & $0.28^{+0.07}_{-0.08}$ & $0.10^{+0.35}_{-0.60}$ & $0.30^{+0.86}_{-0.62}$ 
		\\[3pt]
		III with fixed SZE params 
		& $2.66^{+0.17}_{-0.11}$ & $1.60^{+0.14}_{-0.16}$ & $2$ & $0.26^{+0.05}_{-0.05}$ & $-0.01^{+0.33}_{-0.42}$ & $0.60^{+0.79}_{-0.75}$
		\\[3pt]
        \hline
        \multicolumn{1}{l}{\lmib\ Relation}   & & $B_\mathrm{\obsx,SS}={4\over3}$ & $C_\mathrm{\obsx,SS}={7\over3}$\\
		I: $\obsx(M,z)\propto \Mfiveoo^{B_{\obsx}}E(z)^{C_\obsx}$ 
		& $15.4^{+2.8}_{-3.3}$ & $2.15^{+0.24}_{-0.19}$ & $1.90^{+0.55}_{-0.53}$ & $0.29^{+0.09}_{-0.13}$ & -- & --
		\\[3pt]
		II: $\obsx(M,z)\propto \Mfiveoo^{B_{\obsx}}E(z)^{7\over3}(1+z)^{\gamma_\obsx}$ 
		& $14.8^{+3.5}_{-2.7}$ & $2.19^{+0.21}_{-0.17}$ & $7\over3$ & $0.29^{+0.08}_{-0.13}$ & $-0.14^{+0.62}_{-0.57}$ & --
		\\[3pt]
		III: as II with $B'_\obsx=B_{\obsx} + \delta_{\obsx}\ln\left(\frac{1+z}{1+\zpiv} \right)$ 
		& $13.8^{+3.2}_{-3.9}$ & $2.12^{+0.23}_{-0.18}$ & $7\over3$ & $0.31^{+0.08}_{-0.12}$ & $-0.26^{+0.58}_{-0.60}$ & $1.53^{+0.31}_{-1.11}$
		\\[3pt]
		III with fixed SZE params 
		& $14.94^{+0.65}_{-1.01}$ & $2.24^{+0.13}_{-0.15}$ & $7\over3$ & $0.22^{+0.08}_{-0.10}$ & $-0.17^{+0.43}_{-0.33}$ & $1.67^{+0.28}_{-0.97}$
		\\[3pt]
        \hline
        \multicolumn{1}{l}{\lmeb\ Relation}   & & $B_\mathrm{\obsx,SS}={4\over3}$ & $C_\mathrm{\obsx,SS}={7\over3}$\\
		I: $\obsx(M,z)\propto \Mfiveoo^{B_{\obsx}}E(z)^{C_\obsx}$ 
		& $10.2^{+2.6}_{-2.1}$ & $1.89^{+0.17}_{-0.18}$ & $2.01^{+0.53}_{-0.44}$ & $0.29^{+0.07}_{-0.12}$ & -- & --
		\\[3pt]
		II: $\obsx(M,z)\propto \Mfiveoo^{B_{\obsx}}E(z)^{7\over3}(1+z)^{\gamma_\obsx}$ 
		& $10.7\pm 2.3$ & $1.88^{+0.19}_{-0.17}$ & $7\over3$ & $0.27^{+0.07}_{-0.13}$ & $-0.26^{+0.53}_{-0.43}$ & --
		\\[3pt]
		III: as II with $B'_\obsx=B_{\obsx} + \delta_{\obsx}\ln\left(\frac{1+z}{1+\zpiv} \right)$ 
		& $10.4^{+2.4}_{-2.2}$ & $1.86^{+0.21}_{-0.16}$ & $7\over3$ & $0.28^{+0.07}_{-0.09}$ & $0.02^{+0.48}_{-0.58}$ & $0.76^{+0.76}_{-0.71}$
		\\[3pt]
		III with fixed SZE params 
		& $9.93^{+0.58}_{-0.49}$ & $1.90^{+0.13}_{-0.18}$ & $7\over3$ & $0.25^{+0.07}_{-0.06}$ & $-0.18^{+0.48}_{-0.32}$ & $0.80^{+0.93}_{-0.57}$
		\\[3pt]
        	\hline
	\end{tabular}
\end{table*}
\section{Scaling Relation Constraints}
\label{sec:sclrel}

In this section, we describe the results of the fits and compare them to the self-similar expectation and to previous results in the literature.  We present the scaling relations involving \Tx, then followed by \Micm, \Yx\ and \Lx.  For all X-ray observables aside from \Micm\ we present both core-included and -excised scaling relations.  

Best-fit parameters and uncertainties are presented in Table~\ref{tab:sclrel}, where the parameter constraints for each specific X-ray observable are presented in separate, delineated vertical subsections of the table.  Within each table subsection the first line identifies the scaling relation and presents the self-similar expectation for the mass and redshift trends.  Thereafter, the best-fit parameters are presented for the scaling relation Forms I, II, III and then III with fixed SZE scaling relation parameters.  From left to right in the table we present the scaling relation and then the parameters for the normalization $A_\obsx$, mass trend $B_\obsx$, redshift trend  $C_\obsx$ parametrized using $E(z)$, log-normal intrinsic scatter $\sigma_{\ln\obsx}$ of the observable at fixed \Mfiveoo, departure from self-similar redshift scaling $\gamma_\obsx$, and redshift evolution of the mass trend $\delta_\obsx$.

\subsection{\tm\ Relation}
\label{sec:tm}

Before cluster mass measurements were available, the emission-weighted ICM temperature was viewed as the most robust mass proxy available and was therefore employed in early studies of cluster scaling relations \citep{smith79,mushotzky97,mohr97,mohr99}.  Early attempts to constrain the \Tx--mass relation using hydrostatic masses were carried out first for low temperature clusters using spatially resolved spectroscopy from \textit{ROSAT} \citep{david93} and then later for clusters with a broad range of temperature using the \textit{ASCA} observatory \citep{finoguenov01}.  

By combining the virial condition ($GM/R\sim T$) and the definition of the virial radius ($\Rfiveoo\sim [\Mfiveoo/\rho_{\mathrm{crit}}]^{1/3}$) one can show that the self similar expectation for the \tm\ relation is
\begin{equation}
 \Tx \propto {\Mfiveoo}^{2/3}\ E(z)^{2/3} \, .
\end{equation}
As noted in Section~\ref{sec:fitting}, we examine the scaling relations 
with and without the core region, and for three different scaling relation functional forms. 

\begin{figure*}
\centering
\includegraphics[width=0.95\textwidth]{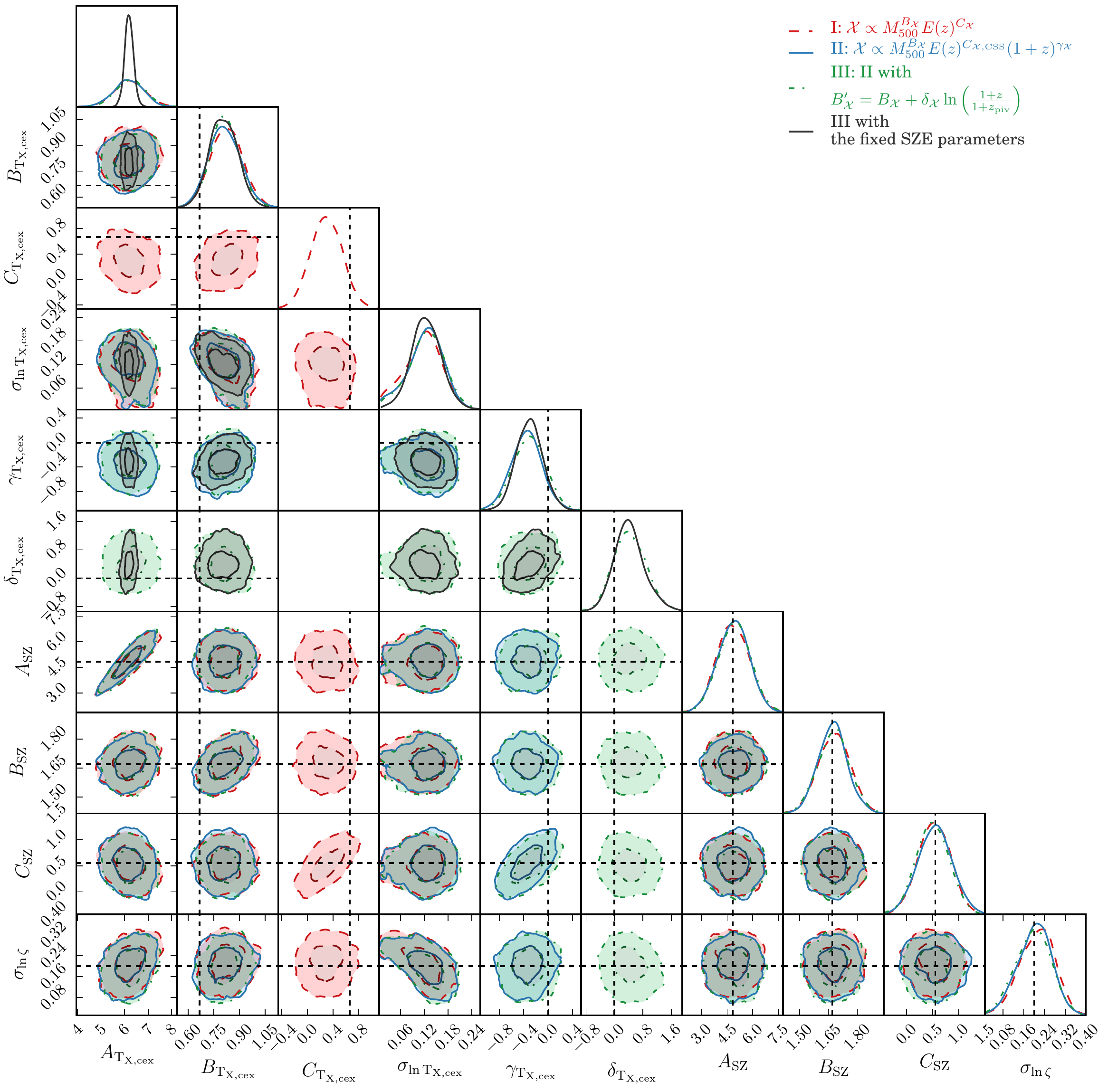}
\caption{
Parameter constraints on the core-excised \tme\ relation.
The parameter constraints while using Forms I, II, and III (see equations~\ref{eq:FormA}, \ref{eq:FormB} and \ref{eq:FormC}) while adopting priors on the SZE mass-observable relation (see Table~\ref{tab:priors}) are shown in red, blue and green, respectively.
In addition, we show in black the results fitting of Form III while fixing the SZE scaling relation parameters to their best-fit values \citep{deHaan16}.
Fully marginalized constraints are shown on the diagonal, and---in the case of the X-ray parameters--are also presented in Table~\ref{tab:sclrel}. 
The off-diagonal plots show joint constraints with $1\sigma$ and $2\sigma$ confidence contours.
Parameters include the normalization $A_{\Txe}$, power law index in mass $B_{\Txe}$ and redshift $C_{\Txe}$, deviation of the redshift trend from the self-similar prediction $\gamma_{\Txe}$, the variation of the mass trend as a function of redshift $\delta_{\Txe}$, and the intrinsic log-normal scatter in observable at fixed mass $\sigma_{\ln\Txe}$.
The parameters of  the SZE observable--mass relation are the normalization $A_{\mathrm{SZ}}$, mass trend $B_{\mathrm{SZ}}$, redshift trend $C_{\mathrm{SZ}}$, and log-normal intrinsic scatter $\sigma_{\ln\zeta}$. 
The dashed lines mark self-similar expectation for the X-ray observable and best-fit values for the SZE mass--observable relation.
}
\label{fig:triangle_txe}
\vspace{2mm}
\end{figure*}
\begin{figure}
\centering
\includegraphics[width=0.45\textwidth]{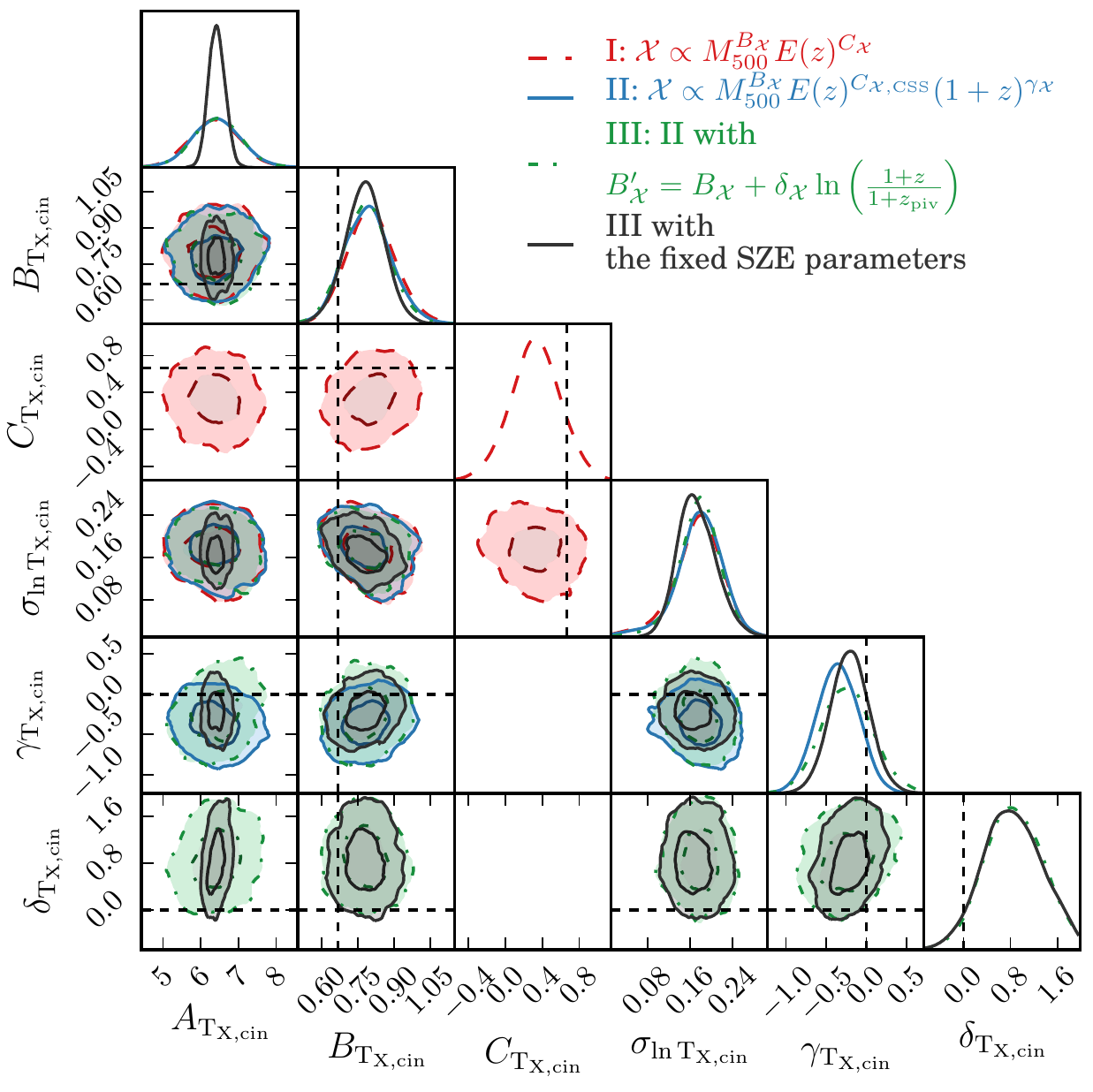}
\caption{Similar to Figure ~\ref{fig:triangle_txe} but containing constraints for the case of the core-included \tmi\ relation. 
}
\label{fig:triangle_txi}
\vspace{2mm}
\end{figure}

\subsubsection{Parameter constraints}

We present the parameters associated with the \tmi\ and \tme\ relations relations in Table~\ref{tab:sclrel}. The marginalized posteriors of the parameters and joint parameter confidence regions using the core-excised and -included observables are contained in Figures~\ref{fig:triangle_txe} and \ref{fig:triangle_txi}.
Here we provide the best-fit \tmi\ and \tme\ relations for the Form II scaling relation.  For the core-included X-ray emission-weighted mean temperature, 
\begin{eqnarray}
\label{eq:txi}
\Txi &= 
&6.41^{+0.64}_{-0.66}~\mathrm{keV} 
\left(\frac{\Mfiveoo}{\Mpiv}\right)^{0.80^{+0.09}_{-0.12}} \nonumber \\
& & \left(\frac{E(z)}{E(\zpiv)}\right)^{\frac{2}{3}} \left(\frac{1+z}{1+\zpiv }\right)^{-0.36^{+0.27}_{-0.26}},
\end{eqnarray}
with the intrinsic scatter of $0.18^{+0.05}_{-0.04}$.
For the core-excised X-ray temperature \Txe, the best-fit relation is
\begin{eqnarray}
\label{eq:txe}
\Txe &= 
&6.09^{+0.76}_{-0.51}~\mathrm{keV}
\left(\frac{\Mfiveoo}{\Mpiv}\right)^{0.80^{+0.11}_{-0.08}} \nonumber \\
&&\left(\frac{E(z)}{E(\zpiv)}\right)^{\frac{2}{3}}\left(\frac{1+z}{1+\zpiv }\right)^{-0.33^{+0.23}_{-0.28}},
\end{eqnarray}
with intrinsic scatter of $0.13^{+0.05}_{-0.04}$.  As for all relations, the mass and redshift pivots are $\Mpiv=6.35\times10^{14}\Msun$ and $\zpiv=0.45$.

The mass trend parameters of the \tme\ relations using forms I, II, and III (see equations~\ref{eq:FormA}, \ref{eq:FormB} and \ref{eq:FormC}) are
$0.83^{+0.09}_{-0.10}$, 
$0.80^{+0.11}_{-0.08}$ and 
$0.81^{+0.09}_{-0.08}$, respectively, showing 
consistency at better than the $1\sigma$ confidence level.
All derived mass trends are consistent with the self-similar expectation at $\approx1.6\sigma$ level.  Note that there is neither a significant redshift-dependence in the mass trend 
($\delta_{\Txe}=0.35^{+0.53}_{-0.41}$) 
nor a strong deviation 
($\gamma_{\Txe}=-0.33^{+0.23}_{-0.28}$) from a self-similar redshift trend.
In addition, as can be seen in Figure~\ref{fig:triangle_txe}, there is a mild degeneracy between the slope and redshift evolution, such that the mass trend can be pushed back closer to the self-similar expectation with a stronger deviation of the redshift evolution from self-similarity.
Fixing the SZE $\zeta$--mass relation does not change the best-fit parameters but reduces the uncertainty on the normalizations $A_{\Tx}$ by a factor of between two and three (see Table~\ref{tab:sclrel}).  This indicates that only the normalizations of the \tm\ relations are dominated by the systematic uncertainties in the SZE-based halo masses.

This behavior is clearly visible in Figure~\ref{fig:triangle_txe}, where we show the joint confidence contours and fully marginalized posterior distributions for each of the \tme\ relation variables for all three forms of the relation. In the figure, the self-similar parameter expectations are marked with dashed red lines. The preference for the mass trend to deviate by $\approx2\sigma$ and the redshift trend to deviate by $\approx1\sigma$ can be seen both in the joint parameter constraint panels and the fully marginalized single parameter distributions. Evidence for parameter covariance is clearest in the parameter pair $C_\Txe$ and $\delta_\Txe$. We include the four \zm\ relation parameters to show the strong positive correlations among the corresponding parameters $A_\mathrm{SZ}$ and $A_\Txe$, $B_\mathrm{SZ}$ and $B_\Txe$, $C_\mathrm{SZ}$ and $C_\Txe$. Also a strong negative correlation among the scatter parameters $\sigma_{\ln\zeta}$ and $\sigma_{\ln\Txe}$ are indicated by the same parameters.  This is as expected and follows from the importance of the SZE-based masses in the \tme\ relation and the fact that the quadrature sum of the scatter in \zm\ and \tme\ is constrained by the measurement-error corrected scatter in the data about the best-fit \tme\ relation.  In all other cases that follow, we exclude the \zm\ parameters from the plots to conserve space, but the correlations persist, as expected.  Finally, this plot makes clear (black lines) that the improvement from fixing the \zm\ scaling relation parameters at their best-fit values is a dramatic decrease in the uncertainties of $A_\Txe$ but only a modest impact on the other parameters.

For the core-included \tmi\ relation, the mass and redshift trends as well as the normalization are consistent with those for the core-exclude case within the quoted $1\sigma$ uncertainties. The discernible difference in the two relations comes from intrinsic scatter.  The log-normal intrinsic scatter for the core-excised observable at fixed mass $\sigma_{\ln\Txe}$ is $\approx0.12$, approximately a factor of 1.5 smaller than in the case of the core-included observable.  Figure~\ref{fig:triangle_txi} contains the joint and single parameter constraints for this relation.

\begin{figure*}
\centering
\includegraphics[width=0.46\textwidth]{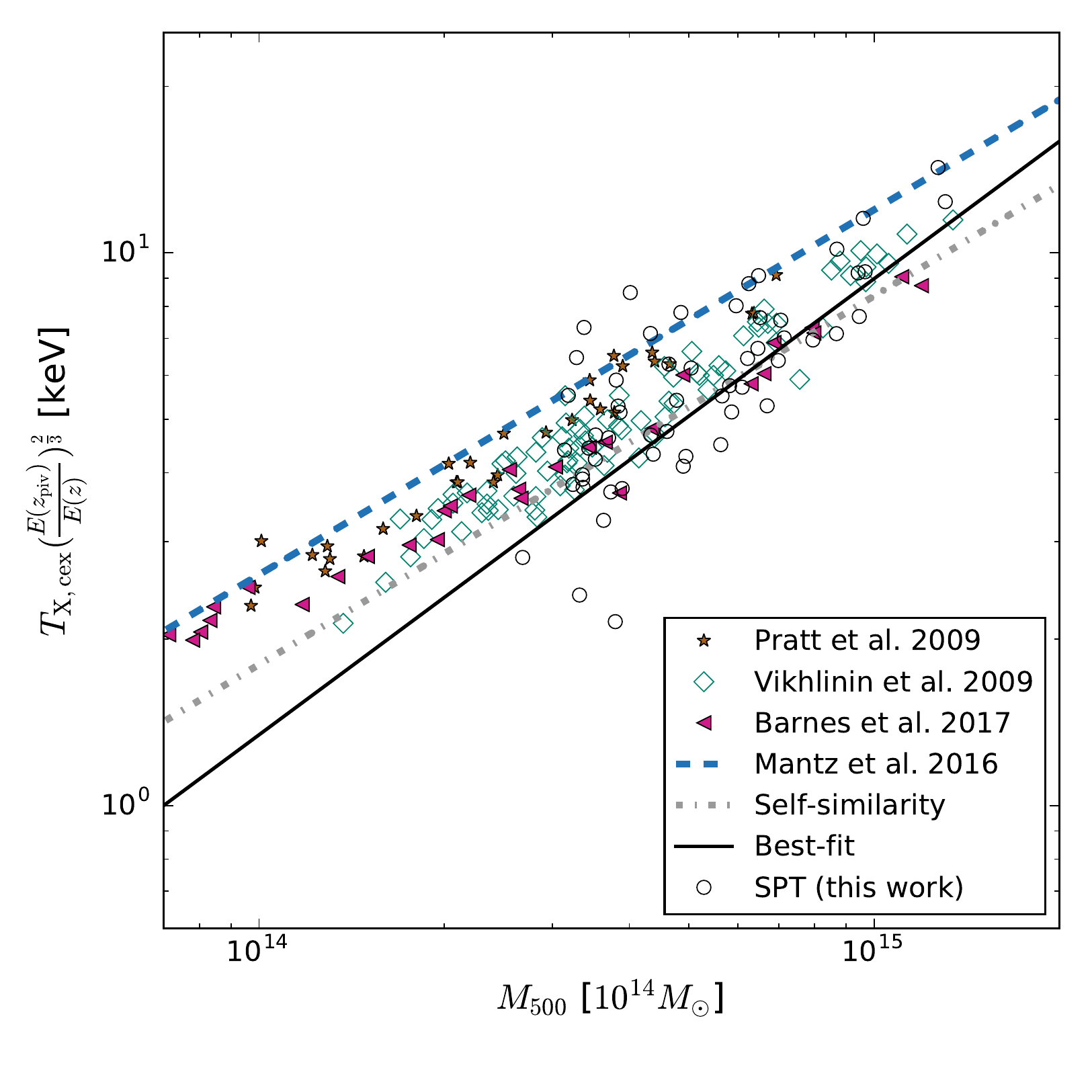}
\includegraphics[width=0.46\textwidth]{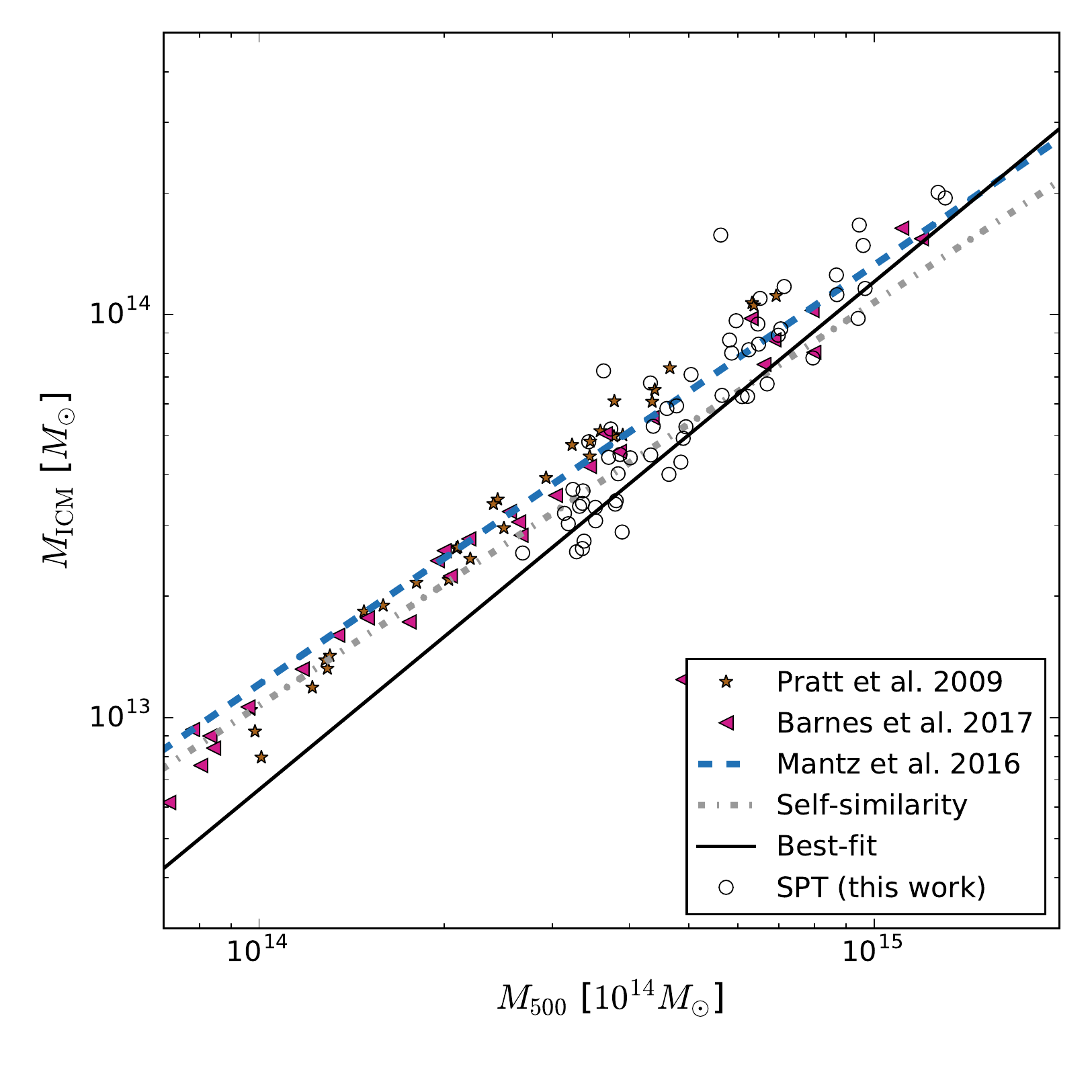}
\includegraphics[width=0.46\textwidth]{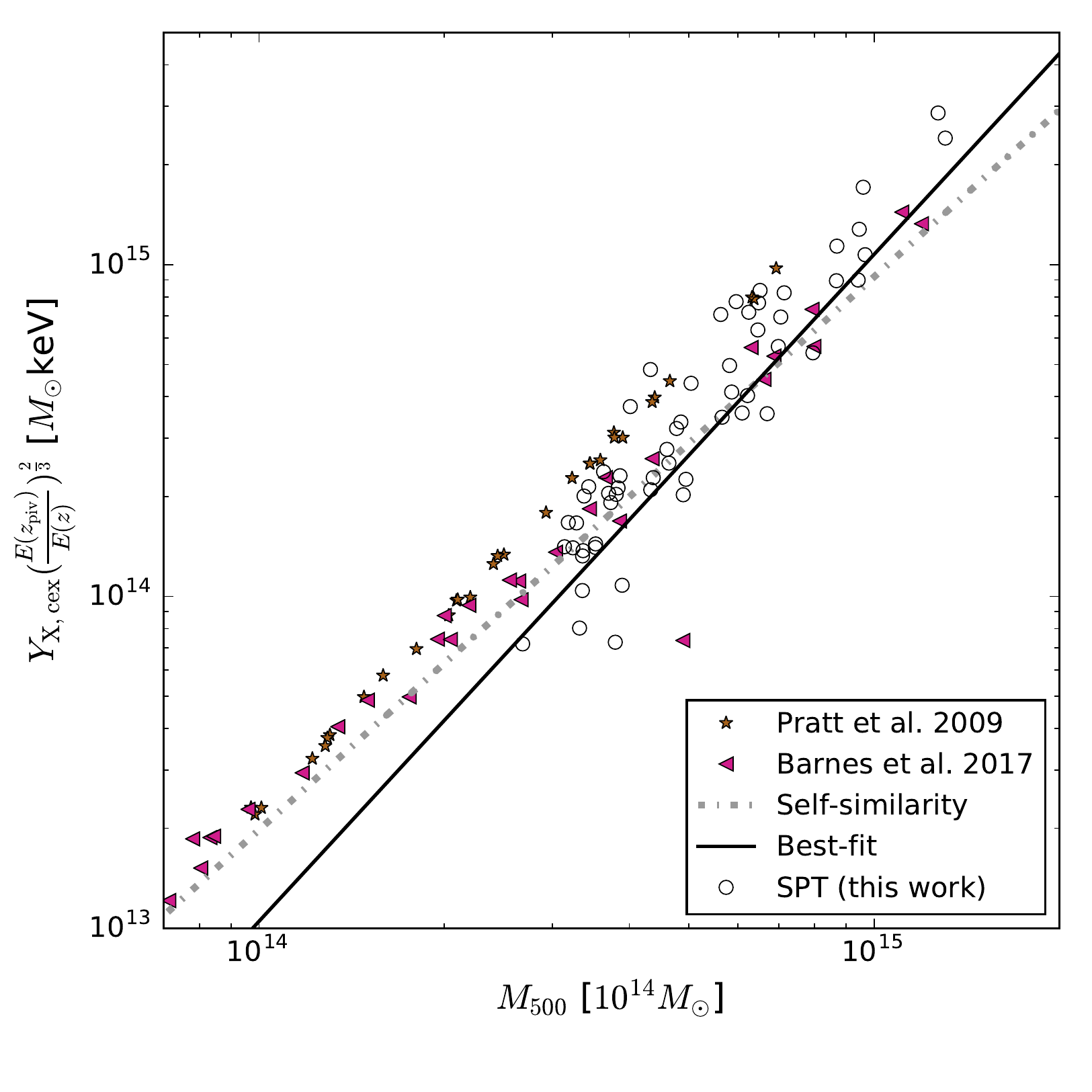}
\includegraphics[width=0.46\textwidth]{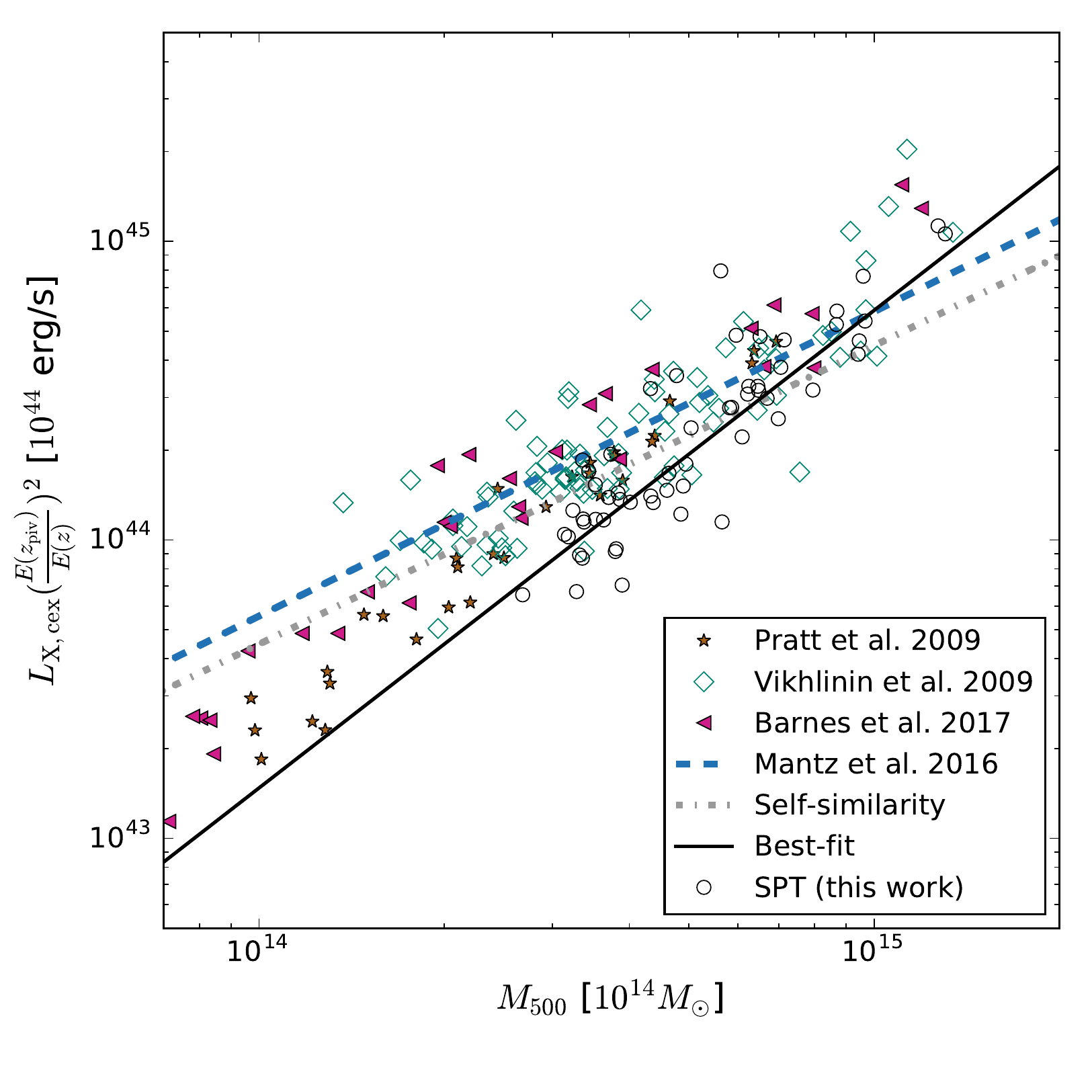}
\caption{Core-excised soft band X-ray luminosity, core-excised \Tx, \Micm\ and \Yx\ are compared with observational data from \citet{vikhlinin09a,pratt09} and cosmological hydrodynamical simulations of the massive C-Eagle clusters from \citet{Barnes2017}. Over-plotted dashed and continuous lines indicate the best-fit scaling relation from \cite{mantz16} and the self-similar expectation with the best-fit normalization reported in Table \ref{tab:sclrel} at the pivotal mass \Mpiv, respectively. 
The black curves indicate the best-fit relations of the SPT clusters (see Table~\ref{tab:sclrel}).
Note that the SPT clusters (open circles) are not the same as the black dots in Figure~\ref{fig:mtrends} because the different re-normalization (i.e., the redshift scaling without the best-fit $\gamma_{\mathcal{X}}$) is applied here.
Our results are broadly consistent with the expectation of scaling relations in simulated clusters. 
}
\label{fig:comps}
\vspace{2mm}
\end{figure*}

\subsubsection{Comparison to previous results}


 
We show the mass and redshift trends in the \Tx\ scaling relations in the top panels of Figs.~\ref{fig:mtrends} and \ref{fig:ztrends}, respectively.  In both figures, the core-excised measurements are on the left and core-included on the right.  In the case of the mass trends, all X-ray observable measurements are corrected to the pivot redshift $\zpiv=0.45$ using the best-fit redshift trend from the Form II scaling relation, and in the case of the redshift trends, all are corrected to the pivot mass $\Mpiv=6.35\times10^{14}\Msun$ using the best-fit mass trend from Form II.  Also shown are the self-similar expectations (red dashed line) and the gray region corresponds to the one sigma allowed region for the relation. One of the outlier clusters in the sample is SPT-CLJ0217-5245, whose temperature is high compared to the expected temperature from the luminosity scaling relations (see Section~\ref{sec:lm}). We note that the noise dominated spectrum of this cluster makes it challenging to determine its temperature from the shape of the continuum Bremsstrahlung emission.

Our analysis shows steeper mass trends than those measured before. \citet{vikhlinin09a}  found a self-similar slope ($B_{T_x} = 0.65\pm0.03$) in the \tme\ relation for X-ray selected clusters observed with \chandra\ in the redshift range of $0.02<z<0.9$ and with hydrostatic mass measurements in the range $10^{14}\Msun\lesssim\Mfiveoo\lesssim10^{15}\Msun$, while a similar mass slope of $B_{T_X} = 0.67\pm0.07$ was reported in \citet{arnaud05} covering the \xmm\ observations of low redshifts clusters at $z\lesssim0.15$ with a hydrostatic mass range of $9\times10^{13}\Msun\lesssim\Mfiveoo\lesssim8.4\times10^{14}\Msun$. Our result for the mass slope is $1.6\sigma$ and $1.1\sigma$ away, respectively, from these results. 
In \cite{mahdavi13}, the mass slopes of $0.51^{+0.42}_{-0.16}$ and $0.70^{+0.11}_{-0.08}$ were derived using weak lensing and hydrostatic masses, respectively, for a sample of 50  galaxy clusters  at $z\lesssim0.5$ with a similar mass range to those we study here; no tension with our results is seen.
A recent result based on the 100 brightest clusters selected in the XXL survey \citep{lieu16} gave slopes of $\approx0.56^{+0.12}_{-0.10}$, and this slope became $\approx0.60\pm0.05$ if combined with the lower mass groups.  Given their preference for a shallower than self-similar relation, these results are in tension at 1.5 and 2.0$\sigma$ with ours. In \cite{mantz16}, a mass trend of $\approx0.66\pm0.05$ was reported for X-ray selected clusters with redshift range of $0.07<z<1.06$ and mass range of $3\times10^{14}\Msun\leq\Mfiveoo\lesssim2\times10^{15}\Msun$. This result is in $1.4\sigma$ tension with ours.

In the upper-left panel of Figure~\ref{fig:comps}, we further compare our results of $\Txe$ to the simulated clusters at $z=0.1$ from the C-Eagle cosmological hydrodynamical simulations \citep{Barnes2017}, together with the nearby clusters from \cite{pratt09} and the clusters at $z\lesssim0.5$ from \cite{vikhlinin09a}.
In addition, we over-plot the best-fit relation from \cite{mantz16} (in blue dashed line) and the self-similar prediction with the normalization anchored to our best-fit value (in grey dashed line).

It is important to note that there exist non-negligible systematic differences among these studies, especially in the estimation of cluster masses.
In \cite{pratt09}, the cluster mass is estimated using the $\Yxe$-\Mfiveoo\ relation derived from hydrostatic mass estimates in nearby, relaxed clusters. 
For the sake of consistency, we take the $\Yx$-inferred masses from \cite{vikhlinin09a}. 
We scale up the cluster masses from \cite{pratt09} and \cite{vikhlinin09a} by a factor of $1.12$  to account for the offsets between hydrostatic masses and our masses \citep{bocquet15}.
For \cite{mantz16}, the cluster masses are calibrated using weak lensing, for which we do not expect significant systematic offsets with our results \citep{deHaan16,dietrich17,schrabback18}.
For the simulated clusters in \cite{Barnes2017}, we directly use the true halo masses.

To make the figure, we re-scale each reported \Txe\ from the literature studies to the pivotal redshift \zpiv\  by multiplying by $\left(E(\zpiv)/E(z)\right)^{\frac{2}{3}}$, because we observe that the core-excised temperature is evolving as predicted by the self-similar evolution.
As seen in Figure~\ref{fig:comps}, our results are consistent in terms of normalization and mass trends with the simulations \citep{Barnes2017} and other observed clusters over the common mass range, except that we observe a shift in normalization in comparison with  \cite{mantz16}.

In summary, the previous results show mass trends that are in agreement with the self-similar prediction (i.e., the value of $2/3$), while the fully marginalized posterior of our mass trend parameter is steeper than self-similar at $\approx1.6\sigma$ and in tension with these previous results at a similar or lower level.
One difference between our work and these others is that we simultaneously fit the mass and redshift trends of the scaling relation, exploiting the fact that our SZE-selected sample is approximately mass selected out to redshift $\approx1.4$.  Most of these previous analyses have assumed self-similar redshift evolution, because their datasets tend to cover very different mass ranges at low and high redshift, introducing strong degeneracies in the mass and redshift trend parameters.  Thus, our sample provides the first direct constraint on the deviation from self-similarity for massive clusters out to $z\approx1.4$ that accounts for both mass and redshift trends.  Only the analysis of larger samples with improved halo mass estimates will allow us to definitively determine departures from self-similarity in the mass trends of the \tm\ scaling relations.

\subsection{\mm\ Relation}
\label{sec:mm}

The \mm\ scaling relation and its redshift trends has important implications for ICM mass fractions and baryon fraction within clusters, because a majority of the baryons reside within the ICM \citep{lin03b,chiu16a}. The expression for the self-similar scaling of \mm\ relation is:
\begin{equation}
\Micm \propto \Mfiveoo.
\end{equation}
That is, in the simplest Universe with no feedback or radiative processes the ICM mass fraction would be expected to be identical in halos of all mass and at all redshifts.

\begin{figure}
\centering
\includegraphics[width=0.45\textwidth]{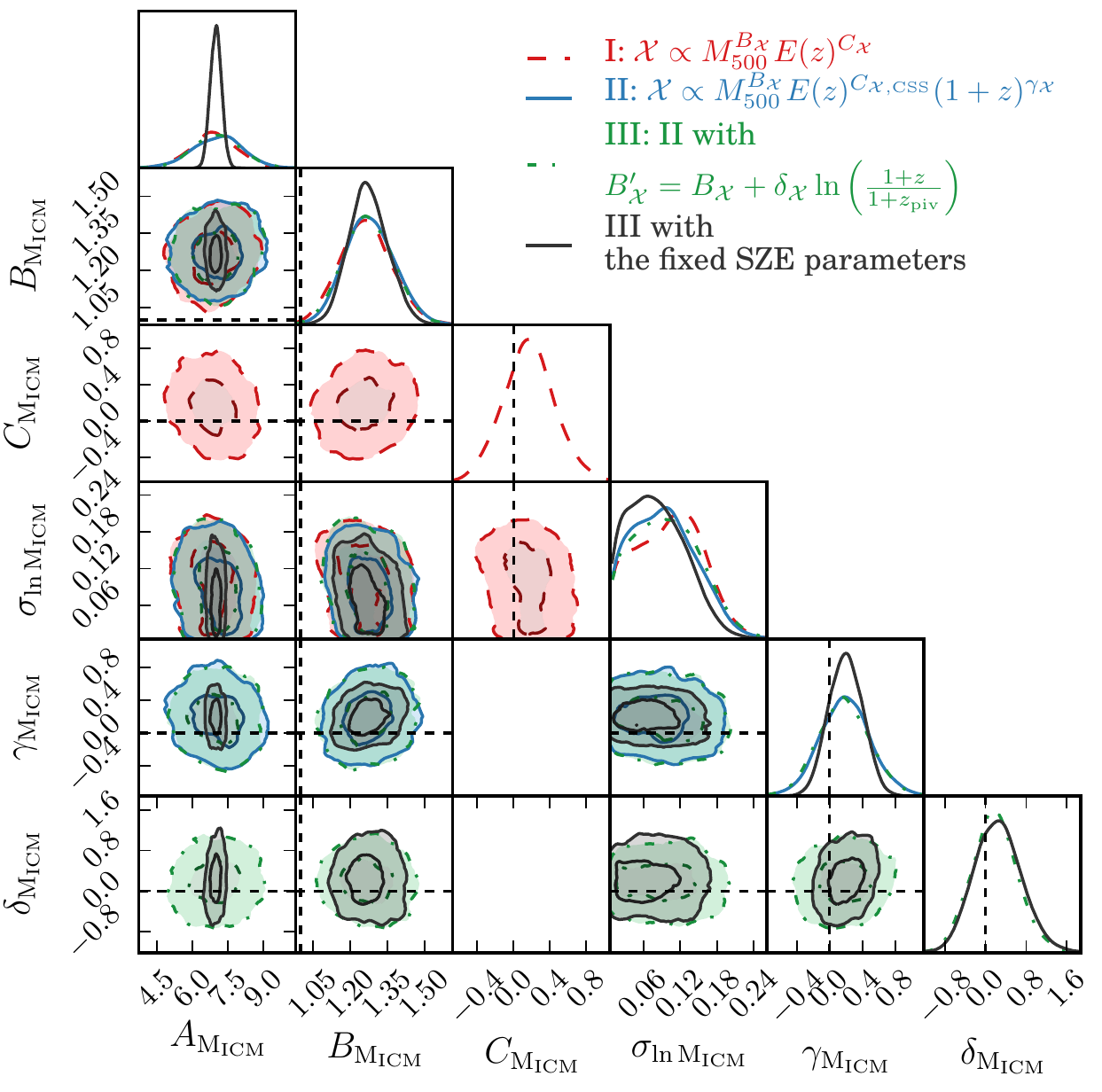}
\caption{Similar to Figure~\ref{fig:triangle_txe} but containing constraints for the case of \mm\ scaling relations.}
\label{fig:triangle_micm}
\vspace{2mm}
\end{figure}

\subsubsection{Parameter constraints}

We present the best-fit parameters of the \mm\ relations using the scaling relation forms I, II, and III (equations~\ref{eq:FormA}, \ref{eq:FormB} and \ref{eq:FormC}) in Table~\ref{tab:sclrel}, and the marginalized posteriors of the single and joint parameter constraints are presented in Figure~\ref{fig:triangle_micm}.  We do not present core-excised values for the ICM mass, because the central core region contains only a negligible portion of the ICM.  The best-fit \mm\ scaling relation using Form II is
\begin{eqnarray}
\label{eq:micm}
\Micm &= 
&7.37^{+0.76}_{-1.35}\times10^{13}\Msun 
\left(\frac{\Mfiveoo}{\Mpiv}\right)^{1.26^{+0.12}_{-0.09}} \times \nonumber \\
&&
\left(\frac{1+z}{1+\zpiv }\right)^{0.18^{+0.30}_{-0.31}} \, ,
\end{eqnarray}
with intrinsic scatter of $0.10^{+0.04}_{-0.07}$.  As before, the mass and redshift pivots are $\Mpiv=6.35\times10^{14}\Msun$ and $\zpiv=0.45$.

We find that the mass trend parameter is $B_{\Micm}=1.26^{+0.12}_{-0.09}$, which is steeper than the self-similar scaling at the $\approx2.9\sigma$ level.
Although the uncertainties are large, there are no significant redshift trends observed.
The data provide no evidence for a redshift-dependent mass slope, given that $\delta_{\Micm}$ of $0.16^{+0.47}_{-0.44}$.
The normalization $A_{\Micm}$ of $7.37^{+0.76}_{-1.35}\times10^{13}\Msun$ suggests an ICM mass fraction of $\approx(16.0\pm2)\%$ at the pivot mass and redshift.
A consistent picture is suggested by all three functional forms.  Furthermore, fixing the SZE parameters $r_{\zeta}$ does not shift the best-fit parameters, but reduces the uncertainty of the normalization by a factor of four and the uncertainties on the mass and redshift trends by a factor of two.

\subsubsection{Comparison to previous results}
We show the redshift and mass trends of \Micm\ in the second row from the top of Figs.~\ref{fig:mtrends} and \ref{fig:ztrends}, respectively.  As for the case of the other X-ray observables shown in this plot, we scale the measurements to the pivot redshift $\zpiv=0.45$ or pivot mass $\Mpiv=6.35\times10^{14}\Msun$ using the best-fit redshift and mass trends from the Form II relation (see Table~\ref{tab:sclrel}).

Our measured mass trends are consistent with that found by \citet[][$B_{\Micm}=1.38\pm0.36$]{zhang12}, where a sample of 19 clusters ($z<0.1$ and $2\times 10^{14}\Msun\lesssim\Mfiveoo\lesssim 2\times10^{15}\Msun$) selected by their X-ray fluxes was studied, and also in the study of the 100 brightest galaxy clusters and groups at redshift range of 0.05--1.1 and and mass range of $10^{13}-10^{15}\Msun$) selected from the XXL survey \citep[$B_{\Micm}=1.21^{+0.11}_{-0.10}$;][]{eckert16}.  The mass trends derived from low-redshift clusters \citep[][$1.24\pm 0.06$ and $1.21\pm 0.03$, respectively]{arnaud07,pratt09} using hydrostatic masses are also consistent with our measurement.  Our derived mass trends are in good agreement with those derived based on the SPT clusters observed with \chandra\ \citep[][$B_{\Micm}=1.33\pm0.07$]{chiu16b,chiu17}.
The agreement between our results and previous \chandra-based works of SPT-selected clusters indicates that \mm\ relation is relatively insensitive to the instrumental systematics.

In two other works, mass trends more consistent with self-similar behavior have been found. Our results are in tension with the \citet[][$B_{\Micm}=1.04\pm0.10$]{mahdavi13} weak lensing analysis at $\approx1.6\sigma$ and with \citet[][$B_{\Micm}=1.004\pm0.015$]{mantz16} analysis of massive, RASS selected clusters at $\approx2.9\sigma$.  The tension with the \citet{mantz16} results is particularly strong, but in general we find that our results are in excellent agreement with those from past studies carried out either with weak lensing or hydrostatic masses.  Most previous studies have been carried out over a narrower range of lower redshifts.  And indeed, as already noted, our data provide no evidence for a redshift dependent mass slope.

Similarly, In Figure~\ref{fig:comps} we also compare our results with simulated clusters \citep{Barnes2017}, together with \cite{pratt09}, \cite{vikhlinin09a} and \cite{mantz16}.
Note that in addition to shifting the hydrostatic mass based halo masses as described in the previous section, here we also scale up the \Micm\ measurements in the literature by $3.8\%$ because \Micm\ is increasing linearly with cluster radius (i.e., $\Micm\propto\Rfiveoo\propto{\Mfiveoo}^{\frac{1}{3}}$).
In the case of the ICM mass, our results show good agreement with the simulations and with the previous results, although with a steeper mass trend in comparison to \cite{mantz16} (see the discussion above).

It is worth noting that the intrinsic scatter $\sigma_{\ln\Micm}$ in \Micm\ at fixed halo mass is at the $\approx10\%$ level, indicating that the ICM mass is among the highest quality cluster mass proxies available.

\subsection{\ym\ Relation}
\label{sec:ym}

The X-ray estimated integrated pressure $Y_{\mathrm{X}}$ is of interest because of relatively low intrinsic scatter, its connection to the SZE observable and its relative insensitivity to the influence of feedback from AGN and star formation
\citep{kravtsov06a,nagai07,bonamente08,vikhlinin09a,andersson11,benson13}.  
The self-similar expectation of the $Y_{\mathrm{X}}$ to mass scaling relations is;
\begin{equation}
Y_{\mathrm{X}} \propto \Mfiveoo^{5/3}\ E(z)^{2/3},
\end{equation}
which results from \Yx\ being the product of \Micm\ and \Tx\ together with the dependence of the \tm\ relation on the evolution of the critical density.

\begin{figure}
\centering
\includegraphics[height=0.37\textheight]{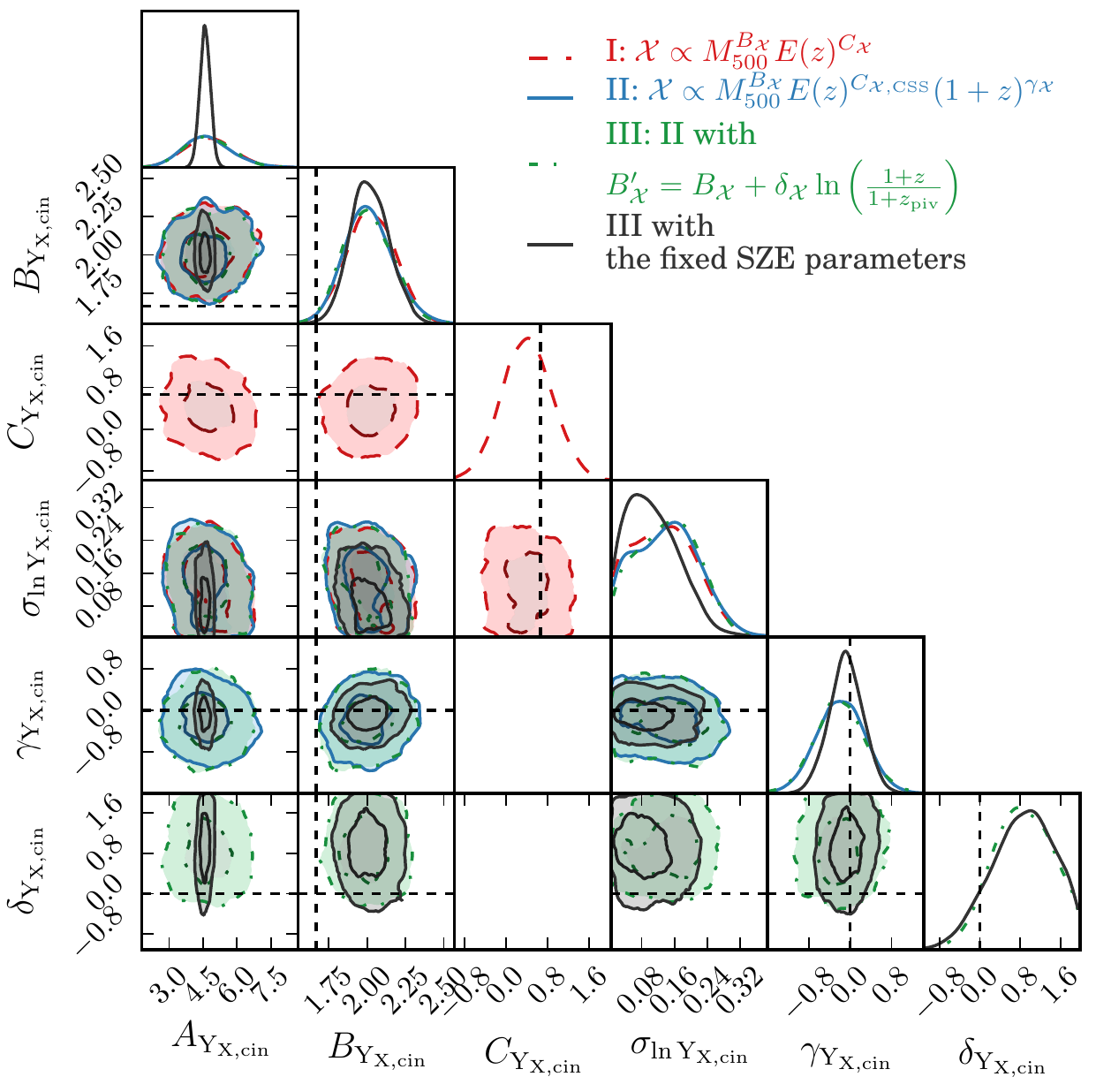}
\includegraphics[height=0.37\textheight]{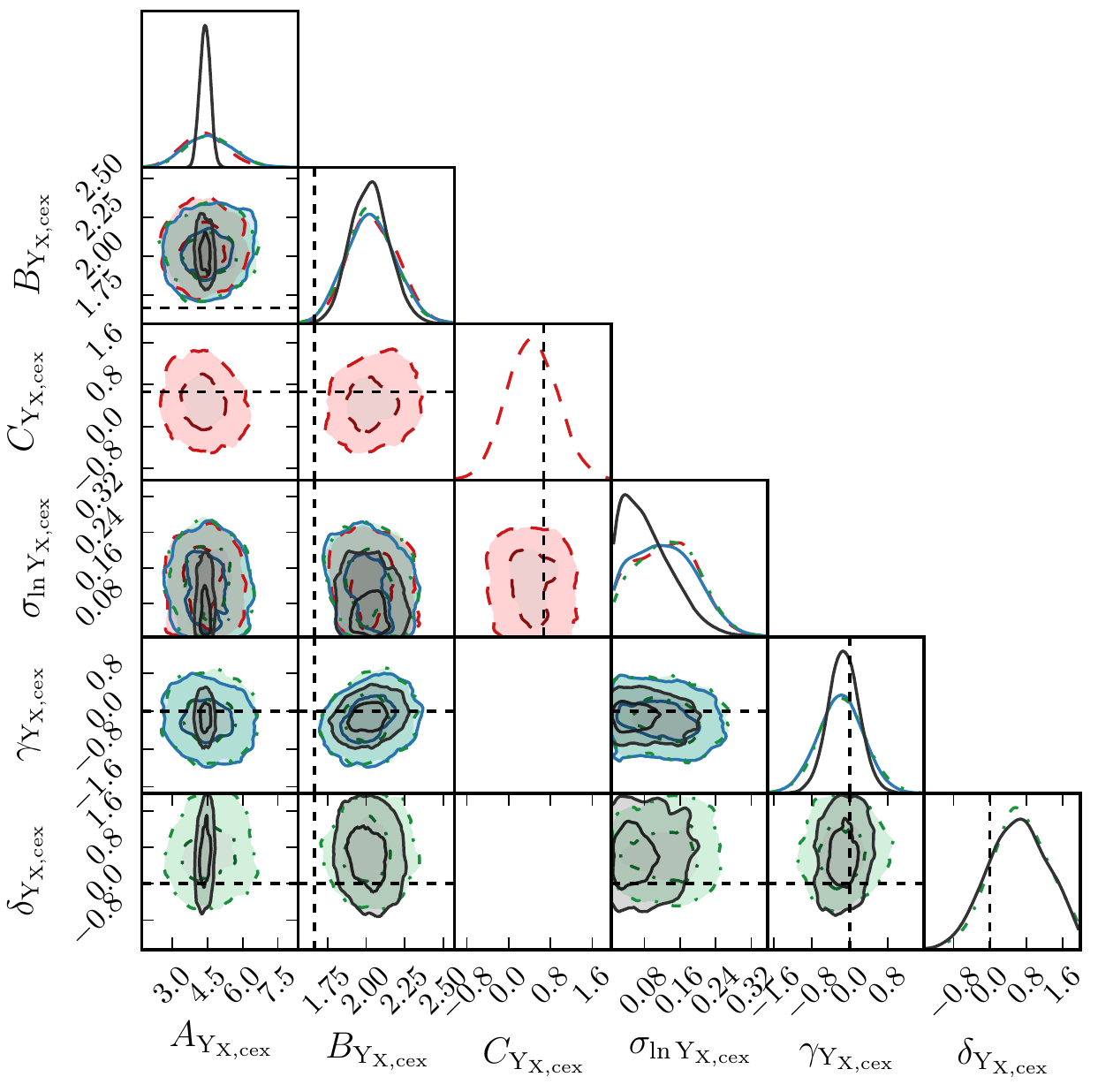}
\caption{Similar to Figure~\ref{fig:triangle_txe} but containing constraints for the cases of the core-excised \ymi\  (the upper panel) and core-included \yme\ (the lower panel) scaling relations.}
\label{fig:triangle_yx}
\vspace{2mm}
\end{figure}

\subsubsection{Parameter constraints}

Similar to previous sections, we present the \ym\ relation derived using both the core-included \Yxi\ and -excised \Yxe\ observables for the scaling relations forms I, II, and III (equations~\ref{eq:FormA}, \ref{eq:FormB} and \ref{eq:FormC}, respectively).
The best-fit parameters and uncertainties of the scaling relations are listed in Table~\ref{tab:sclrel} for both \Yx\ observables, and the marginalized posteriors of the single and joint parameters are presented in Figure~\ref{fig:triangle_yx}.

The best-fit \ym\ scaling relation using functional form II in the core-included case is
\begin{eqnarray}
\label{eq:yxi}
\Yxi &= 
&4.6^{+1.1}_{-1.1}\times10^{14}~\Msun\mathrm{keV}
\left(\frac{\Mfiveoo}{\Mpiv}\right)^{1.99^{+0.17}_{-0.15}} \nonumber \\
&&\left(\frac{E(z)}{E(\zpiv)}\right)^{\frac{2}{3}}
\left(\frac{1+z}{1+\zpiv }\right)^{-0.21^{+0.50}_{-0.45}},
\end{eqnarray}
with intrinsic scatter of $0.16^{+0.05}_{-0.12}$.
For the core-excised observable \Yxe\ the best-fit relation is
\begin{eqnarray}
\label{eq:yxe}
\Yxe &= 
&4.50^{+1.00}_{-1.10}\times10^{14}~\Msun\mathrm{keV}
\left(\frac{\Mfiveoo}{\Mpiv}\right)^{2.02^{+0.16}_{-0.17}} \times \nonumber \\
&&\left(\frac{E(z)}{E(\zpiv)}\right)^{\frac{2}{3}}
\left(\frac{1+z}{1+\zpiv }\right)^{-0.17^{+0.47}_{-0.50}} \, ,
\end{eqnarray}
with intrinsic scatter $0.11^{+0.07}_{-0.08}$.  As for all other cases, the mass and redshift pivots are $\Mpiv=6.35\times10^{14}\Msun$ and $\zpiv=0.45$.

For \ym\ relations, we observe the mass trend $B_{\Yx}$ that is in tension with the self-similar prediction at the $\approx2\sigma$ level for the core-included and -excised X-ray observables.  On the other hand, the redshift trends for all three functional forms are consistent with self-similarity within the quoted $1\sigma$ uncertainty.  There is no evidence for a redshift-dependent mass trend.  Fixing the SZE parameters $r_{\zeta}$ leads to no major parameter shifts, but does reduce the parameter uncertainties on the normalization by a factor of about four and, interestingly, leads to a reduction in the estimate of the intrinsic scatter.  With intrinsic scatter at the $\approx10\%$ level as with the \Micm, the \Yx\ observable with or without core excision offers an outstanding single cluster mass proxy.

\subsubsection{Comparison to previous results}

We show the redshift and mass trends of \Yx\ in the third row from the top of Figs.~\ref{fig:mtrends} and \ref{fig:ztrends}, respectively.  As for the case of the other X-ray observables shown in this plot, we scale the measurements to the pivot redshift $\zpiv=0.45$ or pivot mass $\Mpiv=6.35\times10^{14}\Msun$ using the best-fit redshift and mass trends from the Form II relation (see Table~\ref{tab:sclrel}).

Given that we are adopting halo masses from the \zm\ scaling relation calibrated in the analysis of \citet{deHaan16}, we note that the slope of the \ym\ relation was found in that work to favor a scaling steeper than its self similar predicted value (i.e., $\approx$2 vs 1.67).  In this work, we measure X-ray observables (\Tx, \Micm, \Yx, and \Lx) for the SPT-SZ cluster sample using a different set of observations from the XMM-Newton satellite.  While these data are independent of the data used in \citet{deHaan16}, one would nevertheless expect that given the results of that earlier analysis using \chandra\ data,  that we should see a \ym\ relation that is steeper than self-similar, as indeed we do.

In comparison to other previously published results, the constraints on the mass trend of the full sample is steeper than the reported value in \citet[][$B_{\Yx}=1.83\pm0.09$]{arnaud07}, a difference of $\approx1\sigma$. Other studies employing X-ray hydrostatic masses also resulted in shallower slopes \citep[][$B_{\Yx}=1.75\pm0.09$ and $B_{\Yx}=1.67\pm0.08$, respectively]{vikhlinin09a,Lovisari2015}, which also show weak tension with our results at 1.4 and 1.8$\sigma$ significance, respectively.
The weak lensing based study of \citet{mahdavi13} also found a weaker mass trend of $B_{\Yx}=1.79\pm0.22$ that is nonetheless statistically consistent with our results. The tension between our result and the \citet{mantz16} analysis ($B_{\Yx}=1.61\pm0.04$) is at the 2.3$\sigma$ level.  

In Figure~\ref{fig:comps} we also compare our core-excised \Yxe\ with simulated clusters \citep{Barnes2017} and the observations from \cite{pratt09} and \cite{vikhlinin09a}.
Similar to the case of ICM mass, we also scale up the \Yxe\ by $3.8\%$ because of $\Yxe\equiv\Txe\Micm$.
Our results are broadly consistent with both the simulated and observed clusters but with a preference for a slope that is steeper than the self-similar prediction.

As with the \tm\ relations presented previously, our measured \Yx\ mass trends are steeper and exhibit greater tension with self-similar behavior than previous works.  This can be understood as the combination of the \Tx\ and \Micm\ mass trends---each steeper than self-similar---presented in the last two sections.  However, while the \mm\ relation mass trend we measure is in good agreement with previous analyses, it is our \tm\ relation that appears steeper.  Whether this is due to our unique SZE-selection, leading to an approximately mass-limited sample over a very large redshift range, or due to systematic differences in our mass estimates that include Eddington and Malmquist bias corrections that are typically not considered in earlier works, this must be clarified with a larger sample of clusters and with the ongoing improvements in mass calibration of our own sample.

\subsection{\lm\ Relation}
\label{sec:lm}

We extract the X-ray luminosity obtained from the core-included aperture of $<\Rfiveoo$ in the 0.5--2~keV (i.e., the soft-band luminosity $L_{\mathrm{X}}$) and the 0.01:100~keV band (i.e., the bolometric luminosity $L_{\mathrm{X,bol}}$) to study the \lm\ scaling relations.
In previous studies, the \lm\ scaling relations have tended to exhibit larger scatter if cluster cores are included in the analysis \citep{pratt09} due to the complex cool-core phenomenon that impacts the central regions of clusters.  Indeed, it was argued long ago that the primary driver of the $\Lx-\Tx$ relation scatter was this cool core phenomenon \citep{fabian94c}, and with the availability of cluster samples extending to high redshift it was shown to be true out to $z\approx0.8$ \citep{ohara06}.
Therefore, we also additionally extract the X-ray luminosities obtained from the core-excised aperture of $(0.15-1)\Rfiveoo$ in both soft and bolometric bands.
As a result, we derive four \lm\ scaling relations---(1) core-included and soft-band luminosity to mass \Lxi--\Mfiveoo, (2) core-included and bolometric luminosity to mass \Lxib--\Mfiveoo, (3) core-excised and soft-band luminosity to mass \Lxe--\Mfiveoo, and (4) core-excised and bolometric luminosity to mass \Lxeb--\Mfiveoo\ scaling relations.   
The self-similar expectation of the \lm\ scaling relation is
\begin{eqnarray}
 L_{\mathrm{X}} &\propto &\Mfiveoo \ E(z)^{2} \, .  \nonumber \\
 L_{\mathrm{X,bol}} &\propto &\Mfiveoo^{4/3} \ E(z)^{7/3} \, .  \nonumber \\
\end{eqnarray}
for the the soft-band and bolometric luminosities, respectively, where for the soft-band we have assumed that the emissivity is temperature independent \citep[see discussion in][]{mohr99}.

\begin{figure}
\centering
\includegraphics[height=0.37\textheight]{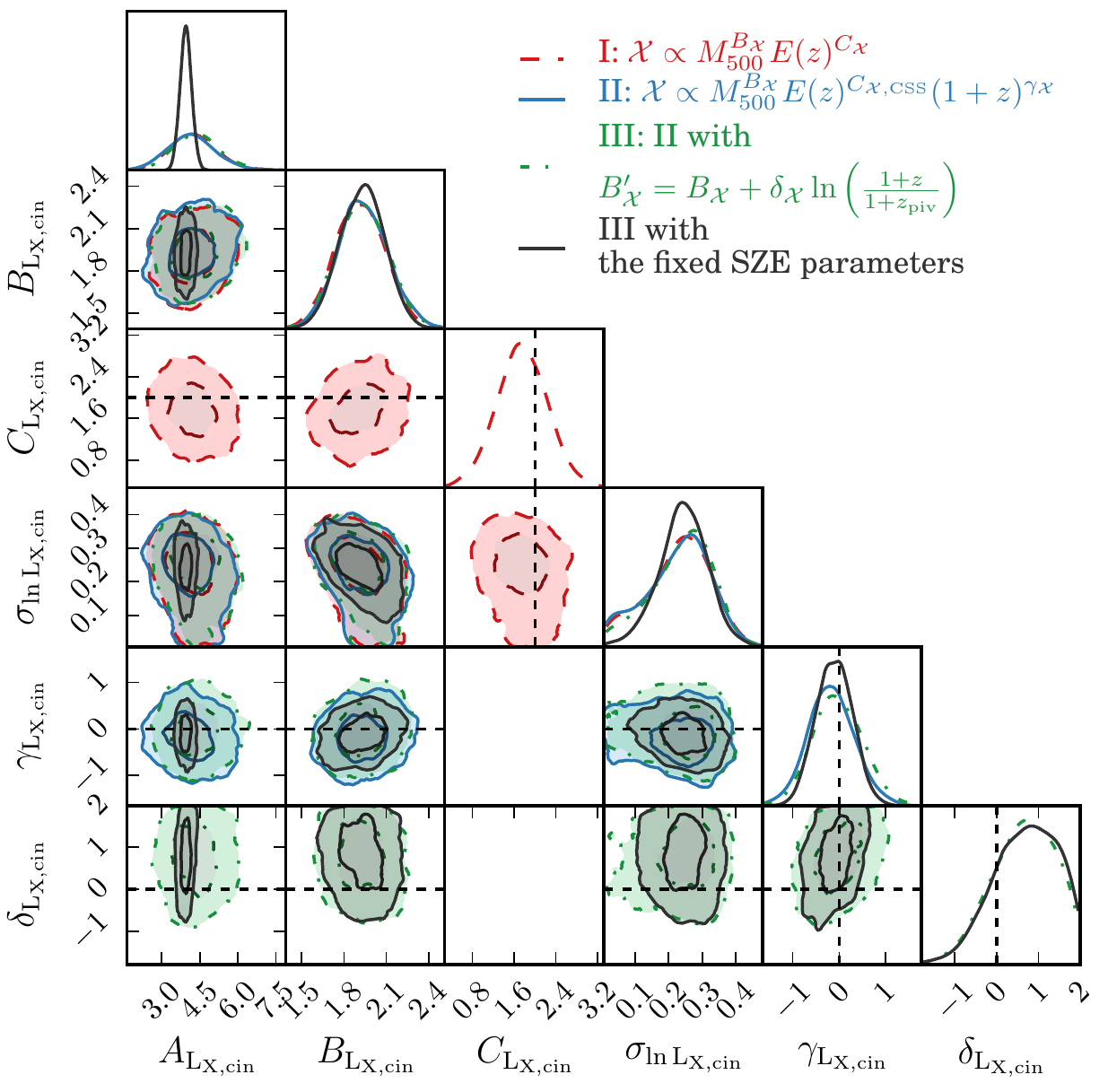}
\includegraphics[height=0.37\textheight]{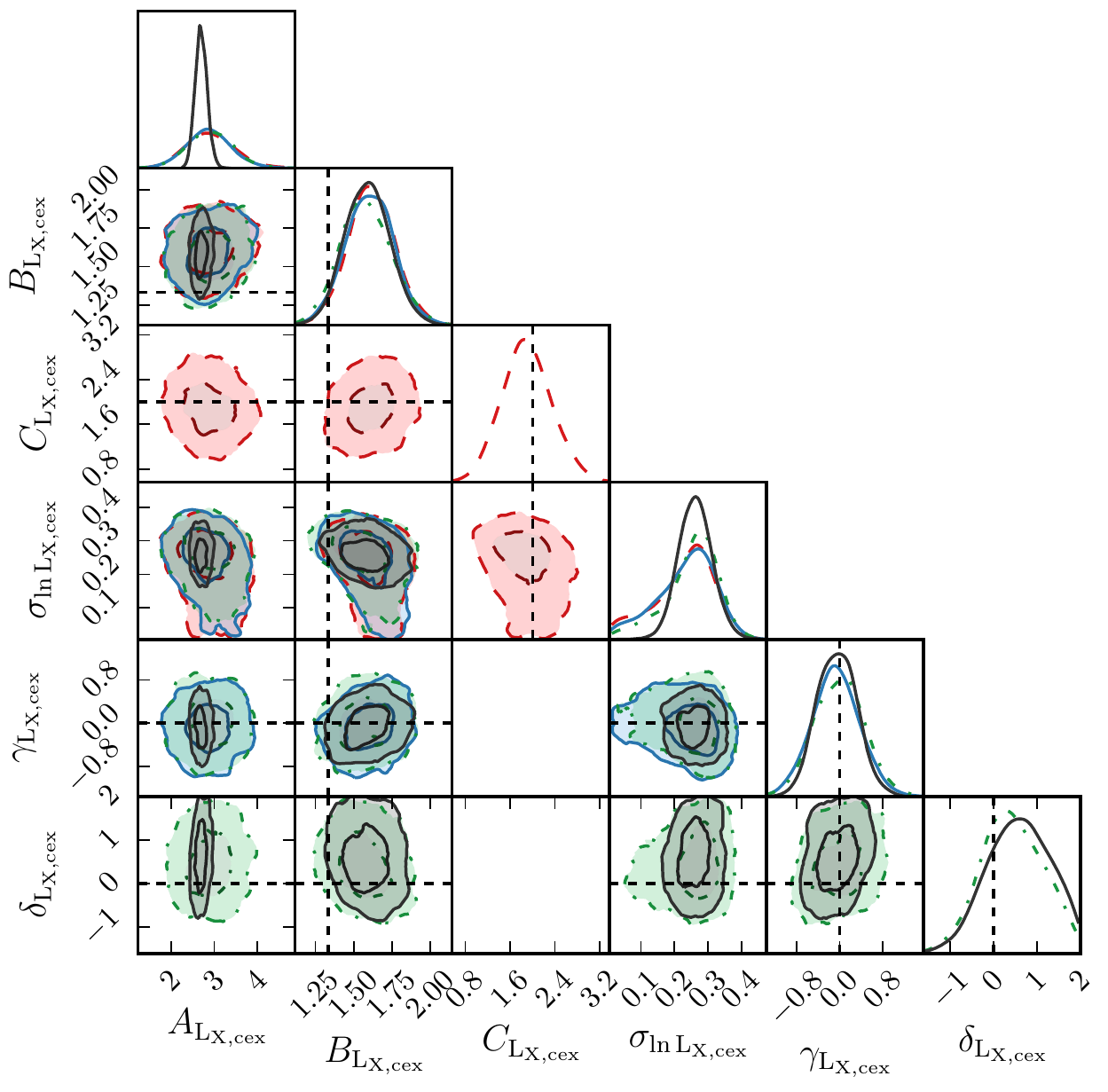}
\caption{Similar to Figure~\ref{fig:triangle_txe} but containing constraints for the cases of the 0.5:2.0~keV core-included luminosity \lmi\ (the upper panel) and core-excised luminosity \lme\ (the lower panel) scaling relations.}
\label{fig:triangle_soft_lx}
\vspace{2mm}
\end{figure}
\begin{figure}
\centering
\includegraphics[height=0.37\textheight]{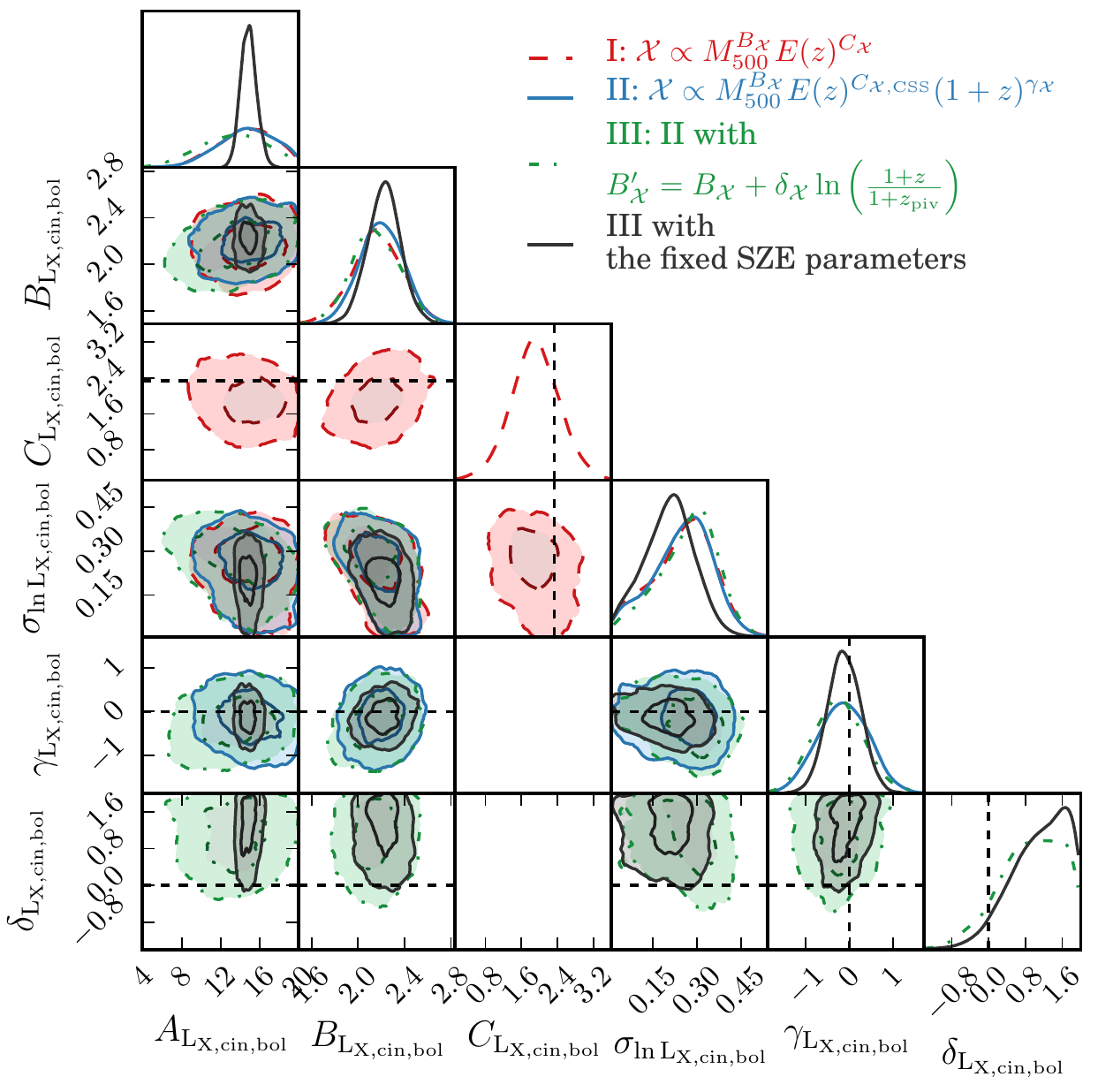}
\includegraphics[height=0.37\textheight]{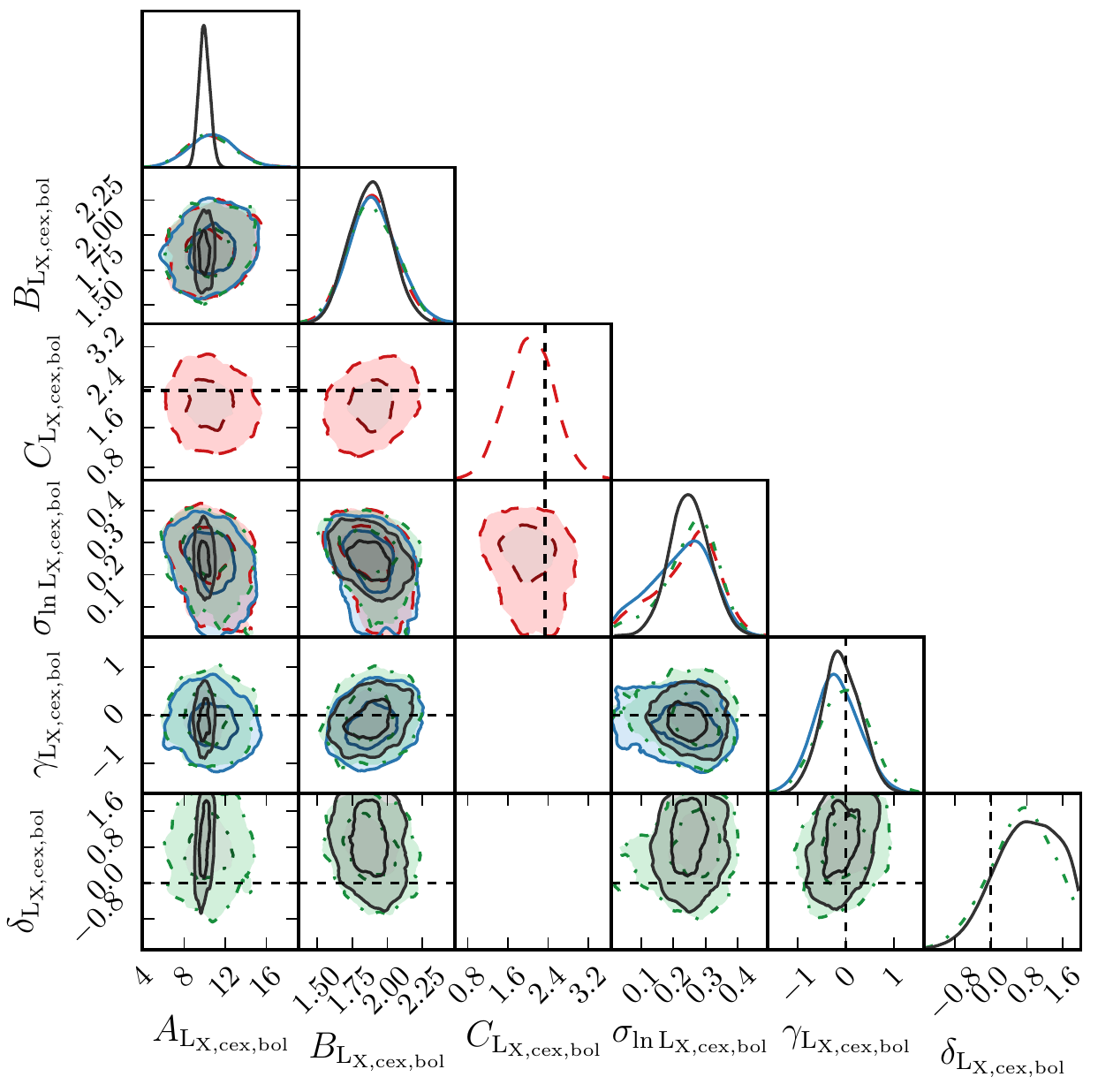}
\caption{Similar to Figure~\ref{fig:triangle_txe} but containing constraints for the cases of the core-included bolometric luminosity \lmi\ (the upper panel) and core-excised bolometric luminosity \lme\ (the lower panel) scaling relations.}
\label{fig:triangle_bol_lx}
\vspace{2mm}
\end{figure}

\subsubsection{Parameter constraints}

The resulting best-fit scaling relation parameters and uncertainties are listed in Table~\ref{tab:sclrel}, and the marginalized posteriors of the single and joint parameters constraints for the core-included and -excised observables appear in Figure~\ref{fig:triangle_soft_lx} (0.5--2.0~keV) and Figure~\ref{fig:triangle_bol_lx} (bolometric).

For the core-included, soft-band 0.5--2.0~keV X-ray luminosity \Lxi, the best-fit relation is
\begin{eqnarray}
\label{eq:lxi-soft}
\Lxi &= 
&4.12^{+0.91}_{-0.94}\times10^{44}\mathrm{erg/s}
\left(\frac{\Mfiveoo}{\Mpiv}\right)^{1.89^{+0.23}_{-0.13}} \nonumber \\
&&\left(\frac{E(z)}{E(\zpiv)}\right)^{2}
\left(\frac{1+z}{1+\zpiv }\right)^{-0.20^{+0.51}_{-0.49}} \, ,
\end{eqnarray}
with intrinsic scatter of $0.27^{+0.08}_{-0.12}$.
For the \lme\ relation, the best-fit is
\begin{eqnarray}
\label{eq:lxe-soft}
\Lxe &= 
&2.84^{+0.53}_{-0.50}\times10^{44}\mathrm{erg/s}
\left(\frac{\Mfiveoo}{\Mpiv}\right)^{1.60^{+0.16}_{-0.15}} \nonumber \\
&&\left(\frac{E(z)}{E(\zpiv)}\right)^{2}
\left(\frac{1+z}{1+\zpiv }\right)^{-0.10^{+0.47}_{-0.42}},
\end{eqnarray}
with intrinsic scatter of $0.27^{+0.07}_{-0.11}$.  
As before, the mass and redshift pivots are $\Mpiv=6.35\times10^{14}\Msun$ and $\zpiv=0.45$.

The soft-band, core-excised \lm\ relation shows a mass trend that is $\approx4\sigma$ higher than the self-similar trend ($B_{\Lx} = 1$), while the core-included relation is steeper and exhibits a tension of $\approx6.8\sigma$ with the self-similar behavior.
The redshift trends for both core-included and -excised luminosities are consistent with the self-similar trend of $C_{\Lx} = 2$.  There is no evidence for a redshift-dependent mass slope in either soft-band \Lx\ measurement. 
Fixing the SZE parameters $r_{\zeta}$ (the black curves in Figure~\ref{fig:triangle_soft_lx}) does not change the overall picture except that the uncertainties of the normalization are reduced by about a factor of four. 

The characteristic luminosities at the pivot mass and redshift for core-included clusters are a factor of $\approx45\%$ higher than the core-excised luminosities, a difference of $\approx 1\sigma$. Interestingly, the scatter of the two relations is similar at $\approx27\%$.

Similarly, for the bolometric luminosities the best-fit \lmib\ and \lmeb\ relations are
\begin{eqnarray}
\label{eq:lxi-bol}
\Lxi &= 
&14.8^{+3.5}_{-2.7}\times10^{44}\mathrm{erg/s}
\left(\frac{\Mfiveoo}{\Mpiv}\right)^{2.19^{+0.21}_{-0.17}} \nonumber \\
&&\left(\frac{E(z)}{E(\zpiv)}\right)^{\frac{7}{3}}
\left(\frac{1+z}{1+\zpiv }\right)^{-0.14^{+0.62}_{-0.57}} \, ,
\end{eqnarray}
and
\begin{eqnarray}
\label{eq:lxe-bol}
\Lxe &= 
&10.7^{+2.3}_{-2.3}\times10^{44}\mathrm{erg/s}
\left(\frac{\Mfiveoo}{\Mpiv}\right)^{1.88^{+0.19}_{-0.17}} \nonumber \\
&&\left(\frac{E(z)}{E(\zpiv)}\right)^{\frac{7}{3}}
\left(\frac{1+z}{1+\zpiv }\right)^{-0.26^{+0.53}_{-0.43}},
\end{eqnarray}
with intrinsic scatter of $0.29^{+0.08}_{-0.13}$ and $0.27^{+0.07}_{-0.13}$, respectively.
The same pivot mass and redshift as before are used.

As expected, the bolometric luminosity relations have steeper mass trends than those of the soft band luminosities.  Similar to the soft band, the bolometric luminosity to mass scaling relations have mass trends that are steeper than self-similar ($B_{\Lx}={4\over3}$) with a significance of $\approx3.2\sigma$ and $\approx5.1\sigma$ for the core-excised and core-included luminosities, respectively.  The redshift trends of the scaling relations are all consistent with the self-similar trend $C_{\Lx}={7\over3}$, and there is a preference for a redshift dependent mass trend in the core-included luminosity scaling relation.  Fixing the SZE parameters $r_{\zeta}$ does not result in significant differences except by decreasing the uncertainties of the normalization by a factor of three to five, and it also slightly reduces the scatter in the core-included relation. Both relations exhibit intrinsic scatter at around the $27\%$ level, which is comparable to that in the soft band.

\subsubsection{Comparison to previous results}

We show the redshift and mass trends of \Lx\ in the two bottom rows in Figures ~\ref{fig:mtrends} and \ref{fig:ztrends}, respectively.  As for the case of the other X-ray observables shown in this plot, we scale the measurements to the pivot redshift $\zpiv=0.45$ or pivot mass $\Mpiv=6.35\times10^{14}\Msun$ using the best-fit redshift and mass trends from the Form II relation (see Table~\ref{tab:sclrel}).

Our core-excised bolometric luminosities are consistent with the bolometric luminosities reported from \xmm\ observations of the low-{\it z} REXCESS clusters \citep[with a slope of $1.77\pm0.05$]{pratt09}. Additionally, our core-excised soft-band luminosities from \chandra\ and \xmm\ observations of the 15 SPT selected clusters \citep[][with a slope of $1.45\pm0.29$]{andersson11}, \chandra\ observations of massive clusters \citet[][with a slope of $1.61\pm 0.14$]{vikhlinin09b}, \chandra\ observations of 115 clusters \citep[][with a slope of $1.63\pm0.08$]{maughan07}, and the \xmm\ observations of HIFLUGCS sample \citep[][with a slope of $1.61\pm0.19$]{Lovisari2015} at $z<0.05$. We  note that all these results in the literature depart from the self-similar expectation. However, the slope of the mass trend of the core-excised soft-band luminosity ($B_{\Lx}\approx1.60\pm0.15$) is steeper than the value reported in \citet[][$B_{\Lx}=1.02 \pm 0.09$]{mantz16} at the $3.4\sigma$ level. Our slope is consistent with the low-redshift ($z<0.2$) HIFLUGCS Cosmology (HICOSMO) sample \citep[$B_{\Lx}=1.35\pm0.07$;][]{schellenberger15} at $\approx1.5\sigma$ level. Overall, in terms of mass trends our study demonstrates a much steeper than self-similar mass trend in agreement with most previously published analyses.

Our constraints on the redshift trend of the core-excised, soft-band \lm\ is $L_{\Lx}\propto E(z)^{1.72^{+0.53}_{-0.46}}$, which is in good agreement with that found by \citet[][$C_{\Lx}=1.82\pm0.35$]{mantz16}.  \citet{vikhlinin09b} reports a redshift trend of $C_{\Lx}=1.85\pm0.40$, which is also consistent with our results.  In addition, our soft band measurements follow a similar trend with that seen in the C-Eagle cosmological hydrodynamical simulations of clusters \citep{Barnes2017}.

In Figure ~\ref{fig:comps}, we over-plot our results of core-excised soft-band luminosity \Lxe\ with the ones from simulated clusters \citep{Barnes2017} and other observational studies \citep{pratt09,vikhlinin09a,mantz16}.
Although our SPT clusters are sampling the relatively high-mass end, our results show no significant tension in the mass trend with previous work extending to the low mass regime. 
With the exception of the analysis of \cite{mantz16}, the \Lxe\ from both simulated and observed clusters all show steeper mass trends with respect to the self-similar prediction (the grey dashed line). 
We note also that the  scatter in the simulated C-Eagle clusters is 0.30, which is larger than, but statistically consistent with, our measurement of $\sigma_{\ln\Lx}=0.27^{+0.07}_{-0.11}$.  This is also true for the values of 0.17 and 0.25 found in the REXCESS \citep{pratt09} and HIFLUGCS samples \citep{Lovisari2015}.  An interesting element of our result is that the scatter is similar in both the core-included and -excised luminosity measurements.
\begin{table*}\scriptsize
        \caption{\scriptsize
        This table holds the best-fit scaling relation parameters when priors on the SZE \zm\ relation
        parameters are taken from \citet[][Table 3, results column 2]{deHaan16}. The Table layout is the 
        same as in Table~\ref{tab:sclrel}.
        }
        \label{tab:sclrel_col2}
		\centering
        \begin{tabular}{lcccccccccc}
        \hline
        	 Scaling Relation  & $A_{\obsx}$ & $B_{\obsx}$ & $C_{\obsx}$ & $\sigma_{\ln\obsx}$ 
        	                   & $\gamma_{\obsx}$ & $\delta_{\obsx}$
        	 				   \\[3pt]
        \hline
        \multicolumn{1}{l}{\tmi\ Relation} & & $B_\mathrm{\obsx,SS}={2\over3}$ & $C_\mathrm{\obsx,SS}={2\over3}$ \\
		I: $\obsx(M,z)\propto \Mfiveoo^{B_{\obsx}}E(z)^{C_\obsx}$ 
		& $6.50\pm 0.66$ & $0.78^{+0.10}_{-0.09}$ & $0.44^{+0.27}_{-0.26}$ & $0.18^{+0.04}_{-0.05}$ & -- & --
\\[3pt]
		II: $\obsx(M,z)\propto \Mfiveoo^{B_{\obsx}}E(z)^{2\over3}(1+z)^{\gamma_\obsx}$ 
		& $6.51^{+0.58}_{-0.70}$ & $0.81^{+0.09}_{-0.10}$ & $2\over3$ & $0.18^{+0.04}_{-0.04}$ & $-0.21^{+0.24}_{-0.22}$ & --
\\[3pt] 
		III: as II with $B'_\obsx=B_{\obsx} + \delta_{\obsx}\ln\left(\frac{1+z}{1+\zpiv} \right)$ 
		& $6.42^{+0.67}_{-0.65}$ & $0.77^{+0.10}_{-0.09}$ & $2\over3$ & $0.17^{+0.05}_{-0.03}$ & $-0.14^{+0.33}_{-0.26}$ & $0.60^{+0.57}_{-0.47}$
\\[3pt]
		III with fixed SZE params
		& $6.44^{+0.24}_{-0.23}$ & $0.79^{+0.08}_{-0.09}$ & $2\over3$ & $0.17^{+0.04}_{-0.03}$ & $-0.05^{+0.25}_{-0.27}$ & $0.81^{+0.49}_{-0.53}$
\\[3pt]
        \hline
        \multicolumn{1}{l}{\tme\ Relation}  & & $B_\mathrm{\obsx,SS}={2\over3}$ & $C_\mathrm{\obsx,SS}={2\over3}$ \\
		I: $\obsx(M,z)\propto \Mfiveoo^{B_{\obsx}}E(z)^{C_\obsx}$ 
		& $6.26^{+0.57}_{-0.68}$ & $0.80^{+0.10}_{-0.08}$ & $0.43^{+0.23}_{-0.22}$ & $0.13^{+0.04}_{-0.06}$ & -- & --
		\\[3pt]
		II: $\obsx(M,z)\propto \Mfiveoo^{B_{\obsx}}E(z)^{2\over3}(1+z)^{\gamma_\obsx}$ 
		& $6.25^{+0.49}_{-0.75}$ & $0.81^{+0.08}_{-0.09}$ & $2\over3$ & $0.14^{+0.04}_{-0.06}$ & $-0.20^{+0.21}_{-0.22}$ & --
		\\[3pt]
		III: as II with $B'_\obsx=B_{\obsx} + \delta_{\obsx}\ln\left(\frac{1+z}{1+\zpiv} \right)$ 
		& $6.18^{+0.63}_{-0.62}$ & $0.81^{+0.10}_{-0.10}$ & $2\over3$ & $0.14^{+0.04}_{-0.05}$ & $-0.17^{+0.28}_{-0.23}$ & $0.35^{+0.41}_{-0.45}$
\\[3pt]
		III with fixed SZE params 
		& $6.14^{+0.22}_{-0.17}$ & $0.81^{+0.08}_{-0.08}$ & $2\over3$ & $0.13^{+0.04}_{-0.04}$ & $-0.13\pm 0.23$ & $0.45^{+0.41}_{-0.48}$ 
		\\[3pt]
        \hline
        \multicolumn{1}{l}{\mm\ Relation}  & & $B_\mathrm{\obsx,SS}=1$ & $C_\mathrm{\obsx,SS}=0$ \\
		I: $\obsx(M,z)\propto \Mfiveoo^{B_{\obsx}}E(z)^{C_\obsx}$ 
		& $6.60^{+1.1}_{-1.0}$ & $1.27^{+0.08}_{-0.11}$ & $0.43^{+0.24}_{-0.23}$ & $0.11^{+0.04}_{-0.08}$ & -- & --
		\\[3pt]
		II: $\obsx(M,z)\propto \Mfiveoo^{B_{\obsx}}E(z)^{0}(1+z)^{\gamma_\obsx}$ 
		& $7.14^{+1.00}_{-1.01}$ & $1.26^{+0.09}_{-0.10}$ & $0$ & $0.11^{+0.04}_{-0.08}$ & $0.39^{+0.25}_{-0.24}$ & --
		\\[3pt]
		III: as II with $B'_\obsx=B_{\obsx} + \delta_{\obsx}\ln\left(\frac{1+z}{1+\zpiv} \right)$ 
		& $6.89^{+1.15}_{-0.91}$ & $1.27^{+0.10}_{-0.10}$ & $0$ & $0.11^{+0.03}_{-0.08}$ & $0.40^{+0.23}_{-0.24}$ & $0.19^{+0.40}_{-0.54}$
 \\[3pt]
		III with fixed SZE params 
		& $6.93^{+0.28}_{-0.19}$ & $1.27^{+0.08}_{-0.08}$ & $0$ & $0.08^{+0.04}_{-0.06}$ & $0.41^{+0.21}_{-0.22}$ & $0.11^{+0.44}_{-0.47}$
		\\[3pt]
        \hline
        \multicolumn{1}{l}{\ymi\ Relation}   & & $B_\mathrm{\obsx,SS}={5\over3}$ & $C_\mathrm{\obsx,SS}={2\over3}$ \\
		I: $\obsx(M,z)\propto \Mfiveoo^{B_{\obsx}}E(z)^{C_\obsx}$ 
		& $4.5^{+1.0}_{-1.1}$ & $2.00^{+0.16}_{-0.14}$ & $0.80^{+0.42}_{-0.35}$ & $0.15^{+0.08}_{-0.09}$ & -- & --
		\\[3pt]
		II: $\obsx(M,z)\propto \Mfiveoo^{B_{\obsx}}E(z)^{2\over3}(1+z)^{\gamma_\obsx}$ 
		& $4.7^{+1.1}_{-1.0}$ & $2.01^{+0.16}_{-0.14}$ & $2\over3$ & $0.15^{+0.07}_{-0.11}$ & $0.15^{+0.39}_{-0.34}$ & --
		\\[3pt]
		III: as II with $B'_\obsx=B_{\obsx} + \delta_{\obsx}\ln\left(\frac{1+z}{1+\zpiv} \right)$ 
		& $4.9\pm 1.1$ & $2.03^{+0.13}_{-0.15}$ & $2\over3$ & $0.17^{+0.06}_{-0.10}$ & $0.32^{+0.39}_{-0.41}$ & $1.16^{+0.44}_{-0.77}$
 \\[3pt]
		III with fixed SZE params 
		& $4.59^{+0.22}_{-0.27}$ & $1.97\pm 0.11$ & $2\over3$ & $0.07^{+0.07}_{-0.06}$ & $0.23^{+0.35}_{-0.33}$ & $0.73^{+0.77}_{-0.55}$
		\\[3pt]
        \hline
        \multicolumn{1}{l}{\yme\ Relation}  & & $B_\mathrm{\obsx,SS}={5\over3}$  & $C_\mathrm{\obsx,SS}={2\over3}$ \\
		I: $\obsx(M,z)\propto \Mfiveoo^{B_{\obsx}}E(z)^{C_\obsx}$ 
		& $4.25^{+0.96}_{-1.02}$ & $1.99^{+0.14}_{-0.16}$ & $0.77^{+0.37}_{-0.42}$ & $0.13^{+0.05}_{-0.10}$ & -- & --
		\\[3pt]
		II: $\obsx(M,z)\propto \Mfiveoo^{B_{\obsx}}E(z)^{2\over3}(1+z)^{\gamma_\obsx}$ 
		& $4.35^{+0.90}_{-1.10}$ & $2.00^{+0.16}_{-0.15}$ & $2\over3$ & $0.05^{+0.12}_{-0.03}$ & $0.09^{+0.44}_{-0.29}$ & -- 
		\\[3pt]
		III: as II with $B'_\obsx=B_{\obsx} + \delta_{\obsx}\ln\left(\frac{1+z}{1+\zpiv} \right)$ 
		& $4.70^{+1.1}_{-1.2}$ & $1.99^{+0.18}_{-0.14}$ & $2\over3$ & $0.13^{+0.05}_{-0.10}$ & $0.35^{+0.38}_{-0.41}$ & $0.47^{+0.63}_{-0.65}$
\\[3pt]
		III with fixed SZE params 
		& $4.31^{+0.30}_{-0.18}$ & $1.98\pm 0.12$ & $2\over3$ & $0.04^{+0.09}_{-0.03}$ & $0.27^{+0.33}_{-0.35}$ & $0.72^{+0.57}_{-0.69}$
		\\[3pt]
        \hline
        \multicolumn{1}{l}{\lmi\ Relation}  & & $B_\mathrm{\obsx,SS}={1}$ & $C_\mathrm{\obsx,SS}={2}$ \\
		I: $\obsx(M,z)\propto \Mfiveoo^{B_{\obsx}}E(z)^{C_\obsx}$ 
		& $4.13^{+0.87}_{-0.98}$ & $1.93^{+0.15}_{-0.18}$ & $2.01^{+0.44}_{-0.37}$ & $0.28^{+0.07}_{-0.12}$ & -- & --
		\\[3pt]
		II: $\obsx(M,z)\propto \Mfiveoo^{B_{\obsx}}E(z)^{2}(1+z)^{\gamma_\obsx}$ 
		& $4.15^{+1.10}_{-0.81}$ & $1.91^{+0.18}_{-0.15}$ & $2$ & $0.25^{+0.08}_{-0.13}$ & $0.20^{+0.41}_{-0.43}$ & --
		\\[3pt] 
		III: as II with $B'_\obsx=B_{\obsx} + \delta_{\obsx}\ln\left(\frac{1+z}{1+\zpiv} \right)$ 
		& $4.33^{+1.11}_{-0.89}$ & $1.85^{+0.21}_{-0.16}$ & $2$ & $0.28^{+0.08}_{-0.11}$ & $0.21^{+0.53}_{-0.44}$ & $0.82^{+0.61}_{-0.94}$
\\[3pt]
		III with fixed SZE params 
		& $3.86^{+0.22}_{-0.20}$ & $1.94\pm 0.15$ & $2$ & $0.23^{+0.08}_{-0.07}$ & $0.33^{+0.42}_{-0.37}$ & $0.79^{+0.83}_{-0.65}$
		\\[3pt]
        \hline
        \multicolumn{1}{l}{\lme\ Relation}   & & $B_\mathrm{\obsx,SS}={1}$ & $C_\mathrm{\obsx,SS}={2}$ \\
		I: $\obsx(M,z)\propto \Mfiveoo^{B_{\obsx}}E(z)^{C_\obsx}$ 
		& $2.83^{+0.53}_{-0.52}$ & $1.57^{+0.17}_{-0.14}$ & $2.17^{+0.34}_{-0.43}$ & $0.26^{+0.07}_{-0.09}$ & -- & --
		 \\[3pt]
		II: $\obsx(M,z)\propto \Mfiveoo^{B_{\obsx}}E(z)^{2}(1+z)^{\gamma_\obsx}$ 
		& $2.82^{+0.61}_{-0.49}$ & $1.63^{+0.13}_{-0.16}$ & $2$ & $0.26^{+0.08}_{-0.09}$ & $0.20^{+0.41}_{-0.34}$ & --
		\\[3pt]
		III: as II with $B'_\obsx=B_{\obsx} + \delta_{\obsx}\ln\left(\frac{1+z}{1+\zpiv} \right)$ 
		& $2.77^{+0.55}_{-0.51}$ & $1.57^{+0.18}_{-0.16}$ & $2$ & $0.28^{+0.07}_{-0.07}$ & $0.24^{+0.51}_{-0.32}$ & $0.47^{+0.93}_{-0.57}$ 
\\[3pt]
		III with fixed SZE params 
		& $2.67^{+0.16}_{-0.15}$ & $1.58\pm 0.15$ & $2$ & $0.26^{+0.06}_{-0.05}$ & $0.25\pm 0.38$ & $0.75^{+0.76}_{-0.74}$
		\\[3pt]
        \hline
        \multicolumn{1}{l}{\lmib\ Relation}   & & $B_\mathrm{\obsx,SS}={4\over3}$ & $C_\mathrm{\obsx,SS}={7\over3}$ \\
		I: $\obsx(M,z)\propto \Mfiveoo^{B_{\obsx}}E(z)^{C_\obsx}$ 
		& $13.8^{+3.6}_{-3.2}$ & $2.21^{+0.15}_{-0.20}$ & $2.28^{+0.46}_{-0.41}$ & $0.29^{+0.09}_{-0.12}$ & -- & --
		\\[3pt]
		II: $\obsx(M,z)\propto \Mfiveoo^{B_{\obsx}}E(z)^{7\over3}(1+z)^{\gamma_\obsx}$ 
		& $14.3\pm 3.2$ & $2.18\pm 0.19$ & $7\over3$ & $0.26^{+0.09}_{-0.13}$ & $0.03^{+0.54}_{-0.37}$ & --
		\\[3pt]
		III: as II with $B'_\obsx=B_{\obsx} + \delta_{\obsx}\ln\left(\frac{1+z}{1+\zpiv} \right)$ 
		& $14.8^{+3.3}_{-2.9}$ & $2.16^{+0.19}_{-0.16}$ & $7\over3$ & $0.28^{+0.09}_{-0.11}$ & $0.19^{+0.44}_{-0.50}$ & $1.14^{+0.69}_{-0.76}$
\\[3pt]
		III with fixed SZE params 
		 & $14.55^{+0.98}_{-0.70}$ & $2.23\pm 0.15$ & $7\over3$ & $0.19^{+0.08}_{-0.10}$ & $0.35^{+0.39}_{-0.42}$ & $1.34^{+0.49}_{-0.74}$
		\\[3pt]
        \hline
        \multicolumn{1}{l}{\lmeb\ Relation}   & & $B_\mathrm{\obsx,SS}={4\over3}$ & $C_\mathrm{\obsx,SS}={7\over3}$ \\
		I: $\obsx(M,z)\propto \Mfiveoo^{B_{\obsx}}E(z)^{C_\obsx}$ 
		& $10.6^{+2.5}_{-2.2}$ & $1.87^{+0.19}_{-0.16}$ & $2.31^{+0.45}_{-0.35}$ & $0.26^{+0.08}_{-0.13}$ & -- & --
		\\[3pt]
		II: $\obsx(M,z)\propto \Mfiveoo^{B_{\obsx}}E(z)^{7\over3}(1+z)^{\gamma_\obsx}$ 
		& $10.5^{+2.1}_{-2.5}$ & $1.87^{+0.20}_{-0.14}$ & $7\over3$ & $0.27^{+0.07}_{-0.14}$ & $0.10^{+0.47}_{-0.35}$ & --
		\\[3pt]
		III: as II with $B'_\obsx=B_{\obsx} + \delta_{\obsx}\ln\left(\frac{1+z}{1+\zpiv} \right)$ 
		& $9.9^{+2.2}_{-2.1}$ & $1.80^{+0.17}_{-0.22}$ & $7\over3$ & $0.30^{+0.07}_{-0.09}$ & $0.23^{+0.45}_{-0.49}$ & $0.73^{+0.78}_{-0.70}$
\\[3pt]
		III with fixed SZE params 
		& $10.06^{+0.54}_{-0.61}$ & $1.86^{+0.15}_{-0.16}$ & $7\over3$ & $0.25^{+0.05}_{-0.07}$ & $0.36^{+0.32}_{-0.49}$ & $0.88^{+0.76}_{-0.57}$
		\\[3pt]
		\hline
      \end{tabular}
\end{table*}

\subsection{SZE-based halo masses with external cosmological priors from BAO}
\label{sec:column2}

Currently, the redshift trend parameter on the SZE \zm\ relation is the least well constrained, and this leads to additional uncertainty in understanding the X-ray observable mass relations.  In \citet[][the second column of Table 3]{deHaan16}, an analysis within the context of a flat $\Lambda$CDM model was undertaken where additional external cosmological priors from BAO were added.  This helped reduce the cosmological parameter space consistent with the SPT cluster sample distribution in $\xi$ and redshift, tightening up \zm\ parameter uncertainties.  In addition, the redshift evolution parameter was shifted upward from $C_\zeta=0.55\pm0.3$ to $C_\zeta=0.80\pm0.15$.  The combination of the shift and reduction in uncertainties have motivated us to present the scaling relations derived using the X-ray observables together with these SZE-based halo masses.  Table~\ref{tab:sclrel_col2} contains the results of these relations.  We recommend that those particularly interested in obtaining precise redshift trends in the scaling relations should use these results.

\section{Conclusions}
\label{sec:concl}
We present here measurements of the X-ray observables in a sample of 59 SZE selected galaxy clusters with redshifts $0.20<z<1.5$ that have been observed with \xmm.  We use these measurements together with SZE-based halo masses \Mfiveoo\ to study the scaling relations between X-ray observables, halo mass and redshift.  A strength of our work is the ability to directly constrain the redshift and mass trends based on an SZE-selected cluster sample spanning a wide range of redshift.  This selection is approximately equivalent to a mass selection, and this sample spans a mass range of $3\times10^{14}\Msun \leq\Mfiveoo\leq1.8\times10^{15}\Msun$. The biasing effects in X-ray selected samples due to the X-ray cool core phenomenon are significantly reduced and perhaps even completely removed.  This simplifies the interpretation of the results from our analysis.

We use the \xmm\ observations to derive X-ray observables \Tx, \Micm, \Yx, and rest frame 0.5--2.0~keV and bolometric \Lx.  For all these observables---save for the \Micm---we extract both core-included and core-excised quantities, where we define the core to be the region within 0.15\Rfiveoo.  The cluster halo masses are derived from the SPT \zm\ scaling relation and are corrected for selection effects, such as Eddington and Malmquist biases as described in detail in other publications \citep{bocquet15}.  As discussed in detail in Section~\ref{sec:priors}, we adopt priors on the \zm\ scaling relation from the \citet{deHaan16} joint cosmology and mass calibration analysis, which have since been validated using weak lensing masses of 32 SPT-SZ clusters \citep{dietrich17} and dynamical masses of 110 SPT-SZ clusters \citep{capasso17}.  These SZE-based halo masses are characteristically uncertain at the $\approx25\%$ level (statistical and systematic uncertainties combined in quadrature).

We fit our data to three different power-law models (see equations~\ref{eq:FormA}, \ref{eq:FormB} and \ref{eq:FormC}) and derive the best-fit normalization $A_\obsx$, mass trend $B_\obsx$, $E(z)$ redshift trend $C_\obsx$, departure from self-similar redshift trend $\gamma_\obsx$, log-normal intrinsic scatter in the X-ray observable at fixed halo mass $\sigma_{\ln\obsx}$, and also a redshift dependence to the mass trend $\delta_\obsx$. While all three scaling relation forms are adequate to fit the data, we recommend that those interested in cosmological studies adopt Form II, because it models the departure from self-similar evolution with redshift using a cosmologically agnostic form $(1+z)^{\gamma_\obsx}$.  We marginalize over the uncertainties in the SZE-based halo masses, adjusting the radius \Rfiveoo\ as appropriate in each iteration in the chain and re-extracting the X-ray observables in a self-consistent manner.  Thus, the final parameter uncertainties of the X-ray observable--mass scaling relations include both measurement and systematic halo mass uncertainties (see Table~\ref{tab:sclrel}).

The halo mass scaling relations for \Tx, \Micm, \Yx\, and \Lx\  are steeper, but statistically consistent (within $2\sigma$ confidence) with the results from the literature.
However, we observe significant departures from the \citet{mantz16} soft band core-excised luminosity at $3.4\sigma$ level, ICM mass at $2.9\sigma$ level, and \Yx\ at $2.3\sigma$ level.
The mass trends we find in all our scaling relations are steeper than the self-similar behavior at $\gtrsim1.6\sigma$ confidence.  In the case of \Micm\ and \Lx\, the mass trends we measure ($\Micm\propto\Mfiveoo^{1.26\pm0.10}$ and $\Lxe\propto\Mfiveoo^{1.60\pm0.15}$ in soft band) are consistent with most previously published results that employ X-ray selected samples and a mix of weak lensing and hydrostatic masses.  However, for \Tx\ and \Yx\ our mass trends ($\Txe\propto\Mfiveoo^{0.80\pm0.09}$ and $\Yxe\propto\Mfiveoo^{2.0\pm0.16}$) are steeper than most previous work at $\approx1.6\sigma$ (see parameter $B_\obsx$ in Tables~\ref{tab:sclrel} and \ref{tab:sclrel_col2}).

In addition, we probe for a redshift-dependent mass trend (Form III, equation~\ref{eq:FormC}) and find that the data currently provide no evidence for such a trend, with the highest significance departure from no evolution being in the core-included \Yx\ and \Lx\ (see parameter $\delta_{\obsx}$ in Tables~\ref{tab:sclrel} and \ref{tab:sclrel_col2}).
 
We examine the redshift trends in all scaling relations, finding no significant departures from the self-similar behavior that arises simply due to the evolution of the critical density with redshift. There is no tension between our results and those from previous studies, although many previous studies were not in a position to examine redshift trends, given the limitations of their samples and the availability of halo mass measurements (see parameter $\gamma_\obsx$ in Tables~\ref{tab:sclrel} and \ref{tab:sclrel_col2}).

We report the intrinsic scatter in X-ray observable at fixed halo mass $\sigma_{\ln\obsx}$ for all scaling relations.  These indicate exquisite scatter at the $\approx10\%$ level for \Micm\ and core-excised integrated pressure \Yxe, somewhat higher scatter of $\approx13\%$ for core-excised temperature \Txe, and scatter of  $\approx27$~percent for X-ray luminosities \Lx\ (see parameter $\sigma_{\ln\obsx}$ in Tables~\ref{tab:sclrel} and \ref{tab:sclrel_col2}).  We do not account for correlated scatter among the SZE and X-ray observables, because previous analyses of larger SPT-SZ selected samples have failed to detect these effects \citep{deHaan16,dietrich17}, and therefore they are too small to have an impact on our results.

In all cases, our baseline results are presented in Table~\ref{tab:sclrel}, and the mass and redshift trends for each observable are highlighted in Figs.~\ref{fig:mtrends} and \ref{fig:ztrends}.  In addition, we present an alternative set of results in Table~\ref{tab:sclrel_col2} that have somewhat better defined redshift trends that come from adopting a calibration of the SZE \zm\ relation that includes external cosmological priors from BAO (see discussion in Section~\ref{sec:column2}).

One of the reasons for the steeper mass trends in \Tx\ and \Yx\ found in this work could be due to calibration differences affecting the temperatures differently in \chandra\ and \xmm. In previous studies of low redshift, high flux clusters, it has been shown that \xmm\ temperature estimates lie below \chandra\ temperatures in a manner that increases as a function of cluster temperature \citep{schellenberger15}. However, our sample contains also high redshift systems where the known calibration differences would have less of an impact.  Moreover, the \xmm\ observations at higher redshift in our sample tend to be lower signal to noise, and in the limit of low signal to noise the background subtraction systematics will tend to be more important than the effective area systematics.  Thus, overall we do not expect that the effective area systematics at high energies between \chandra\ and \xmm\ are playing an important role in the mass trends of the \Tx\ and \Yx\ observables.

Our results are broadly consistent with recent numerical simulations \citet[e.g.][]{Barnes2017} at the 1-2$\sigma$ level. A departure from self-similarity in a scaling relation could well indicate that non-gravitational effects in the galaxy clusters are important, and disagreement between simulated and observed scaling relations provides a direct test of the accuracy of the subgrid physics adopted in the simulations. However, one must always be cautious about halo mass systematics as well.

Another concern is a bias in the calibration of the SZE \zm\ relation, because a bias in the mass trend of the SZE mass--observable relation would indeed be reflected in biased trends in the X-ray observable--mass relations.  Here we note only that this SPT calibrated \zm\ relation offers a unique capability of delivering $\approx25\%$ accurate single cluster masses that have been self-consistently calibrated within a cosmological context that uses the SPT cluster distribution in signal-to-noise and redshift in combination with external mass information.  
Cross-checks of SZE-based masses with weak lensing \citep{dietrich17} and dynamical \citep{capasso17} masses have so far provided no evidence for biases in our masses.
Work continues to improve this calibration using weak lensing information from the Dark Energy Survey \citep[e.g.,][]{stern18}.  
We remind the reader that this work is among the first to extend scaling relation studies to redshifts, which have not yet been covered by the previous X-ray studies. Using SZE-selected clusters and SZE-based halo masses in scaling relations allows us to explore the evolution of massive structures out to higher redshifts.

This work shows the potential of \xmm\ observations in deriving X-ray observables of massive, SZE-selected clusters extending to redshifts $z>1$. With the deployment of the next generation SZE experiments \citep[e.g., SPT-3G, CMB-S4, Advanced ACTPOL;][]{benson14,CMB-S4-17,thornton16} and X-ray surveys with eRosita \citep{merloni12}, a large number of new high-redshift clusters will be discovered.  Moreover, with deep, multi-wavelength optical surveys like DES, it is already possible to use even the shallower RASS survey to probe the $z\approx1$ Universe \citep{klein18}.  X-ray follow-up observations with \xmm\ of these new clusters will provide high quality X-ray spectroscopy for a mass-complete sample at $z>1$ and would enable significant improvements in our understanding of the formation and evolution of the most massive collapsed structures in the Universe.

\section*{Acknowledgements}

Authors thank the anonymous referee and David Rapetti for helpful comments on the draft. We acknowledge the support by the DFG Cluster of Excellence ``Origin and Structure of the Universe'', the DLR award 50 OR 1205 that supported I. Chiu during his PhD project, and the Transregio program TR33 ``The Dark Universe''. The South Pole Telescope is supported by the National Science Foundation through grant PLR-1248097. Partial support is also provided by the NSF Physics Frontier Center grant PHY-1125897 to the Kavli Institute of Cosmological Physics at the University of Chicago, the Kavli Foundation and the Gordon and Betty Moore Foundation grant GBMF 947.  

This paper made use of the package \texttt{ChainConsumer} \citep{hinton16}. This work made use of the IPython package \citep{PER-GRA:2007}, SciPy \citep{jones_scipy_2001}, TOPCAT, an interactive graphical viewer and editor for tabular data \citep{2005ASPC..347...29T}, matplotlib, a Python library for publication quality graphics \citep{Hunter:2007}, Astropy, a community-developed core Python package for Astronomy \citep{2013A&A...558A..33A}, NumPy \citep{van2011numpy}.

\software{%
 \texttt{Astropy} \citep{2013A&A...558A..33A},
 \texttt{ChainConsumer} \citep{hinton16},
 \texttt{emcee} \citep{foreman13},
 \texttt{IPython} \citep{PER-GRA:2007},
 \texttt{Matplotlib} \citep{Hunter:2007},
 \texttt{NumPy} \citep{van2011numpy},
 \texttt{SciPy} \citep{jones_scipy_2001},
  \texttt{Sherpa} \citep{Freeman2001,doe07},
  \texttt{TOPCAT} \citep{2005ASPC..347...29T},
  \texttt{XSPEC} \citep{arnaud96}
}
\bibliographystyle{aasjournal}
\bibliography{literature}

\end{document}